\documentclass[iopam]{iopart}
\pdfoutput=1
\usepackage{hyperref}
\usepackage{array,multirow,graphicx}
\usepackage{iopams}
\usepackage{bm}
\usepackage{color}
\usepackage{epsfig}
\usepackage{mathrsfs}
\usepackage{float}
\renewcommand{\mathbf}{\bm}
\newcommand{\be}{\begin{equation}}
\newcommand{\ee}{\end{equation}}
\newcommand{\ba}{\begin{eqnarray}}
\newcommand{\ea}{\end{eqnarray}}
\newcommand{\barr}{\begin{array}}
\newcommand{\earr}{\end{array}}
\newcommand{\bei}{\begin{itemize}}
\newcommand{\eei}{\end{itemize}}
\newcommand{\rhom}{\rho_{_m}}
\newcommand{\rhol}{\rho_{_{\Lambda}}}
\newcommand{\omegam}{\Omega_{m}}
\newcommand{\omegak}{\Omega_{{\kappa}}}
\newcommand{\omegal}{\Omega_{\Lambda}}
\newcommand{\Hperp}{H_{\perp}}
\newcommand{\Hperpo}{H_{\perp 0}}
\newcommand{\Hparo}{H_{\parallel 0}}
\newcommand{\Hpar}{H_{\parallel}}
\newcommand{\aperp}{a_{\perp}}
\newcommand{\aperpin}{a_{\perp}^{\textnormal{in}}}
\newcommand{\aperpo}{a_{\perp 0}}
\newcommand{\apar}{a_{\parallel}}
\newcommand{\LCDM}{\Lambda\textnormal{CDM}}
\newcommand{\tb}{t_{B}}

\newcommand{\dotaperp}{\dot{a}_{\perp}}
\newcommand{\dotapar}{\dot{a}_{\parallel}}

\newcommand{\rhomtilde}{\widetilde{\rho}_{m}}
\newcommand{\kappatilde}{\widetilde{\kappa}}
\newcommand{\etatilde}{\widetilde{\eta}}
\newcommand{\Xtilde}{\widetilde{X}}
\newcommand{\rtilde}{\widetilde{r}}
\newcommand{\submin}{_{\scriptsize{\textnormal{min}}}}
\newcommand{\tinysubmin}{_{\tiny{\textnormal{min}}}}
\newcommand{\submax}{_{\scriptsize{\textnormal{max}}}}
\newcommand{\supin}{^{\scriptsize{\textnormal{in}}}}
\newcommand{\supout}{^{\scriptsize{\textnormal{out}}}}
\newcommand{\normmax}{\scriptsize{\textnormal{max}}}

\newcommand{\Gpc}{\ensuremath{\textnormal{Gpc}}}

\newcommand{\norm}[1]{\ensuremath{||#1||}}
    
\newlength{\Oldarrayrulewidth}
\newcommand{\Cline}[2]{%
  \noalign{\global\setlength{\Oldarrayrulewidth}{\arrayrulewidth}}%
  \noalign{\global\setlength{\arrayrulewidth}{#1}}\cline{#2}%
  \noalign{\global\setlength{\arrayrulewidth}{\Oldarrayrulewidth}}}

\begin{document}

\title{Evolution of linear perturbations in spherically symmetric dust spacetimes}

\author{S February$^1$, J Larena$^2$, C Clarkson$^1$ and D Pollney$^2$}
\address{$^1$ Astrophysics, Cosmology and Gravity Centre, and, Department of Mathematics and Applied Mathematics, University of Cape Town, Rondebosch 7701, South Africa.}
\address{$^2$ Department of Mathematics, Rhodes University, Grahamstown, 6140 South Africa}
\ead{\mailto{sean.february@uct.ac.za}}

\begin{abstract}
We present results from a numerical code implementing a new method to solve the master equations describing the evolution of linear perturbations in a spherically symmetric but inhomogeneous background. This method can be used to simulate several configurations of physical interest, such as relativistic corrections to structure formation, the lensing of gravitational waves and the evolution of perturbations in a cosmological void model. This paper focuses on the latter problem, i.e. structure formation in a Hubble scale void in the linear regime.  This is considerably more complicated than linear perturbations of a homogeneous and isotropic background because the inhomogeneous background leads to coupling between density perturbations and rotational modes of the spacetime geometry, as well as gravitational waves. Previous analyses of this problem ignored this coupling in the hope that the approximation does not affect the overall dynamics of structure formation in such models. We show that for a giga-parsec void, the evolution of the density contrast is well approximated by the previously studied decoupled evolution only for very large-scale modes. However, the evolution of the gravitational potentials within the void is inaccurate at more than the 10\% level, and is even worse on small scales. 
\end{abstract}

\maketitle

\section{Introduction}
We present results from a numerical code implementing a new method that solves the first order gauge-invariant linear perturbation equations in a Lema\^i tre-Tolman-Bondi (LTB) background. LTB models are spherically symmetric but inhomogeneous dust solutions of the Einstein field equations. Compared to a Friedman-Lema\^itre-Robertson-Walker (FLRW) background, perturbations around a LTB background are complicated by the fact that they cannot be decomposed into the standard scalar vector and tensor modes, as the reduction of symmetry (through radial inhomogeneity) causes these modes to couple~\cite{Clarkson:2009sc}.

The method presented here can be used to model a variety of different astrophysical/cosmological scenarios. For example:
\begin{description}

\item[Relativistic corrections for structure formation.]
Currently structure formation in cosmology is modelled either using non-linear models of Newtonian systems, or relativistically but only in the linear regime. This leaves an important area unexplored: non-linear relativistic aspects of structure formation~-- {see, e.g., \cite{Clarkson:2011zq} for a review}. Certain aspects of this have started to be taken into account in N-body methods which make use of relativistic corrections to the potentials~\cite{Bruni:2013mua,Adamek:2013wja}. Our model can be used to analyse the growth of structure \emph{on top of} a strongly non-linear background --- either an over-density such as a cluster, or a large void, both of which generate large curvature and shear.  The coupling of density perturbations to vector and tensor degrees of freedom can be explored and the errors induced by neglecting this coupling quantified. 

\item[Evolution of perturbations in void models.]
If we were to live in a large, underdense void of a few giga-parsecs in diameter, distant supernovae would appear fainter than expected in a FLRW model, and the dark energy phenomenon could be explained on purely relativistic terms without invoking any new physics (see, e.g., \cite{Enqvist:2006cg,GarciaBellido:2008nz,February:2009pv,Biswas:2010xm} and~\cite{Clarkson:2012bg} for a comprehensive review). Clearly, structure formation places a key constraint on such models by probing their difference from the concordance model, and so serves as a test of the Copernican principle~\cite{February:2012fp}. Structure formation in LTB models has only been quantified for the special case which neglects the coupling of the scalar gravitational potential to vector and tensor degrees of freedom~\cite{Tomita:1997pt,February:2012fp,Zibin:2008vj,Dunsby:2010ts,Alonso:2010zv,Sussman:2013yq}. While this seems reasonable, the accuracy has not been quantified. A recent alternative approach to this problem, based on second-order perturbation theory in FLRW, can be found here: \cite{Nishikawa:2012we}. 

\item[Weak lensing of gravitational waves]
Gravitational waves (GWs) from supermassive black hole mergers act as precise ``standard sirens,'' promising to significantly improve upon standard(-isable) candles such as Type-Ia Supernovae (SN1a) (see e.g.,~\cite{Cutler:2009qv}). However, weak lensing of the GWs by the intervening dark matter distribution distorts the signal, degrading their use as cosmological distance estimators~\cite{Shapiro:2009sr}. A particular problem is that the GW wavelength is comparable to the size of the dark matter halos which produce a portion of the lensing effect. Thus the geometric optics approximation cannot be used to model the expected lensing. By modelling a dark matter halo using a LTB model, and scattering gravitational waves off it using our method, we can hope to quantify the lensing of gravitational waves more accurately. 

\end{description}
In this paper, we focus on the linear evolution of perturbations in Gpc-void cosmological models, and present the results for this case (other scenarios shall be analysed elsewhere). To illustrate the performance of the method and the physics of the evolution of perturbations, we concentrate on polar perturbations in a large-scale cosmological void that is asymptotically Einstein-de Sitter (EdS) and that fits the distance-redshift relations given by SN1a observations as well as age data. We show results for several different spherical harmonic frequencies, considering a variety of initial conditions at different locations throughout the void. We find that couplings can only be neglected if one is interested in the matter density growth function on very large scales. Indeed, for this quantity, the evolution without couplings is accurate at the sub-percent level when the large-scale quadrupole is considered. However,  the metric perturbation that is the closest to the standard ``Newtonian'' gravitational potential gets corrections of up to 10\% even in this case (which implies percent-level inaccuracies to the 2-point correlation function given in~\cite{February:2012fp}).  Furthermore, for higher frequency perturbations even the density contrast suffers more than 10\% inaccuracies. This has significant impact on constraints on LTB models which use such approximations for analyses of structure formation.

\section{Background evolution}
The general unperturbed LTB line element may be written as,
\ba
ds^2 = -dt^2 + X^2(t,r)\,dr^2 + A^2(t,r)\,d\Omega^2\,,
\ea where \cite{Enqvist:2006cg} 
\ba
X(t,r) = \frac{\apar(t,r)}{\sqrt{1-\kappa(r)r^2}}\,,\quad A(t,r) = r\aperp(t,r)\,,\quad \apar \equiv (r\aperp)'\,, 
\ea with a prime being shorthand for $\partial_{r}$. The angular and radial scale factors, $\aperp$ and $\apar$, respectively, are associated with individual expansion rates 
\ba
\Hperp = \frac{\dotaperp}{\aperp} \;\;\,,\; \Hpar = \frac{\dotapar}{\apar}\,,
\ea where an overdot denotes $\partial_{t}$. While we only focus on a dust energy-density component in our analysis, we include the cosmological constant $\Lambda$ in our equations for completeness. The angular component of the expansion then obeys the following Friedmann equation with different curvature, i.e. $\kappa(r)$, on each radial shell: 
\ba
\Hperp^2(t,r) &=& \Hperpo^2\big[\omegam \aperp^{-3} + \omegak\aperp^{-2}+\omegal  \big] \,, \label{LTB_expansion}
\ea where
\ba
&\displaystyle\omegam(r) \equiv \frac{M(r)}{\Hperpo^2(r)}\,,~~\omegak(r) \equiv -\frac{\kappa(r)}{\Hperpo^2(r)}\,,~~\omegal(r)=\frac{\Lambda}{3\Hperpo^2(r)}\,, & \nonumber \\
&~\omegam(r)+\omegak(r)+\omegal(r)=1.&
\ea 
In these expressions, $\Hperpo(r)$ is the angular Hubble parameter today, $M(r)$ a boundary condition related to the matter content within a comoving shell of radius $r$, and we set $\aperpo(r)=1$ by convention. The expansion and shear scalars are given by
\ba
\Theta &\equiv& 2\Hperp + \Hpar\,,\\
\sigma^2 &\equiv& \frac{2}{3}\Big(\Hpar - \Hperp\Big)\,,
\ea 
respectively. The total energy density (including dust and a cosmological constant) may be expressed as 
\be
8\pi G \big(\rhom + \rhol \big) = 3\Hperpo^2\Bigg\{\frac{\omegam}{\apar\aperp^2}\Bigg[1 +  \frac{r}{3}\Bigg(2\frac{\Hperpo'}{\Hperpo} + \frac{\omegam'}{\omegam} \Bigg) \Bigg] + \omegal \Bigg\}\,.
\ee 

We require the solution (i.e. $\aperp$) to \Eref{LTB_expansion}, from which everything else follows. Integrating \Eref{LTB_expansion} we find
\be
 t - \tb(r) = \frac{1}{\Hperpo(r)}\int_0^{\aperp}\frac{dx}{\sqrt{\omegam(r)x^{-1} + \omegak(r)+\omegal(r)x^2}}\,, \label{scale_factor_sol}
\ee where $\tb(r)$ is the bang time function.
\begin{figure}[!t]
\centering
\begin{tabular}{cm{13.5cm}}
  &  \multicolumn{1}{c}{\rule{0pt}{0.3cm}\hspace{-1.4cm}$\rhom(t,r)/\rhom(t,\infty)-1$\hspace{0.5cm}$\Hperp(t,r)/\Hperp(t,\infty)-1$\hspace{0.5cm}$\Hpar(t,r)/\Hpar(t,\infty)-1$}  \\ [0.7ex]
 \multicolumn{1}{c}{} &  \multicolumn{1}{c}{\rule{0pt}{0.3cm}\hspace{-3.9cm}\tiny{}$[B,A]=$\hspace{0.7cm}\tiny{}$[-6.8,0.3]\!\times\!10^{-1}$\hspace{1.9cm}$[2.5\!\!\times\!\!10^{-6}\!,2.7]\!\times\!10^{-1}$\hspace{1.5cm}$[-0.9,2.7]\!\times\!10^{-1}$} \\
\multicolumn{1}{c}{\parbox[t]{3.0mm}{{\rotatebox[origin=c]{90}{\hspace{0.46cm}\Large{Cosmic time (Gyr) $\;\;\longrightarrow$}}}}}  &  \begin{tabular}{c} \rule{0pt}{3.0cm}{\includegraphics[width=0.95\textwidth]{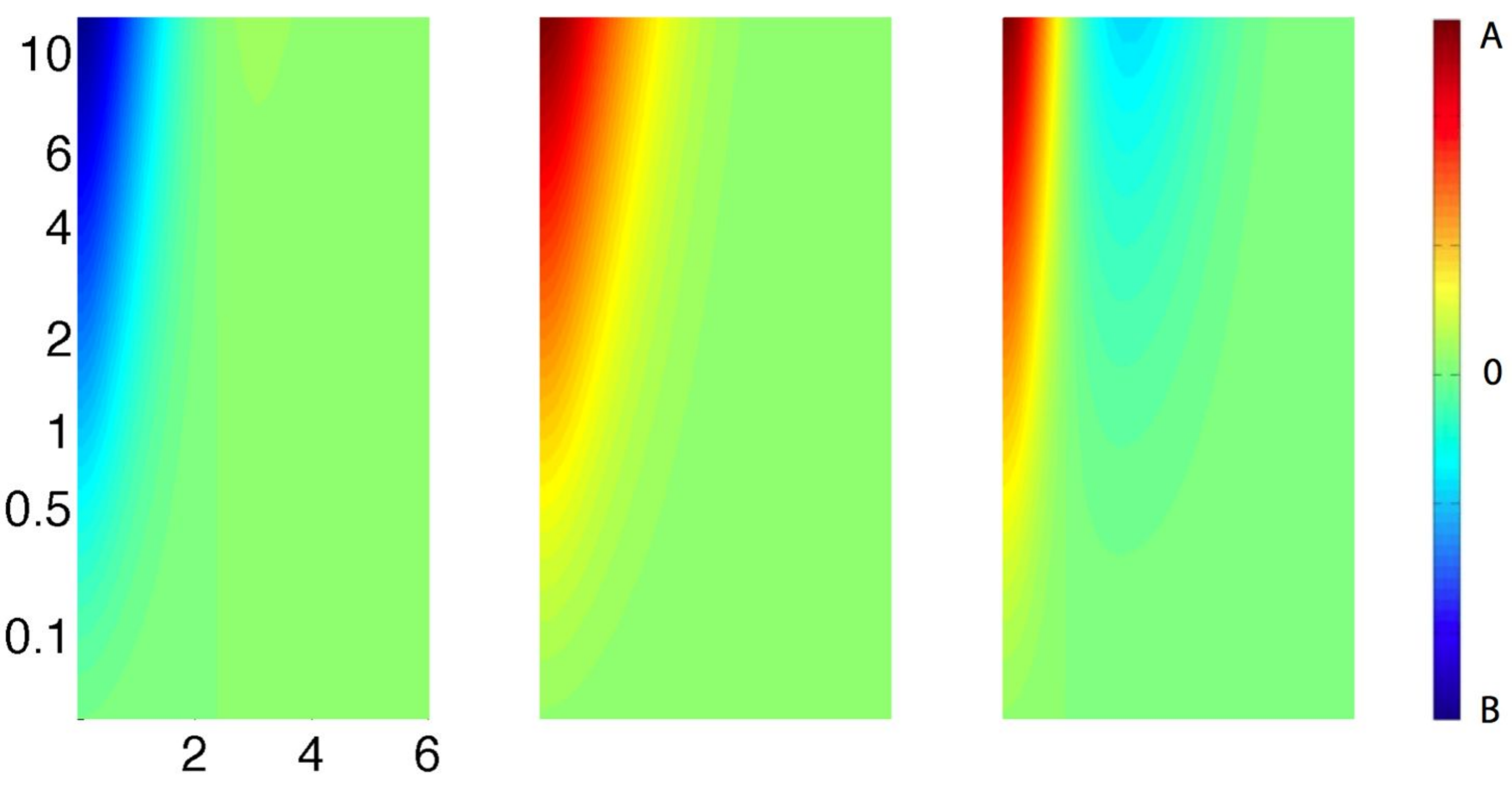}}  \end{tabular} \\
  \multicolumn{1}{c}{} & \multicolumn{1}{c}{\Large{Radial distance (Gpc) $\longrightarrow$}} \\ [-0.5ex]
\end{tabular}
\caption{The spacetime evolution of selected contrasts in the background dynamics, illustrating the growth of the background void over time. \textit{Left:} Contrast in the energy density $\rhom$. \textit{Centre:} Contrast in $\Hperp$. \textit{Right:} Contrast in $\Hpar$. Note that scales of the vertical and horizontal axes apply to all such 2D plots in this paper. The values of $A$ and $B$ (respectively the maximum and minimum of the color scale) can be read at the top of each 2D plot.}
\label{Tab:background}
\end{figure}

In this work we consider a test case to demonstrate the method, and so focus on an open ($\omegak>0$), dust-only background model ($\Lambda=0$) that is asymptotically Einstein-de Sitter, and which is known to comfortably accommodate distances to SN1a (see e.g. \cite{February:2009pv}). Here we model the total (dimensionless) matter density profile today according to:
 \ba
 \omegam(r) = \omegam\supout - (\omegam\supout-\omegam\supin)\exp\Big[{-r^2}/{L^2}\Big]\,,
\ea
where $\omegam\supin=0.2,~\omegam\supout=1.0,~L = 2.0\,{\rm Gpc}\,$ is the length-scale of the inhomogeneity, and $H_0\equiv\Hperpo\supin=\Hparo\supin\;$= 70 km/s/Mpc $=$ 0.23 Gpc$^{-1}$ is the Hubble constant. The superscript ``in'' denotes evaluation at $r=0$. For consistency with an inflationary paradigm, we ignore decaying modes by forcing the bang time function to be uniform throughout space; setting $\tb(r)=0$ is sufficient. In this case, we may write the solution to \Eref{scale_factor_sol} in the following parametric form
\ba
\aperp(t,r)  &=&  \frac{\omegam(r)}{2\omegak(r)}\big[\cosh{2 u(t,r)} - 1\big] \,,\\
 \qquad t  &=& \frac{\omegam(r)}{2\Hperpo(r)}\frac{\big[\sinh{2 u(t,r)} - 2u(t,r)\big]}{\big[ \omegak \big]^{3/2}}  \,,
\ea where
\ba
\Hperpo(r) &= \frac{\omegam(r)}{2t_0}\frac{\big[\sinh{2u_0(r)} - 2u_0(r)\big]}{\big[\omegak(r)\big]^{3/2}}   \,,\\
\;\;\; u_0(r) &= \frac{1}{2}\cosh^{-1}\Big[\frac{2}{\omegam(r)}-1\Big]\,,\\
\;\;\;\;\;\;\;\; t_0 &= \frac{\omegam\supin}{2H_0}\frac{\big[\sinh{2u_0\supin} - 2u_0\supin\big]}{\big[\omegak\supin\big]^{3/2}} \,.
\ea Figures 1 and 2 show the behaviour of various illustrative quantities in this model for the values of the constants quoted above.

\begin{figure}[!t]
\centering
\begin{tabular}{cm{6.0cm}}
  &  \multicolumn{1}{c}{\rule{0pt}{0.3cm}\hspace{-0.2cm}Shear scalar ($\sigma^2$)}  \\ [0.4ex]
 \multicolumn{1}{c}{} &  \multicolumn{1}{c}{\rule{0pt}{0.1cm}\tiny{}\hspace{-3cm}$[B,A]=$\hspace{1cm}\tiny{}$[-8.8,0.0]\!\times\!10^{-2}$} \\[-0.3ex]
\multicolumn{1}{c}{\parbox[t]{2.0mm}{{\rotatebox[origin=c]{90}{\hspace{0.9cm}\Large{Cosmic time (Gyr) }}}}}  &  \begin{tabular}{c} \rule{0pt}{3.0cm}{\includegraphics[width=0.33\textwidth]{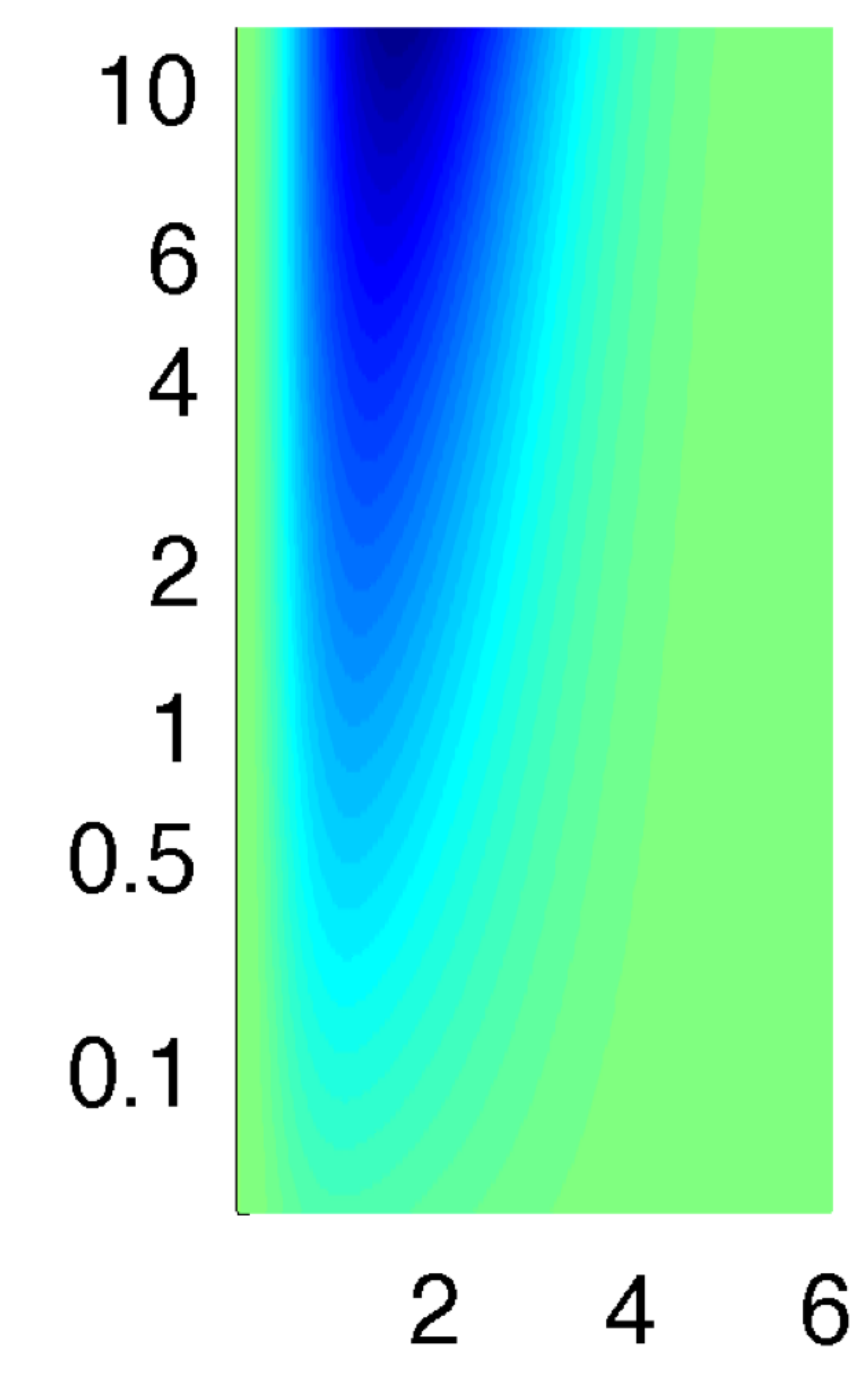}}  \end{tabular} \\
  \multicolumn{1}{c}{} & \multicolumn{1}{c}{\hspace{-0.2cm}\Large{Radial distance (Gpc) }} \\ [-0.5ex]
\end{tabular}
\caption{The spacetime evolution of the shear scalar, $\sigma^2$. In the vicinity of the origin and far outside the void, the spacetime is effectively FLRW (depicted in green). The presence of propagating modes is more apparent in all the variables here. The maximum and minimum values of the colour scale are given in brackets above.}
\label{Tab:background}
\end{figure}

\section{Polar perturbations}
Perturbations on a spherically symmetric background are decomposed into spherical harmonics, where tensorial quantities of degree 1 or 2 split into two parities~-- polar and axial (this is analogous, but not identical, to the usual scalar-vector-tensor split in a FLRW background).
In this paper, we restrict our attention to the polar (even parity) sector since this is where the density perturbation is defined. Furthermore, we only consider modes with spherical harmonic index $l>1$. The modes corresponding to $l=0$ and $l=1$ obey different equations (see \cite{Clarkson:2009sc}) and should be treated separately, although in a similar fashion as far as the numerical integration is concerned. A similar treatment could be straightforwardly applied to the axial (odd parity) sector. Details of the derivation of the perturbation equations we present below can be found in \cite{Clarkson:2009sc}. Let us emphasize here that we are not trying to develop a full analysis of realistic structure formation in a LTB Universe. Rather, we would like to demonstrate that the method we develop can integrate the perturbation equations, allowing us to study a few remarkable features of the evolution of perturbations in a large cosmic void. Note that we have adapted the equations to include the cosmological constant. We have also written them in terms of partial radial derivatives rather than frame derivatives, in readiness for numerical integration.
 
\subsection{Formalism}

The general form of polar perturbations to the background LTB metric, in the Regge-Wheeler (RW) gauge, is expanded in spherical harmonics as:
\ba
\label{metric}
ds^2 \!\!&=&\!\! -\Big[1 + \Big(2\eta - \chi - \varphi\Big)Y \Big]dt^2  - 2\varsigma Y X dtdr \nonumber \\
&&+ \Big[1 + \Big(\chi+\varphi\Big)Y \Big]X^2dr^2 + \Big[1 + \varphi Y\Big]A^2d\Omega^2\,, \nonumber \\
\ea where $\chi$, $\varphi$ and $\varsigma$ are functions of $(l,t,r)$ (independent of $m$ due to the spherical symmetry of the background), $Y=Y^{(l m)}(\theta,\phi)$ are the scalar spherical harmonic functions. For notational convenience, an implicit sum over $(l,m)$ is implied whenever a quantity is multiplied by $Y$. The 4-velocity field in the polar sector is given by (indices which are capitals run over $t,r$, and indices $a,b,c,\ldots$ run over $\theta,\phi$)
\ba
u_{\mu} = \Bigg[\hat{u}_{A} + \Bigg(w \hat{n}_{A} + \frac{1}{2}k_{AB} \hat{u}^B \Bigg)Y, vY_a\Bigg] \,,
\ea where $\hat{u}^A = (1,0)$ and $\hat{n}^A = (0,X^{-1})$, $w$ and $v$, both functions of $(l,t,r)$, are the radial and angular (peculiar) velocities, respectively, $k_{AB}$ is the gauge-invariant metric perturbation, and $Y_a \equiv \nabla_{a}Y$. The  energy-momentum tensor
\ba
{T^{\mu}}_{\nu} = \rhom (1+ \Delta Y)u^{\mu}u_{\nu} \,,
\ea
defines the density contrast $\Delta$. Note that all our perturbation variables, i.e. both metric and fluid perturbations, are automatically gauge-invariant. This is due to the fact that the all perturbations conveniently reduce, in the RW gauge, to the corresponding variables arising from a general gauge (coordinate) transformation (see \cite{Gundlach:1999bt} and references therein).
 
 The 1st-order perturbed Einstein equations for the case $l\ge2$  reduce to:
\ba
  \ddot{\varphi}  &=& - 4\Hperp\dot{\varphi}   - \Hperp\dot{\chi} + \frac{\apar}{X^2\aperp r} \chi'  \nonumber\\
&& + \bigg[\frac{2\kappa}{\aperp^2}  - \Lambda \bigg]\varphi + \frac{3\apar\sigma^2}{X\aperp r} \varsigma 
+ \Bigg[\frac{2\kappa}{\aperp^2} - \Lambda + \frac{l(l+1)-2}{2\aperp^2 r^2}\Bigg] \chi  \,,  \label{varphi_evol_eq} \\ 
\dot{\varsigma}  &=& -2\Hpar\varsigma - X^{-1}\chi' \,,  \label{varsigma_evol_eq} \\
\ddot{\chi}  &=&  X^{-2}\chi''  -  X^{-2}\Bigg[\frac{{\apar}'}{\apar} + \frac{\kappa r + \frac{1}{2}r^2\kappa'}{1-\kappa r^2} +2\frac{\apar}{\aperp r}\Bigg] \chi'  +  3X^{-1}\sigma^2 \varsigma' \nonumber \\
&& -  6\sigma^2\dot{\varphi} - 3\Hpar\dot{\chi} + \Bigg[4\Bigg(\frac{\aperp}{\apar}-1\Bigg)\frac{\kappa}{\aperp^2} +\frac{2r\kappa'}{\aperp \apar} \Bigg](\chi + \varphi)  \nonumber \\
&& +  2X^{-1}\Big[{\Hpar} - 2{\Hperp} \Big]' \varsigma - \Bigg[\frac{l(l+1) - 2}{\aperp^2 r^2} \Bigg]\chi  \label{chi_evol_eq} \,.
\ea 
The three equations \eref{varphi_evol_eq}, \eref{varsigma_evol_eq} and \eref{chi_evol_eq} represent the master equations of our problem, and are the evolution equations for the metric perturbations $\varphi$, $\varsigma$ and $\chi$. Knowing these master variables, one can then obtain $\Delta$, $w$ and $v$, i.e. the behaviour of the matter perturbations. These are given by

\ba
8\pi G \rhom \Delta \!&=&\!  - X^{-2}\varphi'' + X^{-2}\Bigg[\frac{{\apar}'}{\apar} + \frac{\kappa r + \frac{1}{2}r^2\kappa'}{1-\kappa r^2} -2\frac{\apar}{\aperp r}\Bigg] \varphi'  \nonumber \\
&& + 2X^{-1}\Hperp \varsigma' + X^{-2}\frac{\apar}{\aperp r} \chi' + \Theta\dot{\varphi}  + \Hperp \dot{\chi} \nonumber \\
&& +  \Bigg[3\Hperp\Big(\sigma^2 + \Hperp \Big) - \Big(1+2\frac{\aperp}{\apar}\Big)\frac{\kappa}{\aperp^2} - \frac{r\kappa'}{\aperp\apar}\nonumber \\
&& + \frac{l(l+1)}{\aperp^2 r^2}\Bigg]\Big(\varphi + \chi\Big)- \;\Bigg[\frac{l(l+1)-2}{2\aperp^2 r^2}\Bigg] \chi \nonumber\\
&& + X^{-1}\frac{\apar}{\aperp r}\Big(3\sigma^2+4\Hperp\Big) \varsigma  \,,  \label{poiss_LTB} \\
8\pi G \rhom w  &=& X^{-1}\Bigg[\dot{\varphi}' - \left(3\sigma^2 - \Hpar\right)\varphi' - \frac{\apar}{\aperp r}\dot{\chi} + \Hperp\chi'\Bigg] \nonumber\\
&&+ \Bigg[\frac{3}{2}\Hperp\Big(\sigma^2 + \Hperp \Big)  -\bigg(\frac{\aperp}{\apar} - \frac{1}{2}\bigg)\frac{\kappa}{\aperp^2}  \nonumber \\
&& -  \frac{r\kappa'}{2\aperp\apar} + \frac{l(l+1)}{2\aperp^2 r^2} - \frac{\Lambda}{2} \Bigg] \varsigma \,,\\ 
 8\pi G \rhom v &=& \dot{\varphi} + \frac{1}{2}\dot{\chi} + \frac{1}{2X}\varsigma'  + \Hpar\Big(\varphi  + \chi\Big) \,. \label{ang_vel_LTB}
\ea
The conservation of the perturbed energy-momentum, i.e. $\nabla_{\mu}{T^{\mu}}_{\nu}=0$, implies that our solutions must satisfy:
\ba
C_{\Delta} &\equiv& \dot{\Delta} +\frac{3}{2}\dot{\varphi} + \frac{1}{2}\dot{\chi} + X^{-1}\Big(w + \varsigma/2\Big)' \nonumber \\
&&+ X^{-1}\Big[\frac{{\rhom}'}{\rhom} + 2\frac{\apar}{\aperp r} \Big]\Big(w + \varsigma/2\Big) - \frac{l(l+1)}{\aperp^2 r^2}v\, \label{delta_constraint} = 0 \,,\\
C_{w} &\equiv& \dot{w} - \frac{1}{2X}\varphi' + \Hpar\Big(w + \varsigma/2\Big) = 0\,, \label{w_constraint}\\
C_{v} &\equiv& \dot{v} - \frac{1}{2}\Big(\varphi + \chi\Big) = 0 \,. \label{v_constraint}
\ea
These three equations can thus be seen as constraints that a solution to the previous system, i.e. Eq.'s \Eref{varphi_evol_eq}$-$\Eref{chi_evol_eq} along with \Eref{poiss_LTB}$-$\Eref{ang_vel_LTB}, must satisfy. These constraints can be used to test the accuracy of the numerical implementation. 

\subsection{Weyl curvature}
In studying perturbations of a LTB background, it is interesting to consider the evolution of the Weyl curvature tensor, as it describes genuine relativistic effects. In particular the magnetic part of the Weyl tensor is zero in the background and is a gauge-invariant tensor sourced by purely relativistic effects --- frame dragging (vector modes in FLRW perturbation theory, for example) and gravitational waves.

In the background $H_{\mu\nu}$ is zero, and the only non-zero background parts for $E_{\mu\nu}$ are:
\ba
\hat{E}_{rr} \!&=&\! X^{2}\left[\Hperp \sigma^2 + \frac{2}{3}\left(\frac{\aperp}{\apar} - 1\right)\frac{\kappa}{\aperp^2} + \frac{1}{3}\frac{r\kappa'}{\aperp\apar} \right]\,, \\
\hat{E}_{ab} \!&=&\! -\frac{1}{2}\bigg(\frac{A}{X}\bigg)^2\hat{E}_{rr}\gamma_{ab}\,.
\ea

The non-zero perturbed parts for the electric and magnetic Weyl tensors are:
\ba
\delta E_{rr} &=& -\frac{1}{3}\Bigg\{ \varphi'' - \Bigg[\frac{{\apar}'}{\apar} + \frac{\kappa r + \frac{1}{2}r^2\kappa'}{1-\kappa r^2} + \frac{\apar}{\aperp r}\Bigg] \varphi'  - 2X\Hperp \varsigma'\nonumber\\
&& - \frac{\apar}{\aperp r}\chi' - \frac{3}{2}\sigma^2\dot{\varphi}  - X^2\Hperp\dot{\chi}  + X^2\bigg[\bigg(\frac{\aperp}{\apar} -1 \bigg)\frac{\kappa}{\aperp^2}   \nonumber \\
&&     - 3\Hperp\sigma^2 + \frac{r\kappa'}{\aperp \apar} + \frac{l(l+1)}{2\aperp^2 r^2}\bigg](\varphi  + \chi)  \nonumber \\
&& - 2X\frac{\apar}{\aperp r}(3\sigma^2 - \Hpar)\varsigma  - X^2 \bigg[ \frac{l(l+1)-2}{\aperp^2 r^2} \bigg]\chi \Bigg\} \,, \\
\delta E_r \!&=&\! -\frac{1}{2}\bigg[ \varphi' - \frac{\apar}{\aperp r}\big(\varphi + \chi \big) - X\Hperp \varsigma \bigg]  \,,
\ea
\ba
\delta E_{(T)} \!&=&\! -\frac{1}{2}\bigg(\frac{A}{X}\bigg)^2\delta E_{rr}  + \frac{A^2}{3}\bigg[\frac{3}{2}\Hperp\sigma^2 + \bigg(\frac{\aperp}{\apar} - 1\bigg) \frac{\kappa}{\aperp^2} \nonumber \\
&& + \frac{1}{2}\frac{r\kappa'}{\apar\aperp}  \bigg]\chi  \,, \\
\delta E_{(TF)} \!&=&\! -\frac{1}{2}\Big(\varphi + \chi\Big)\,,
\ea and 
\ba
\overline{\delta H}_r \!&=&\! -\frac{1}{4}\varsigma' - \frac{1}{4}X\dot{\chi}  - \frac{3}{4}X\sigma^2\big(\varphi+\chi\big) + \frac{1}{2}\frac{\apar}{\aperp r}\varsigma \,,\\
\overline{\delta H}_{(TF)} \!&=&\! - \frac{1}{2}\varsigma \,, \label{deltaHTF}
\ea respectively.
Here the trace ($T$) and trace free ($TF$) parts arise from a decomposition on the 2-sphere. The magnetic part of the Weyl tensor has a parity opposite to the electric part, and carries a bar to denote that the axial part of it appears in the polar equations (and vice versa). 

\section{Numerical Implementation}
We use a method of lines approach~\cite{leveque1992numerical} to solving the system \Eref{varphi_evol_eq}$-$\Eref{chi_evol_eq}, whereby the spatial domain is discretized by standard finite differences and integrated pointwise in time using a 4th-order Runge-Kutta solver. The system can be recast in terms of dimensionless variables through the following transformations:
\ba
\rtilde &\equiv& H_0 r\,,\\
\etatilde &\equiv& H_0\eta \,=\, H_0 \int \frac{dt}{\aperp\supin(t)}\,,
\ea where the last equality, motivated by the standard form, defines the conformal time $\eta$ for central observers. Then, using
\ba
\partial_{t} &=& \frac{H_0}{\aperp\supin}\partial_{\etatilde} \,,\\
\partial_{r} &=& H_0 \partial_{\rtilde}\,,
\ea we have
\ba
H_{\perp/\parallel} &=& \frac{H_0}{\aperp\supin}\widetilde{\mathcal{H}}_{\perp/\parallel} \,,\\
\kappa &=& H_0^2\kappatilde\,,\\
\rhom &=& \Bigg(\frac{H_0}{\aperp\supin}\Bigg)^2 \rhomtilde\,,\\
X &=& \aperp\supin\Xtilde \,.
\ea{}
We also introduce the dimensionless angular peculiar velocity,
\be
\tilde{v} \equiv \frac{\aperp\supin}{H_0}v\,.
\ee
and relate the cosmic time $t$ to the dimensionless conformal time $\etatilde$ by
\ba
t &=& \frac{\omegam\supin}{2\left[\omegak\supin\right]^{3/2}}\left[\sinh\left(\etatilde\sqrt{\omegak\supin}\right) - \etatilde\sqrt{\omegak\supin} \right] \,.
\ea
Thus, to evolve the system from some initial time ($t\submin$) until today ($t_0$),
we compute the corresponding initial and final values for $\etatilde$, i.e. $\etatilde\submin$ and $\etatilde_{0}$, respectively, using:
\ba
\etatilde(t) &=& \frac{2}{\sqrt{\omegak\supin}}u\supin(t) \,.
\ea

\subsection{Discretisation}
We introduce discretized time and space coordinates $\tilde{\eta}_{i}$
and $\tilde{r}_{j}$ given by
\ba
  \etatilde_{i} &=& \etatilde\submin + i{\Delta \etatilde}/\alpha\,,{}
  \label{discrete-eta}\\
  \rtilde_{j} &=& \rtilde\submin + j{\Delta \rtilde}/\alpha\,,{}
  \label{discrete-r}
\ea{}
where ${\Delta\etatilde}$ and ${\Delta \rtilde}$ are grid spacings in
$\eta${} and $r$, respectively, and $i=0...N_{\tilde{\eta}}$ and
$j=0...N_{\tilde{r}}$. The factor $\alpha$ determines the grid resolution
relative to an $\alpha=1$ baseline (with the number of grid points
$N_{\tilde{\eta}}$ and $j=0...N_{\tilde{r}}$ increased proportionally
to cover the same domain). For our system of equations, the
Courant-Friedrics-Lewy (CFL) condition requires that
\begin{equation}
\frac{{\Delta \etatilde}}{{\Delta \rtilde}} \lesssim \Xtilde\,,
\end{equation}
for numerical stability, where $\Xtilde\epsilon\{1,1.47\}$. For our simulations we use 
\begin{equation}
  \frac{{\Delta \etatilde}}{{\Delta \rtilde}} = 0.98\alpha^{-1} \,.
\end{equation}

Spatial derivatives are calculated using 2nd-order finite difference
operators. For some quantity $Q_{i,j}$ evaluated at time
$\tilde{\eta}_{i}$ and position $\tilde{r}_{j}$ at the baseline resolution,
\ba
H_0^{-1}Q'_{i,j} &=& \frac{Q_{i,j+1} - Q_{i,j-1}}{2\Delta \tilde{r}} + \mathcal{O}(\Delta \tilde{r}^2) \,,
\label{eq:fd-1}\\
H_0^{-2}Q''_{i,j} &=& \frac{Q_{i,j+1} - 2Q_{i,j} + Q_{i,j-1}}{\Delta \tilde{r}^2} + \mathcal{O}(\Delta \tilde{r}^2)
\label{eq:fd-2} \,.
\ea{}
We evaluate the RHS of Eqs.~\Eref{varphi_evol_eq}$-$\Eref{chi_evol_eq} on a $\etatilde=\textnormal{constant}$ slice and evolve forward in time using a standard 4th-order Runge-Kutta integrator. The overall scheme is 2nd-order accurate due to the spatial finite differencing used.

\begin{figure}[b]
  \begin{center}
   \includegraphics[width=0.492\textwidth]{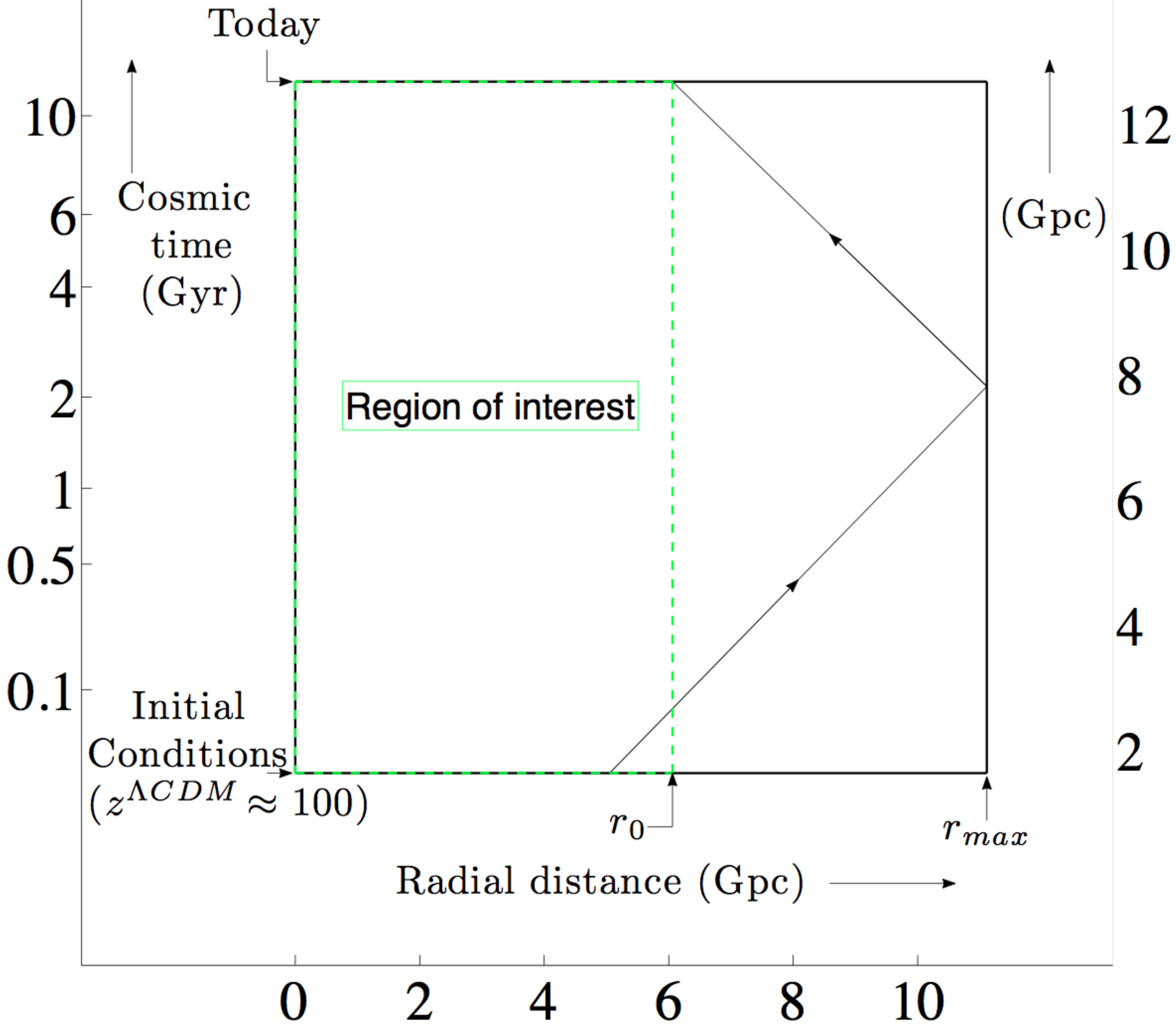}~~~~~~~~~~\includegraphics[width=0.34\textwidth]{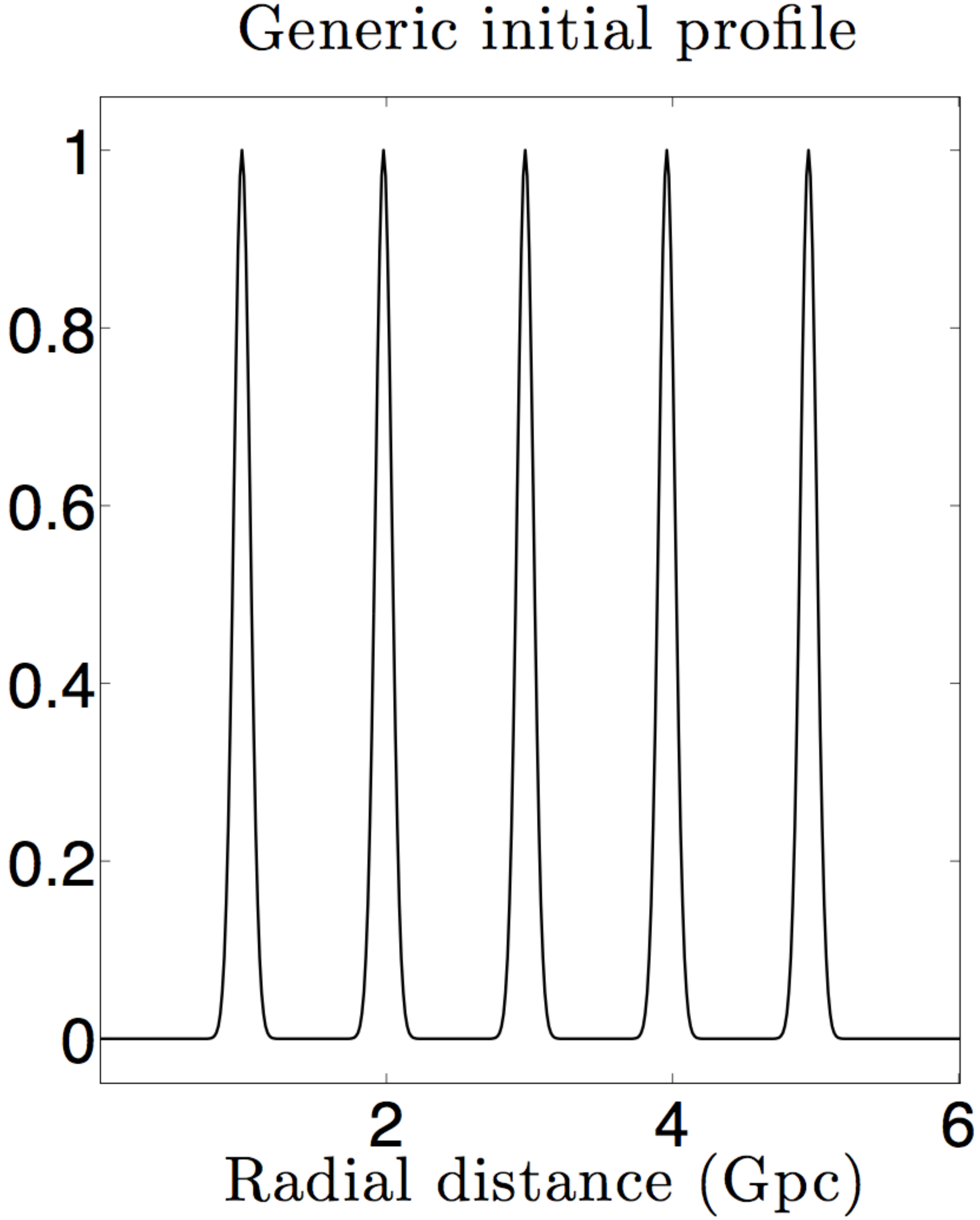}
  \end{center}
  \caption{An illustration of the spacetime domain of the model (left), along with the generic form of the initial conditions used (right), according to Eq.~\Eref{ICprofile}.}
    \label{fig:numericalgrid}
\end{figure}

\subsection{Initial and boundary conditions}

For this work, we have used a generic set of initial conditions for each of the three master variables:
\ba
Q_{0,j} &=& \sum^5_{k=1}\exp\left(-\frac{(r_{j}-p_{k})^2}{s^2} \label{ICprofile}
\right)\,,\\
\dot{Q}_{0,j} &=& 0 \,, \label{ICvelocity}
\ea
where $p_{k} \equiv 0.99\times\{1,2,3,4,5\}$ Gpc is an array of five equally spaced points between $r\submin$ and our desired region of interest  $r_{0}$, and $s\equiv 0.08$ Gpc sets the width of each pulse $-$ see the right panel of Fig. \ref{fig:numericalgrid}. The conditions are set at $\etatilde\submin=0.42$ (corresponding to a time $t\submin=0.018$ Gyr, or a redshift $z\approx100$ in a fiducial $\LCDM$ cosmology). These initial data are not meant to represent a realistic physical state of perturbations in the early Universe; they simply allow us to test the method and to extract the physical behavior of perturbations.

Regularity conditions determine the variables in the neighbourhood of
the origin according to the prescription of~\cite{Gundlach:1999bt}. Near $r=0$, we require (for all $l \ge 2$):
\ba
\chi \equiv r^{l+2} \hat{\chi}\,,\;\; \varphi \equiv r^{l}\hat{\varphi}\,,\;\; \varsigma \equiv r^{l+1} \hat{\varsigma}\,,
\ea where the hatted variables are all polynomials of even power in
$r$. Using
\ba
\hat{\chi} = \sum_{n=0}^{\infty}a_n r^{2n}\,, \quad \hat{\varphi} = \sum_{n=0}^{\infty}b_n r^{2n}\,, \quad \hat{\varsigma} = \sum_{n=0}^{\infty}c_n r^{2n}\,, 
\ea
we find that
\ba
\chi_{,r} &=& \sum_{n=0}^{\infty}(l+2n + 2)a_n r^{l+n+1}\,, \\ 
\chi_{,rr} &=& \sum_{n=0}^{\infty}(l+2n + 2)(l+2n + 1)a_n r^{l+n}\,, \\
\varphi_{,r} &=& \sum_{n=0}^{\infty}(l+2n)b_n r^{l+2n-1}\,, \\
\varsigma_{,r} &=& \sum_{n=0}^{\infty}(l + 2n + 1)c_n r^{l+2n}\,,
\ea
which vanish at $r=0=r\submin$. Fixing the value of all variables to zero at the origin is sufficient for regularity\footnote{Note that while our choice of initial conditions are not exactly zero at the origin, they are well below the machine precision.}

We require an additional boundary condition at the outer edge of the
computational domain, $r\submax$. This boundary condition is
necessarily artificial since we do not compactify the spatial coordinate.
We place the boundary at a distance from the origin such that it is causally disconnected from the region where the initial perturbations (\ref{ICprofile}) and (\ref{ICvelocity}) are set for the duration of the evolution. This prevents artificially reflected signals from influencing the evolution of these perturbations. We can estimate an appropriate distance by tracing null geodesics inward from the outer boundary. Using the background LTB line element, radial null geodesics are given by
\be
  \frac{d\etatilde}{d\rtilde} =  \Xtilde\,,
\ee
where our characteristics approach 45$^{\circ}$ at large distances in our coordinates. An appropriate value for $\rtilde\submax$ which is sufficiently removed from the measurement domain is
\ba
\tilde{r}\submax = \tilde{r}_0 +
  \frac{1}{2}\int_{\etatilde{i}}^{\etatilde_{0}}d\etatilde\, \Xtilde^{-1}\,,
  \label{R_boundary}
\ea
where $r_{0}$ is the outer boundary of the domain in which we would like to analyse the behaviour of perturbations between times $\etatilde\submin$ and $\etatilde_{0}$. Since we are working in a single spatial dimension, this grid extension to remove outer boundary effects is not overly costly in terms of memory or computation time, though in the future it may be useful to consider a logarithmic radial coordinate. Given that the spacetime in our model is effectively homogeneous above $r=5 \,\Gpc$, we choose a conservative region of interest of $6\,\Gpc$. 

\begin{figure*}[!ht]
\centering
\begin{tabular}{|m{6.5cm}|m{6.5cm}|}
\Cline{0.3pt}{1-2}
 \multicolumn{1}{|c|}{\rule{0pt}{0.36cm}\hspace{0.4cm}$\mathcal{C}^{(n)}_{\Delta}\!\times\!(n/8)^2$\hspace{0.3cm}$\mathcal{C}^{(n)}_{w}\!\times\!(n/8)^2$\hspace{0.4cm}$\mathcal{C}^{(n)}_{\tilde{v}}\!\times\!(n/8)^2\!$}  & \multicolumn{1}{|c|}{\rule{0pt}{0.36cm}\hspace{0.4cm}$\mathcal{C}^{(n)}_{\Delta}\!\times\!(n/8)^2$\hspace{0.3cm}$\mathcal{C}^{(n)}_{w}\!\times\!(n/8)^2$\hspace{0.4cm}$\mathcal{C}^{(n)}_{\tilde{v}}\!\times\!(n/8)^2\!$}   \\ 
 \Cline{0.3pt}{1-2}
 \multicolumn{1}{|c|}{} & \multicolumn{1}{c|}{}\\[-2.0ex]
 \multicolumn{1}{|c|}{\small{}\hspace{0.6cm}$\mathbf{l = 2}$} & \multicolumn{1}{c|}{\small{}\hspace{0.6cm}$\mathbf{l = 10}$}\\[-1.0ex]
\hspace{-0.35cm}\begin{tabular}{c} \rule{0pt}{4.1cm}{\includegraphics[width=0.55\textwidth]{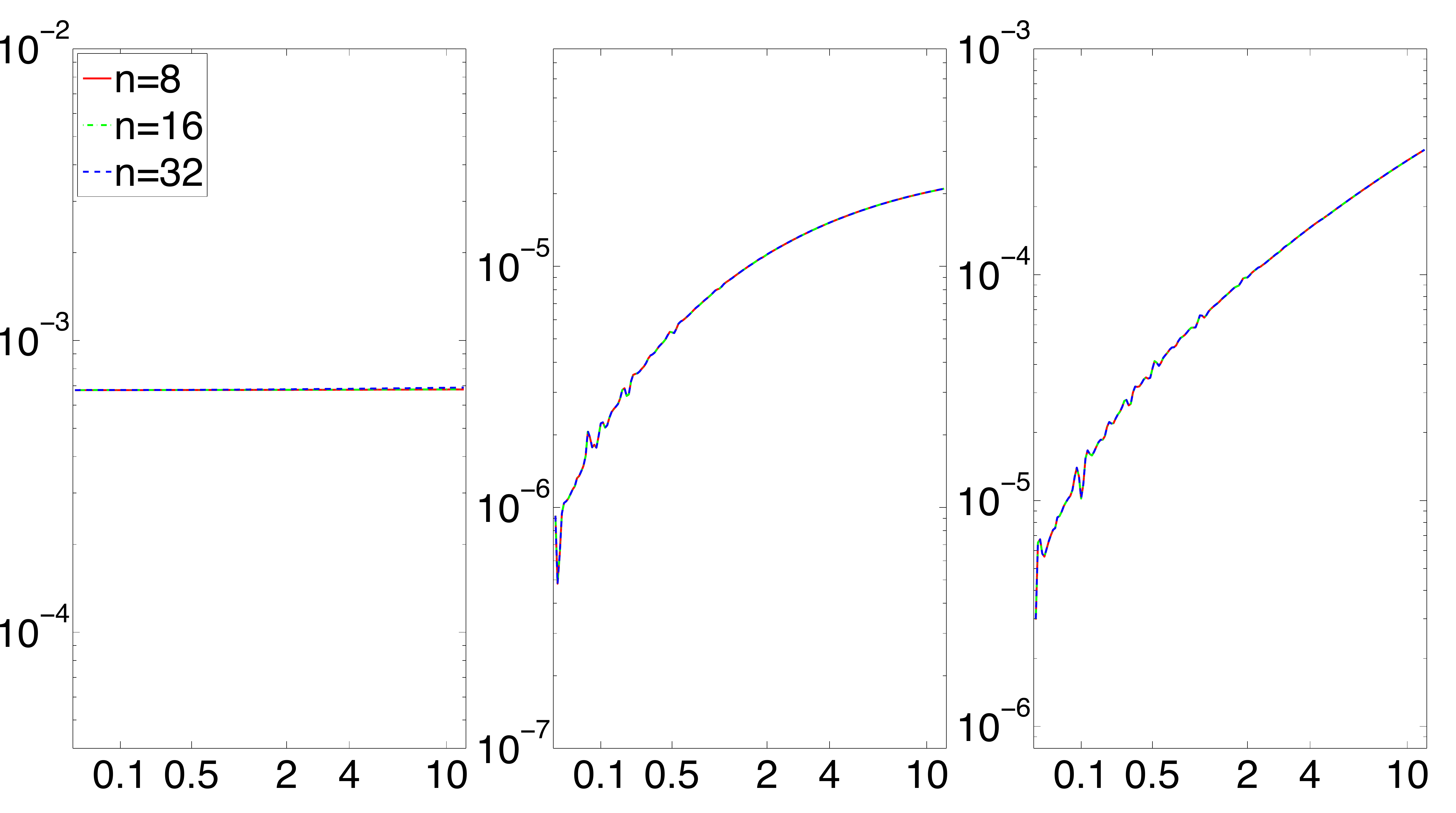}}  \end{tabular} & \hspace{-0.35cm}\begin{tabular}{c} \rule{0pt}{4.1cm}{\includegraphics[width=0.55\textwidth]{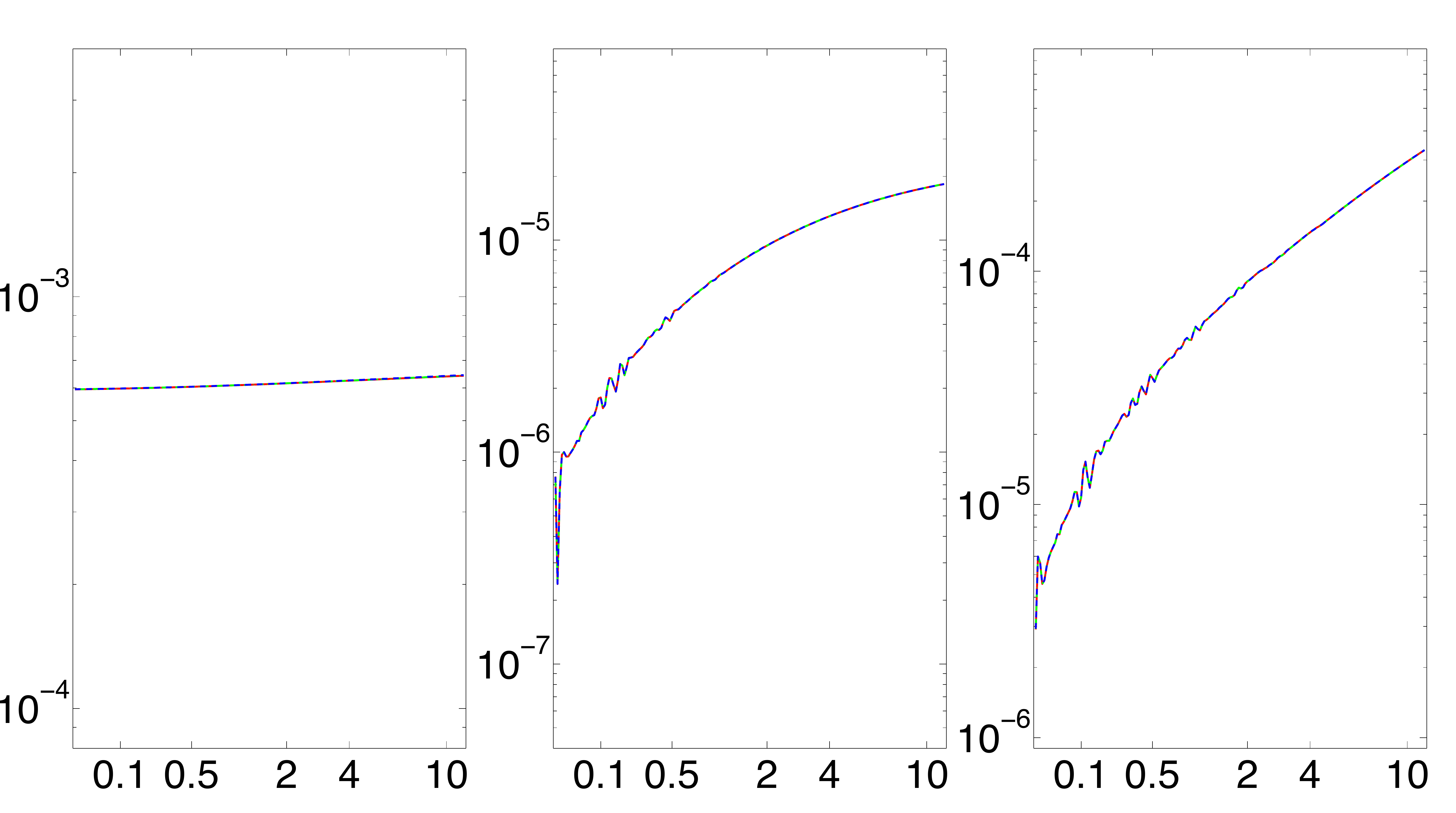}}  \end{tabular} \\ 
\Cline{0.3pt}{1-2}
 \multicolumn{1}{|c|}{} & \multicolumn{1}{c|}{}\\[-2.0ex]
 \multicolumn{1}{|c|}{\small{}\hspace{0.5cm}$\mathbf{l = 200}$} & \multicolumn{1}{c|}{\small{}\hspace{0.4cm}$\mathbf{l = 1000}$}\\[-1.0ex]
\hspace{-0.35cm}\begin{tabular}{c} \rule{0pt}{4.1cm}{\includegraphics[width=0.55\textwidth]{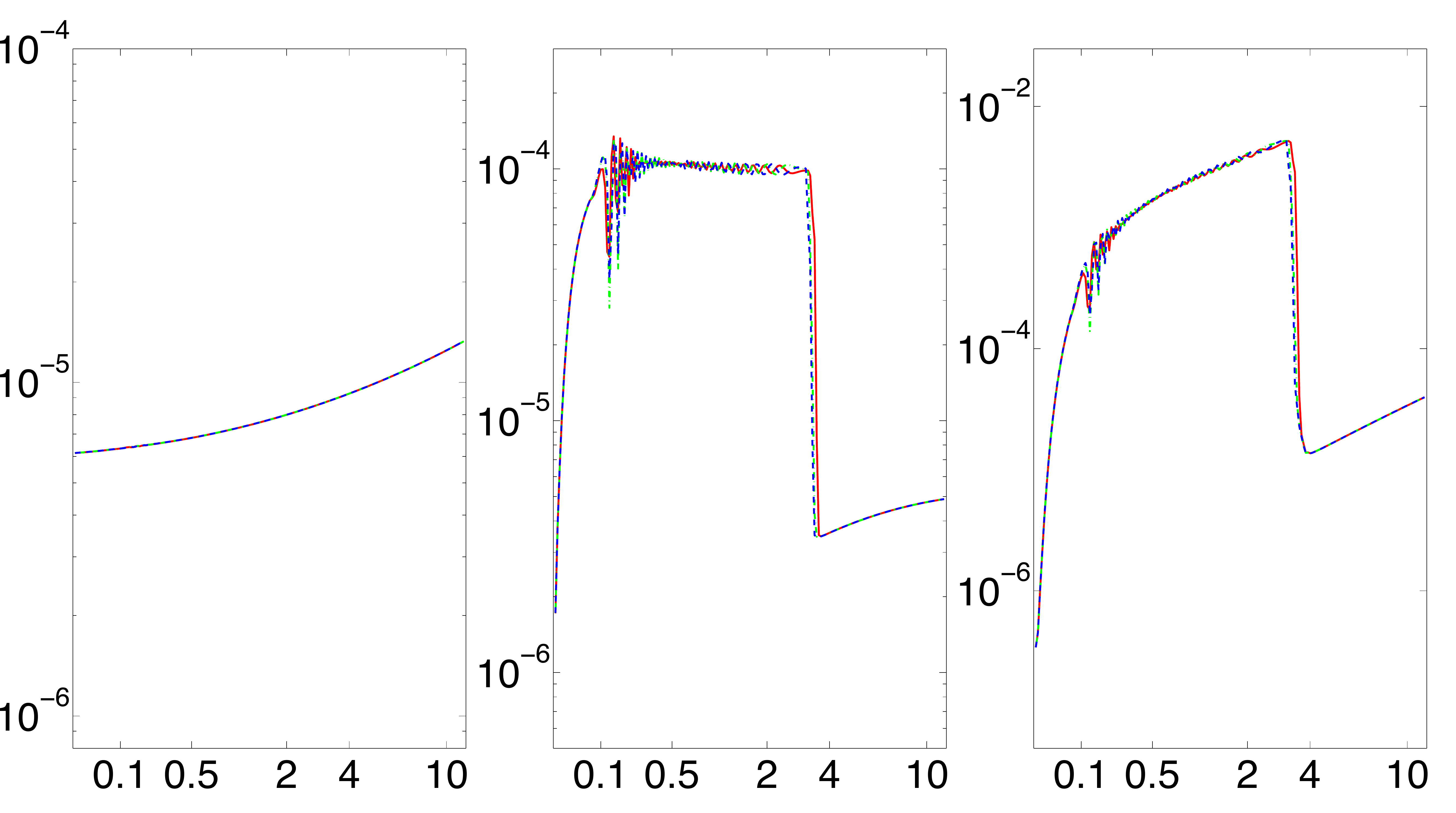}}  \end{tabular}  & \hspace{-0.35cm}\begin{tabular}{c} \rule{0pt}{4.1cm}{\includegraphics[width=0.55\textwidth]{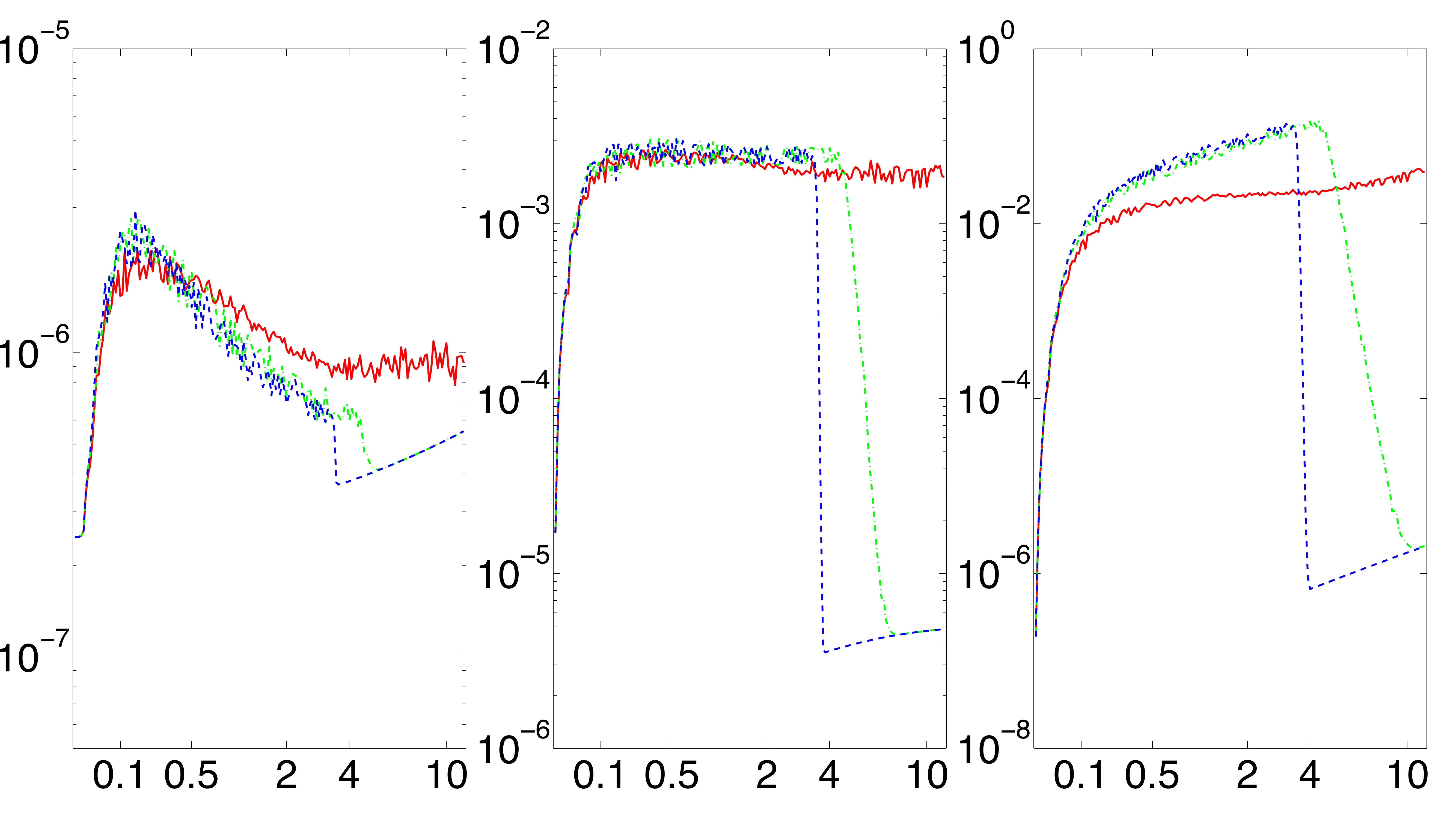}}  \end{tabular} \\  
\Cline{0.3pt}{1-2}
\multicolumn{2}{c}{}\\[-2ex]
\multicolumn{2}{c}{Cosmic time (Gyr) $\longrightarrow$} \\ [-3.0ex]
\end{tabular}
\caption{Amplitude of the measure of error $\mathcal{C}^{(n)}$ over time for each constraint equation, in the case of an initialised $\varphi$, for $l=2$, 10, 200, 1000. Starting from a reference resolution of $n=8$, we include curves of double ($n=16$) and four times ($n=32$) the resolution. We multiply the higher resolution errors by factors of 4 and 16, respectively, indicating 2nd-order convergence where the curves line up. We see clear convergence in each case, though for $l=1000$, the resolution must be pushed to at least $n=16$ for the curves to align satisfactorily. The sudden drop in error seen in some plots around $t=$ 4 Gyr is associated with the exiting of the initial pulses from the measurement domain.}
\label{Tab:convergence} 
\end{figure*}

\section{Convergence Tests}

To establish the accuracy of the discretization, we carry out a
standard convergence test. We check the 2nd-order convergence empirically
by carrying out a series of runs of the same initial data at successively
doubled resolution, corresponding to $\alpha=n$, $\alpha=2n$, and
$\alpha=4n$ in Eqs.~\Eref{discrete-eta} and~\Eref{discrete-r}. The rate
of convergence for a variable $Q$ is given by
\begin{equation}
  \beta^{(n)}_{Q} = \log_2\left| \frac{
    \norm{Q^{(n)}}-\norm{Q^{(2n)}}}{\norm{Q^{(2n)}} - \norm{Q^{(4n)}}}
   \right| \,,{}
   \label{convergence}
\end{equation}
where \norm{Q^{(n)}} is the $L_{2}$-norm on the fixed-resolution grid,
\begin{equation}
  \norm{Q^{(n)}} \equiv \frac{1}{N}
    \left(\sum_{j=1}^{N}(Q^{(n)}_{i,j})^2\right)^{1/2}\,,
\end{equation} with $N$ ($<N_{\rtilde}$) the number of spatial grid-points stored for analysis within the range $0\le \rtilde \le \rtilde_0$ .
We use the following dimensionless measure to quantify how well the constraints are satisfied:
\be
\mathcal{C}^{(n)}_{Q}(\etatilde)\equiv  \frac{||C^{n}_{Q}(\etatilde)||_2}{||\dot{Q}^{n}(\etatilde)||_2} \,,
\ee where $Q\,\epsilon\, \{\Delta,w,\tilde{v}\}$, $C_{Q}$ is one of \Eref{delta_constraint}$-$\Eref{v_constraint}, and we estimate $\dot{Q}$ via a centered difference, i.e.
\be
\frac{\aperpin(\etatilde)}{H_0} \dot{Q}^n(\etatilde) = \bigg[\frac{Q^n(\etatilde+\Delta\etatilde) - Q^n(\etatilde-\Delta\etatilde)}{2\Delta \etatilde}\bigg]\,.
\ee 
\label{}

For all of our evolution variables and constraints, we observe the
expected 2nd-order convergence rate ($\beta=2$). Examples are plotted
in Fig.~\ref{Tab:convergence}, which shows how well the constraint equations perform for various multipole moments in the case of an initial $\varphi$. Using a reference resolution of $n=8$, we include curves of double ($n=16$) and four times ($n=32$) the resolution, multiplying each by a factor of 4 and 16, respectively. When the curves line up, then $\beta=2$, i.e. we obtain second order convergence as expected. It is clear that, in the case $l=1000$, one must go to a resolution of at least $n=16$ to obtain the correct convergence. We have used a base resolution of $H_0^{-1}\Delta\rtilde=1.375$ Gpc throughout, typically with $4\le\alpha\le32$.

\section{Results}

\subsection{Evolution of perturbations}
To study the evolution of perturbations in a LTB cosmological void, we concentrate, for the sake of illustration, on the evolution of the perturbation variables for the spherical modes $l=2$ and $l=10$. For each $l$, we consider three distinct cases: we initialise the profile of any one of $\varphi$, $\varsigma$ and $\chi$ according to \Eref{ICprofile} while setting the others to zero, and apply \Eref{ICvelocity} to all variables (except $\varsigma$ since it satisfies a 1st-order PDE). Here, cases 1, 3 and 5 correspond to $l=2$, and cases 2, 4 and 6 correspond to $l=10$.
\begin{figure*}[!b]
\begin{tabular}{cc|m{6.0cm}|m{6.0cm}|}
\Cline{0.3pt}{3-4}
  & &  \multicolumn{1}{c|}{\rule{0pt}{0.3cm}\hspace{0.2cm}$\varphi$\hspace{2.0cm}$\varsigma$\hspace{2.0cm}$\chi$}  & \multicolumn{1}{c|}{\rule{0pt}{0.3cm}\hspace{0.2cm}$\varphi$\hspace{2.0cm}$\varsigma$\hspace{2.0cm}$\chi$} \\ \Cline{0.3pt}{2-4} 
\multicolumn{1}{c}{}  & \multicolumn{1}{|c|}{} &  \multicolumn{1}{c|}{\rule{0pt}{0.3cm}\hspace{-0.4cm}\tiny{}$[1.0]$\hspace{1.7cm}$[0.3\!\times\!10^{-2}]$\hspace{0.9cm}$[5.4\!\times\!10^{-3}]$}  & \multicolumn{1}{c|}{\rule{0pt}{0.3cm}\hspace{-0.4cm}\tiny{}$[1.0]$\hspace{1.7cm}$[3.1\!\times\!10^{-3}]$\hspace{0.9cm}$[8.0\!\times\!10^{-4}]$} \\[-0.5ex]
& \multicolumn{1}{|c|}{\parbox[t]{2.0mm}{{\rotatebox[origin=c]{90}{\hspace{0.46cm}Initialising $\varphi$}}}}  & \hspace{-0.38cm} \begin{tabular}{c} \rule{0pt}{3.7cm}{\includegraphics[width=0.48\textwidth]{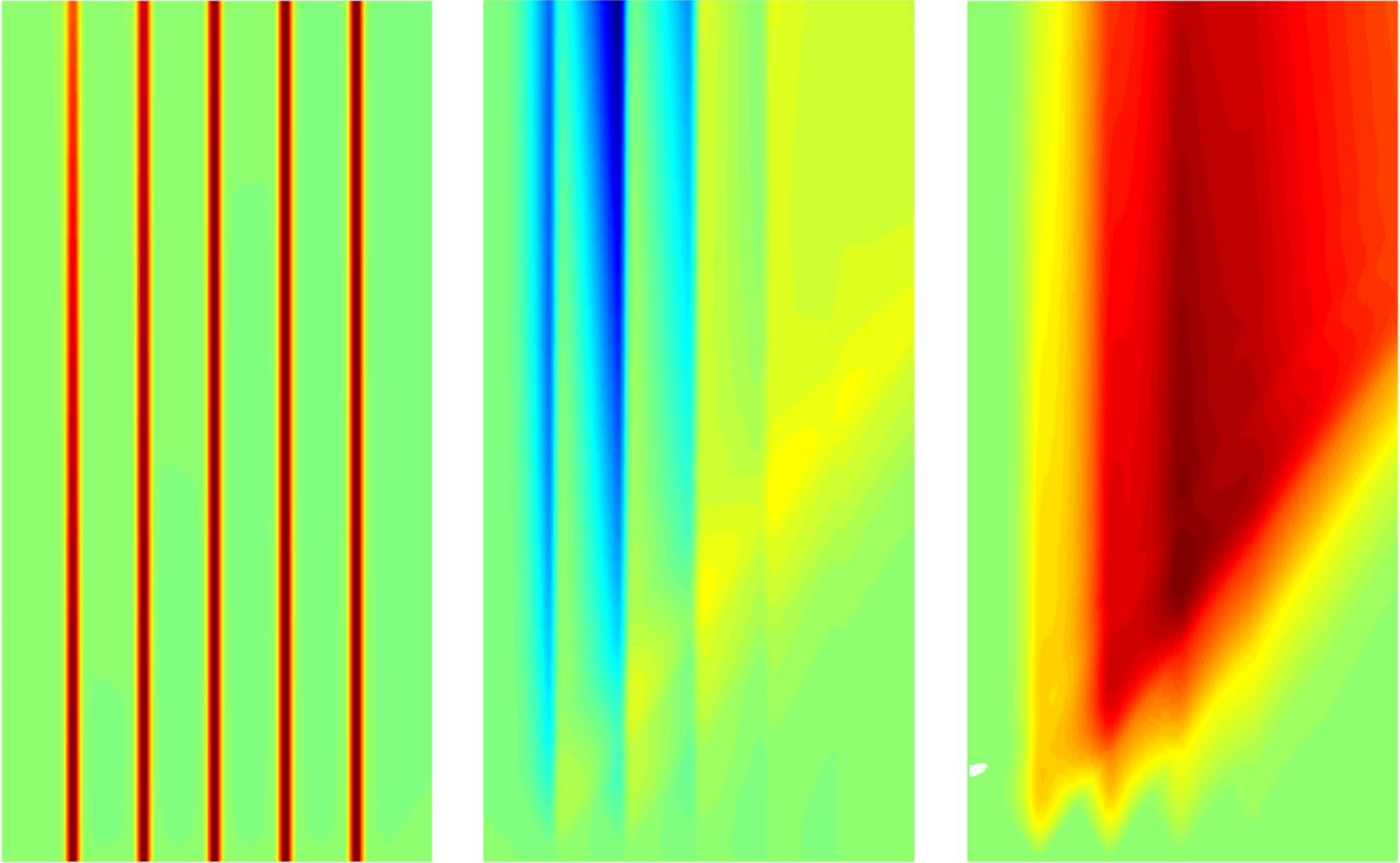}}  \end{tabular} &  \hspace{-0.37cm}  \begin{tabular}{c} \rule{0pt}{3.7cm}{\includegraphics[width=0.48\textwidth]{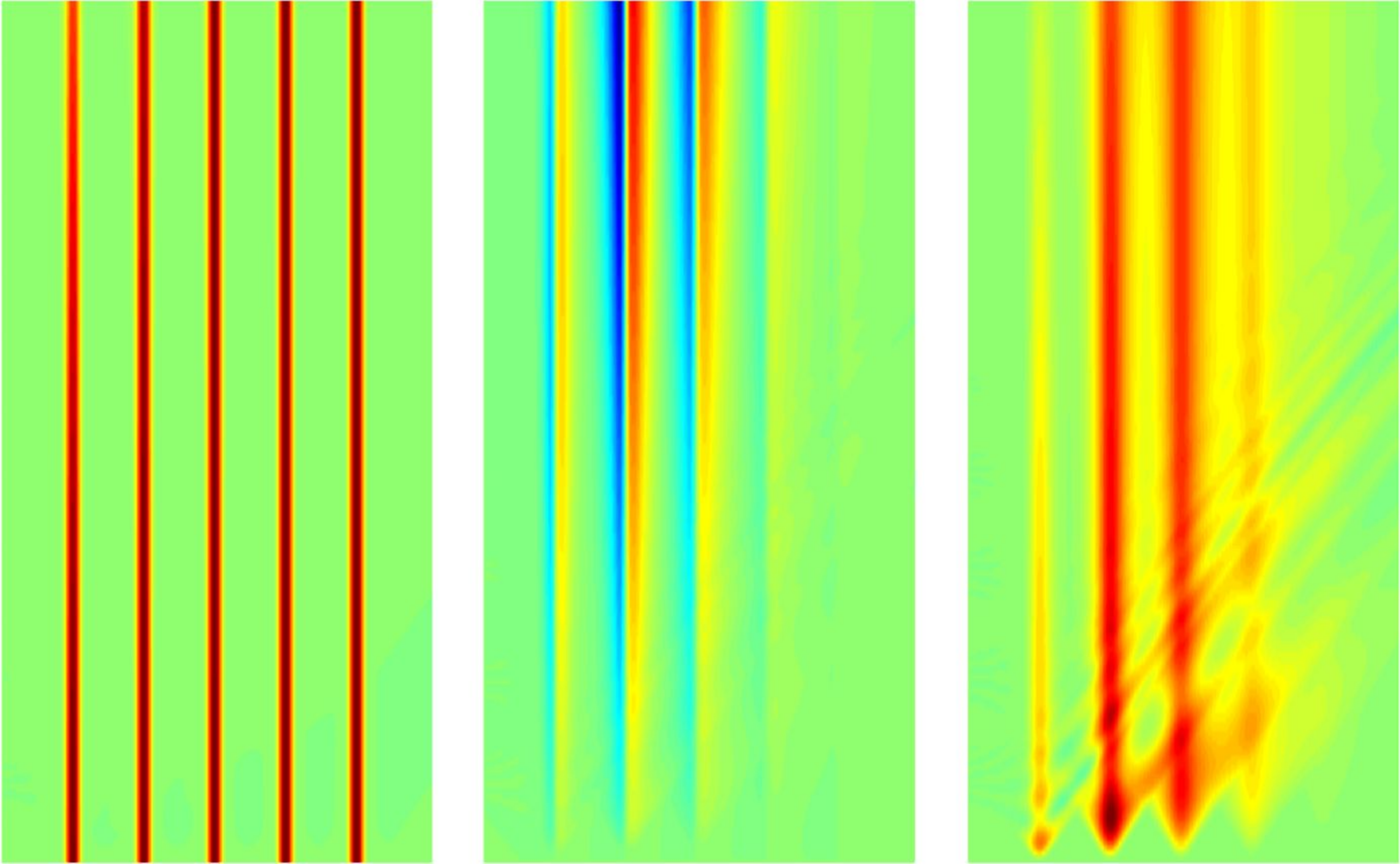}}  \end{tabular} \\[-1.5ex]
\multicolumn{1}{c}{}  & \multicolumn{1}{|c|}{} &  \multicolumn{1}{c|}{\rule{0pt}{0.3cm}\hspace{-0.25cm}\tiny{}$[-2.0\!\times\!10^{-3}]$\hspace{0.7cm}$[-1.1\!\times\!10^{-2}]$\hspace{0.75cm}$[-1.5\!\times\!10^{-6}]$}  & \multicolumn{1}{c|}{\rule{0pt}{0.3cm}\hspace{-0.2cm}\tiny{}$[-3.5\!\!\times\!\!10^{-5}]$\hspace{0.8cm}$[-4.3\!\times\!10^{-3}]$\hspace{0.75cm}$[-0.6\!\times\!10^{-4}]$} \\[-0.5ex]
& \multicolumn{1}{|c|}{}  &   \multicolumn{1}{c|}{\small{\textbf{CASE 1}}} & \multicolumn{1}{c|}{\small{\textbf{CASE 2}}} \\ \Cline{0.3pt}{2-4}
\multicolumn{1}{c}{}  & \multicolumn{1}{|c|}{} &  \multicolumn{1}{c|}{\rule{0pt}{0.3cm}\hspace{-0.3cm}\tiny{}$[0.9\!\times\!10^{-4}]$\hspace{0.9cm}$[1.0]$\hspace{1.7cm}$[5.0\!\times\!10^{-4}]$}  & \multicolumn{1}{c|}{\rule{0pt}{0.3cm}\hspace{-0.3cm}\tiny{}$[2.7\!\times\!10^{-3}]$\hspace{0.9cm}$[1.0]$\hspace{1.7cm}$[3.3\!\times\!10^{-3}]$} \\[-0.5ex]
\begin{tabular}{c} \parbox[t]{0.5mm}{{\rotatebox[origin=c]{90}{\hspace{0.46cm}\large{Cosmic time $\longrightarrow$ }}}} \end{tabular}  & \multicolumn{1}{|c|}{\parbox[t]{2.0mm}{{\rotatebox[origin=c]{90}{\hspace{0.46cm}Initialising $\varsigma$}}}} & \hspace{-0.35cm} \begin{tabular}{c} \rule{0pt}{3.7cm}{\includegraphics[width=0.48\textwidth]{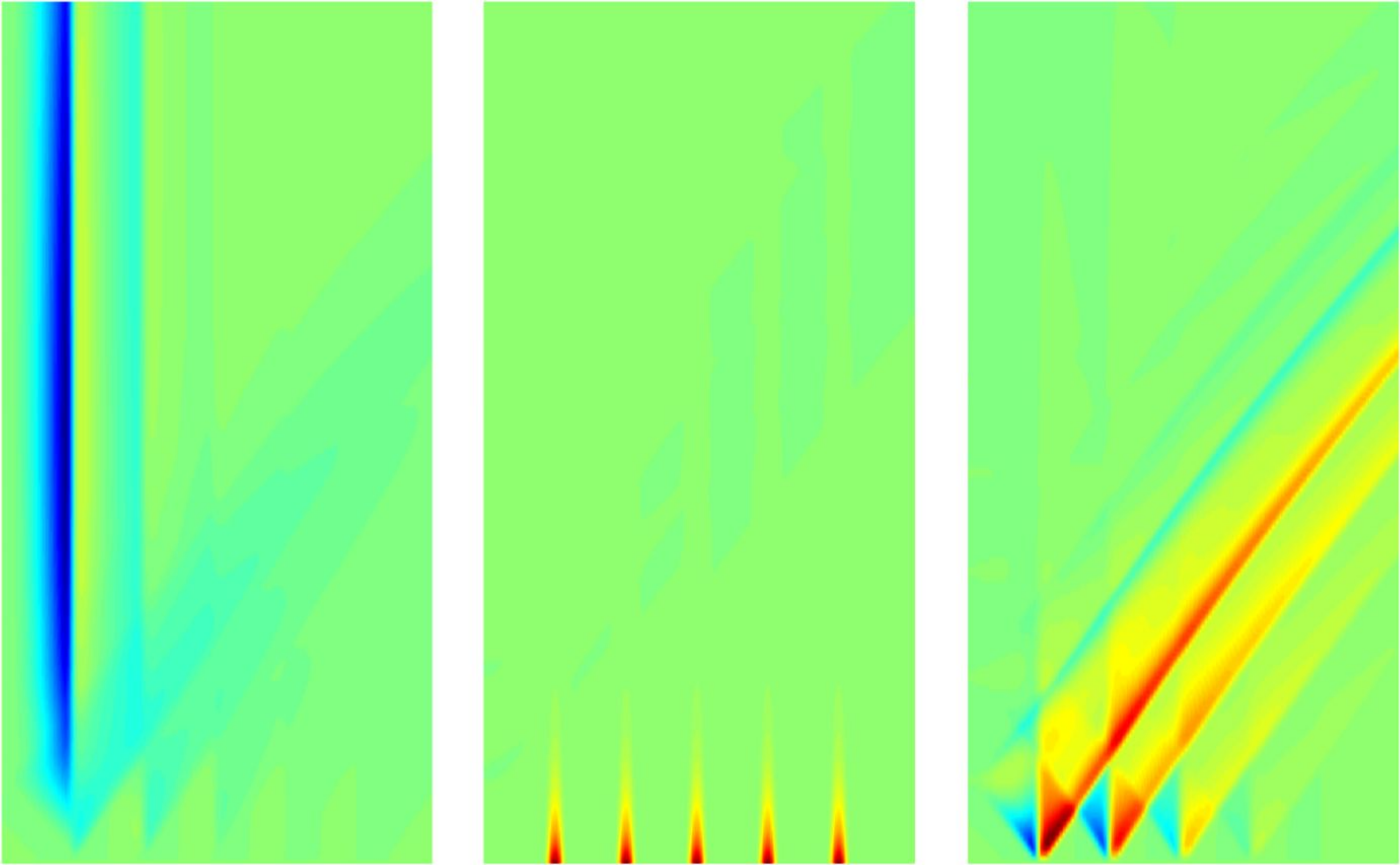}}  \end{tabular} & \hspace{-0.35cm} \begin{tabular}{c} \rule{0pt}{3.7cm}{\includegraphics[width=0.48\textwidth]{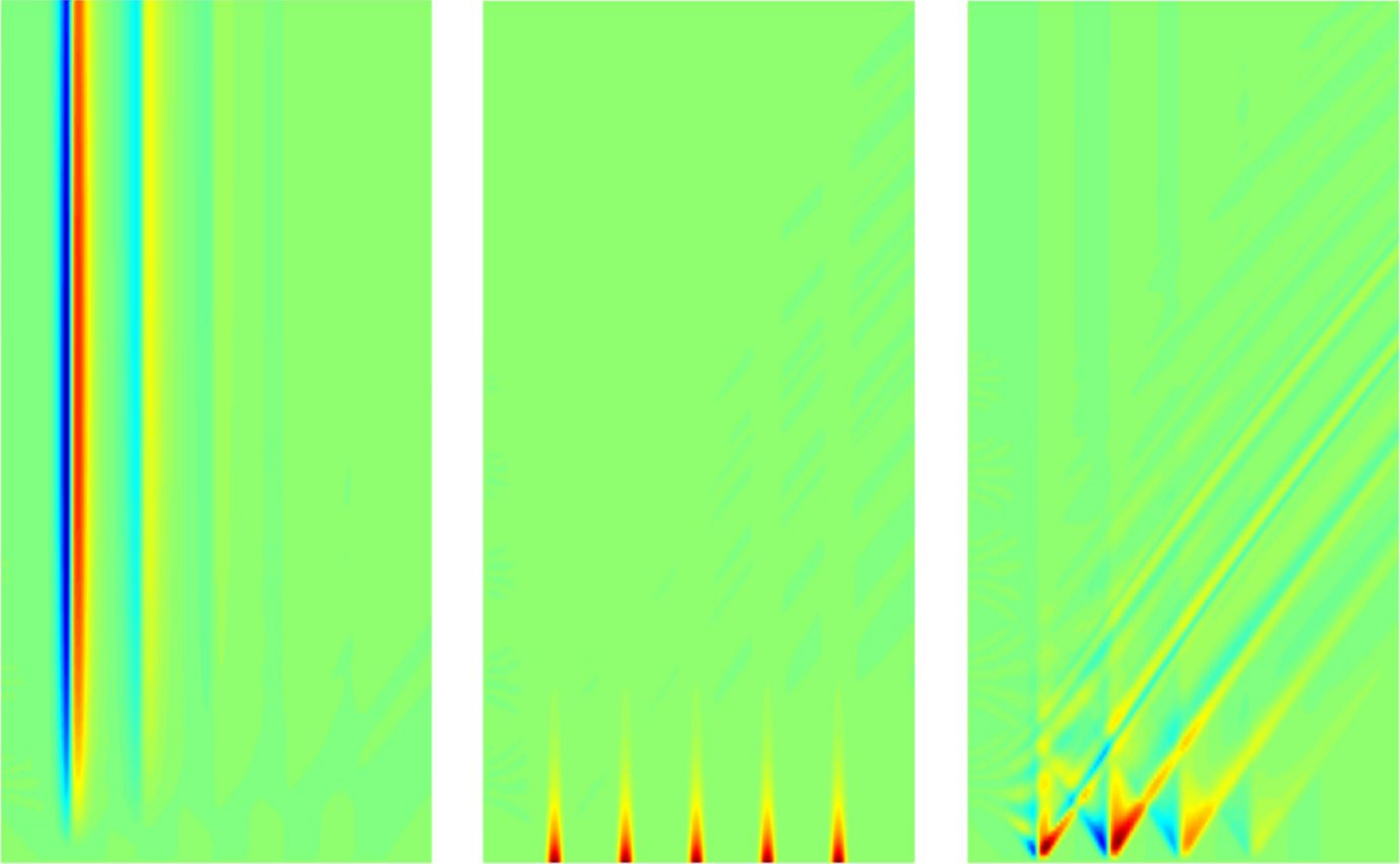}}  \end{tabular} \\[-1.5ex]
\multicolumn{1}{c}{}  & \multicolumn{1}{|c|}{} &  \multicolumn{1}{c|}{\rule{0pt}{0.3cm}\hspace{-0.20cm}\tiny{}$[-7.4\!\times\!10^{-4}]$\hspace{0.7cm}$[-1.1\!\times\!10^{-4}]$\hspace{0.75cm}$[-3.6\!\times\!10^{-4}]$}  & \multicolumn{1}{c|}{\rule{0pt}{0.3cm}\hspace{-0.15cm}\tiny{}$[-4.2\!\times\!10^{-3}]$\hspace{0.75cm}$[-7.3\!\times\!10^{-5}]$\hspace{0.75cm}$[-2.7\!\times\!10^{-3}]$} \\[-0.5ex]
& \multicolumn{1}{|c|}{}  &   \multicolumn{1}{c|}{\small{\textbf{CASE 3}}} & \multicolumn{1}{c|}{\small{\textbf{CASE 4}}} \\ \Cline{0.3pt}{2-4}
\multicolumn{1}{c}{}  & \multicolumn{1}{|c|}{} &  \multicolumn{1}{c|}{\rule{0pt}{0.3cm}\hspace{-1.2cm}\tiny{}$[4.7\!\times\!10^{-1}]$\hspace{0.9cm}$[4.4\!\times\!10^{-1}]$\hspace{0.9cm}$[1.0]$}  & \multicolumn{1}{c|}{\rule{0pt}{0.3cm}\hspace{-1.1cm}\tiny{}$[5.8\!\times\!10^{-1}]$\hspace{0.9cm}$[4.3\!\times\!10^{-1}]$\hspace{0.9cm}$[1.0]$} \\[-0.5ex]
 & \multicolumn{1}{|c|}{\parbox[t]{2.0mm}{{\rotatebox[origin=c]{90}{\hspace{0.46cm}Initialising $\chi$}}}}  &   \hspace{-0.3cm}\begin{tabular}{c} \rule{0pt}{3.7cm}{\includegraphics[width=0.48\textwidth]{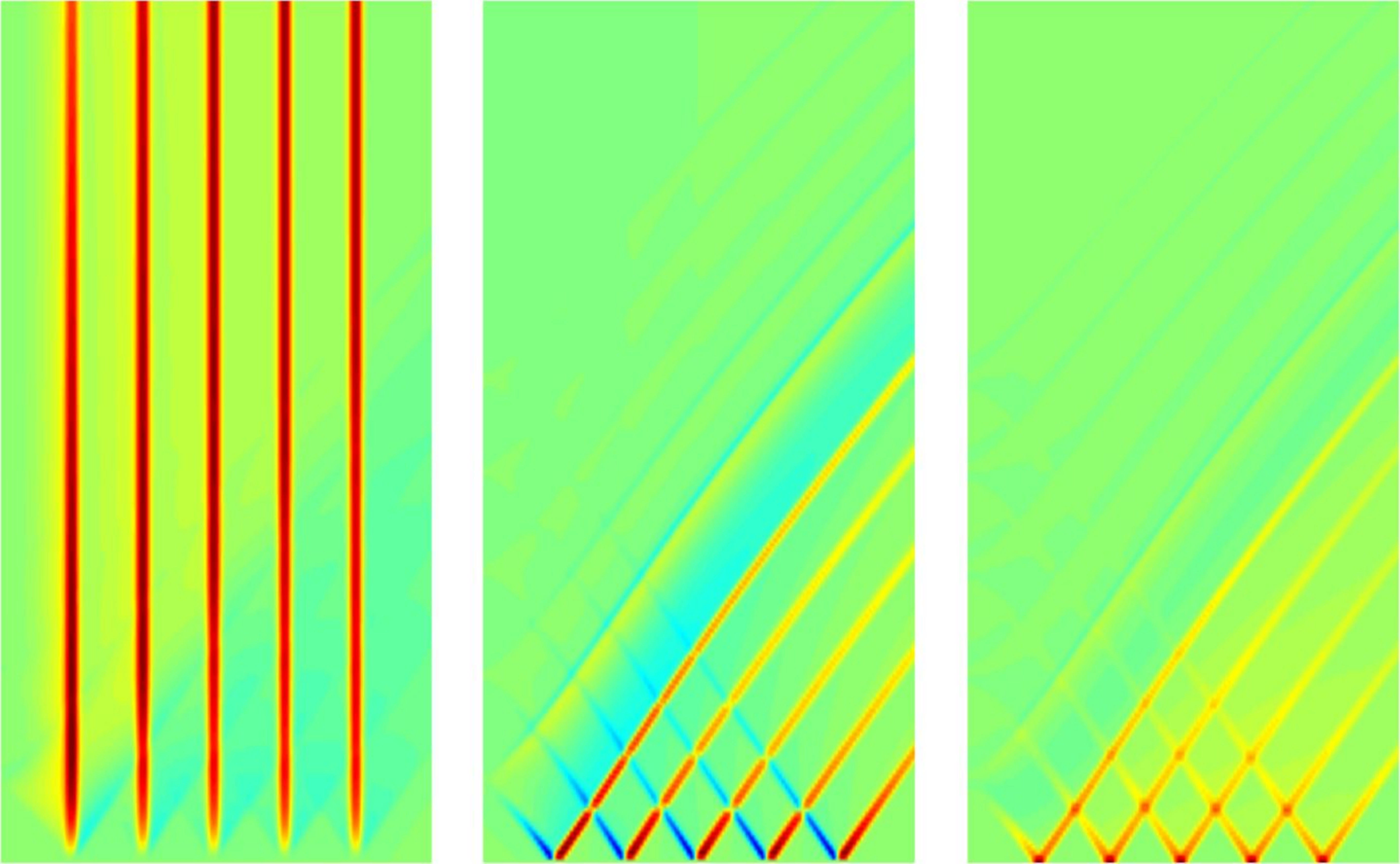}}  \end{tabular} &  \hspace{-0.35cm} \begin{tabular}{c} \rule{0pt}{3.7cm}{\includegraphics[width=0.48\textwidth]{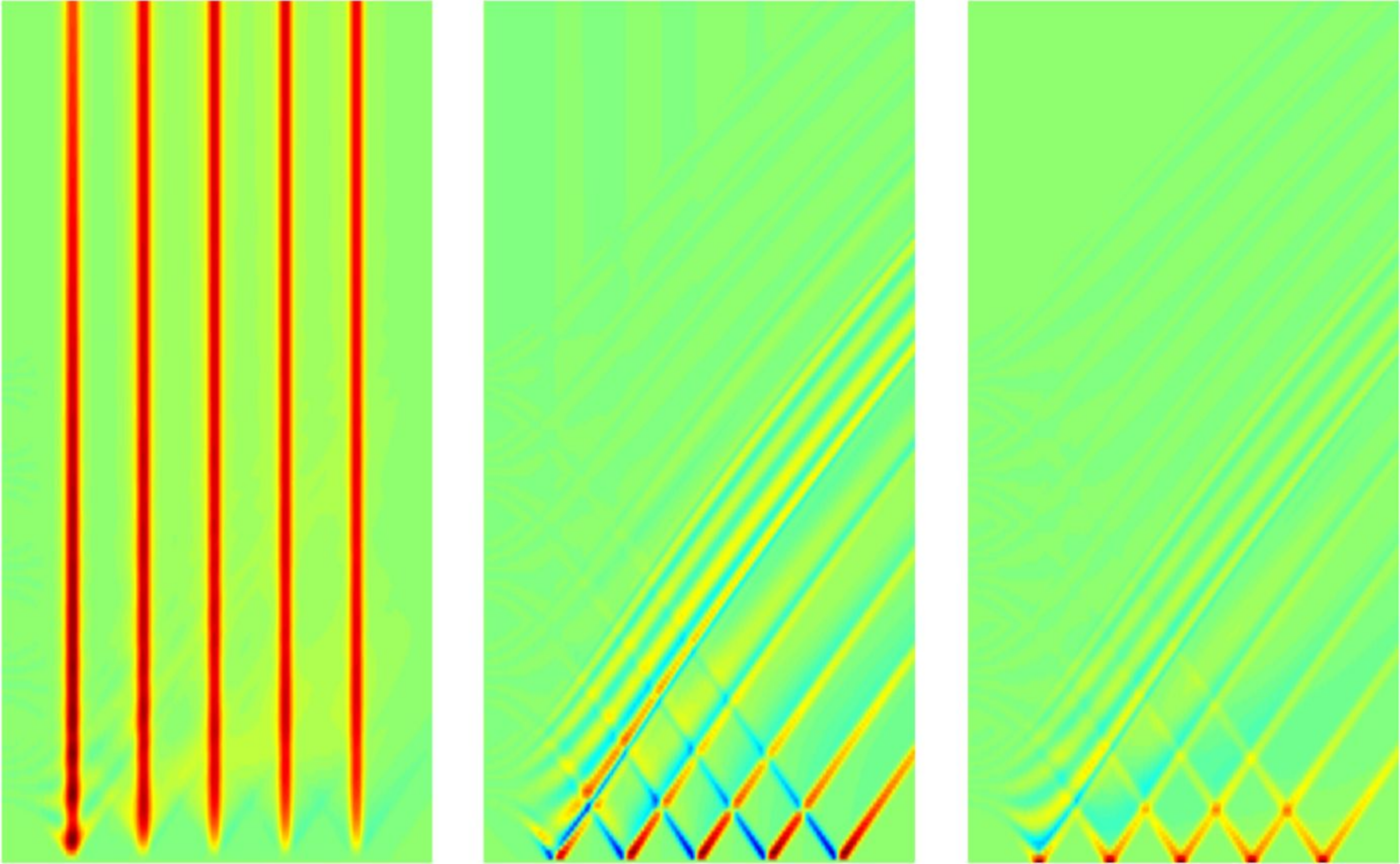}}  \end{tabular} \\ [-1.5ex]
 \multicolumn{1}{c}{}  & \multicolumn{1}{|c|}{} &  \multicolumn{1}{c|}{\rule{0pt}{0.3cm}\hspace{-0.25cm}\tiny{}$[-1.1\!\times\!10^{-1}]$\hspace{0.75cm}$[-4.3\!\times\!10^{-1}]$\hspace{0.75cm}$[-7.8\!\times\!10^{-2}]$}  & \multicolumn{1}{c|}{\rule{0pt}{0.3cm}\hspace{-0.85cm}\tiny{}$[-0.6\!\times\!10^{-1}]$\hspace{0.75cm}$[-4.3\!\times\!10^{-1}]$\hspace{0.75cm}$[-0.4]$} \\[-0.5ex]
 & \multicolumn{1}{|c|}{}  &   \multicolumn{1}{c|}{\small{\textbf{CASE 5}}} & \multicolumn{1}{c|}{\small{\textbf{CASE 6}}} \\ \Cline{0.3pt}{2-4}
  \multicolumn{1}{c}{}  & \multicolumn{1}{c}{} & \multicolumn{2}{c}{}\\[-2ex]
 \multicolumn{1}{c}{}  & \multicolumn{1}{c}{} & \multicolumn{2}{c}{\large{Radial distance $\longrightarrow$}} \\ [-3.0ex]
\end{tabular}
\caption{Spacetime evolution of each of the master variables in the various cases considered. In cases 1 and 2, $\varphi$ excites both $\varsigma$ and $\chi$ to about the sub-percent level. The propagating modes resulting from $\chi$ is visible in $\varsigma$, or alternatively the trace-free part of the magnetic Weyl tensor $\delta \overline{H}_{(TF)}$ \Eref{deltaHTF}, thus clearly showing relativistic degrees of freedom at work. In cases 3 and 4, while an initial $\varsigma$ decays away quickly due to the Hubble friction, it still manages to excite the other two variables, albeit to a low level. The final two cases, 5 and 6, are rather interesting: an initial $\chi$ generated inside a void excites a relatively significant amount of $\varphi$ and $\varsigma$. The presence of propagating modes is more apparent in all the variables here. The maximum and minimum values of the colour scale are indicated respectively in brackets above and below each 2D plot.}
\label{Tab:MasterTab}
\end{figure*}

The evolution of each of the variables is presented in Fig.~\ref{Tab:MasterTab} for each of the $6$ corresponding cases. The resolutions used in all of the plots are typically in the range $32\le n \le 128$, $4\le\alpha\le32$.
\begin{description}
\item[Cases 1 and 2:]  In these cases, we initialise $\varphi$, and set $\varsigma=\chi=0$ initially. We clearly see the ``bleeding'' of the modes due to the coupling.
\begin{figure*}[!t]
\centering
\begin{tabular}{c|m{6.35cm}|m{6.35cm}|}
 \multicolumn{1}{c}{} & \multicolumn{1}{c}{\small{}\textbf{CASE 1}} &  \multicolumn{1}{c}{\small{}\textbf{CASE 2}}\\
\Cline{0.3pt}{2-3}
   &  \multicolumn{1}{c|}{\small{}\hspace{0.05cm}Time evolution\hspace{1.35cm}Profile today}  & \multicolumn{1}{c|}{\small{}\hspace{0.05cm}Time evolution\hspace{1.35cm}Profile today} \\ \Cline{0.3pt}{1-3} 
 \multicolumn{1}{|c|}{$\!\!\varphi\!\!$}  &  \hspace{-0.5cm} \begin{tabular}{c} \rule{0pt}{3.9cm}{\includegraphics[width=0.52\textwidth]{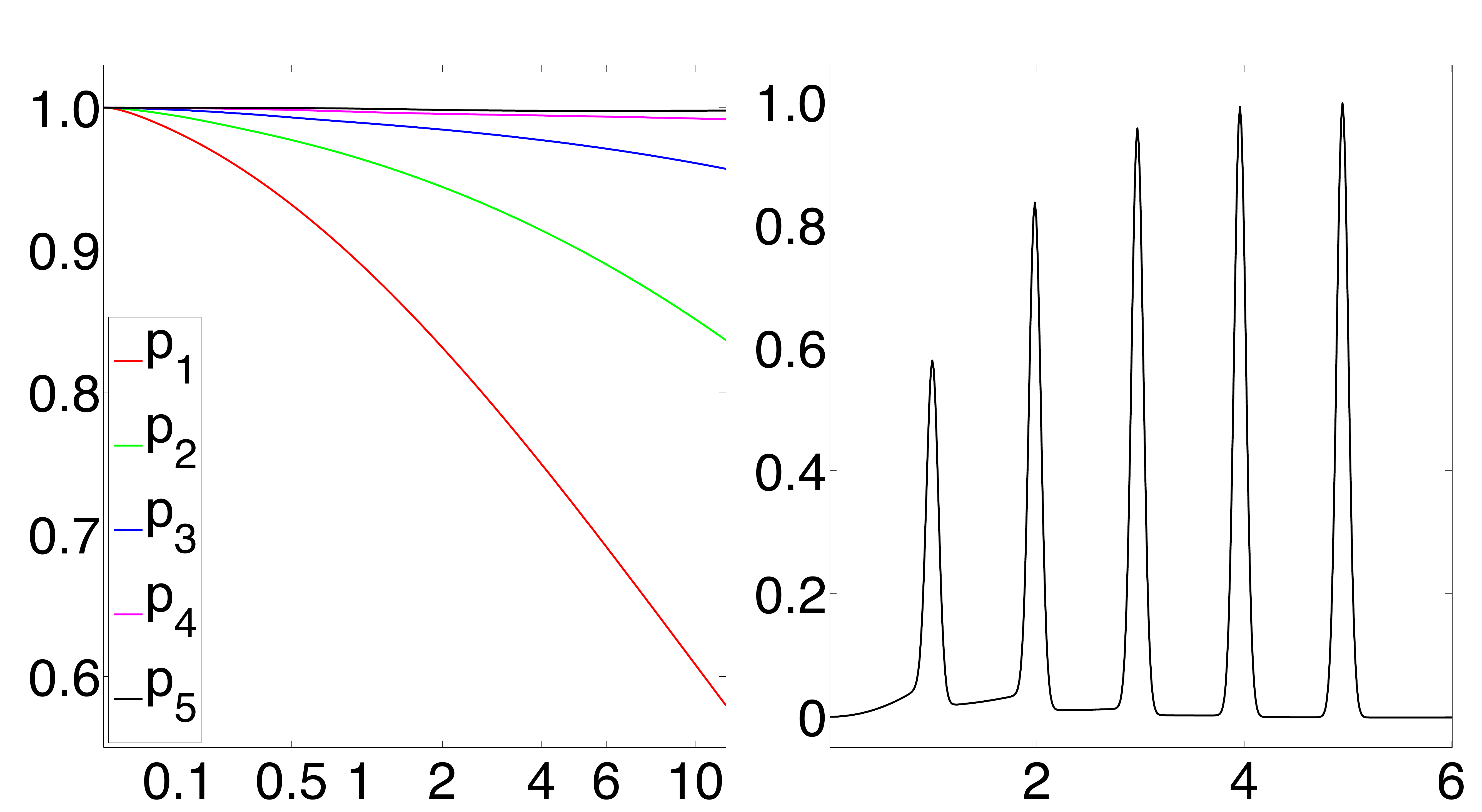}}  \end{tabular} &   \hspace{-0.5cm} \begin{tabular}{c} \rule{0pt}{3.9cm}{\includegraphics[width=0.52\textwidth]{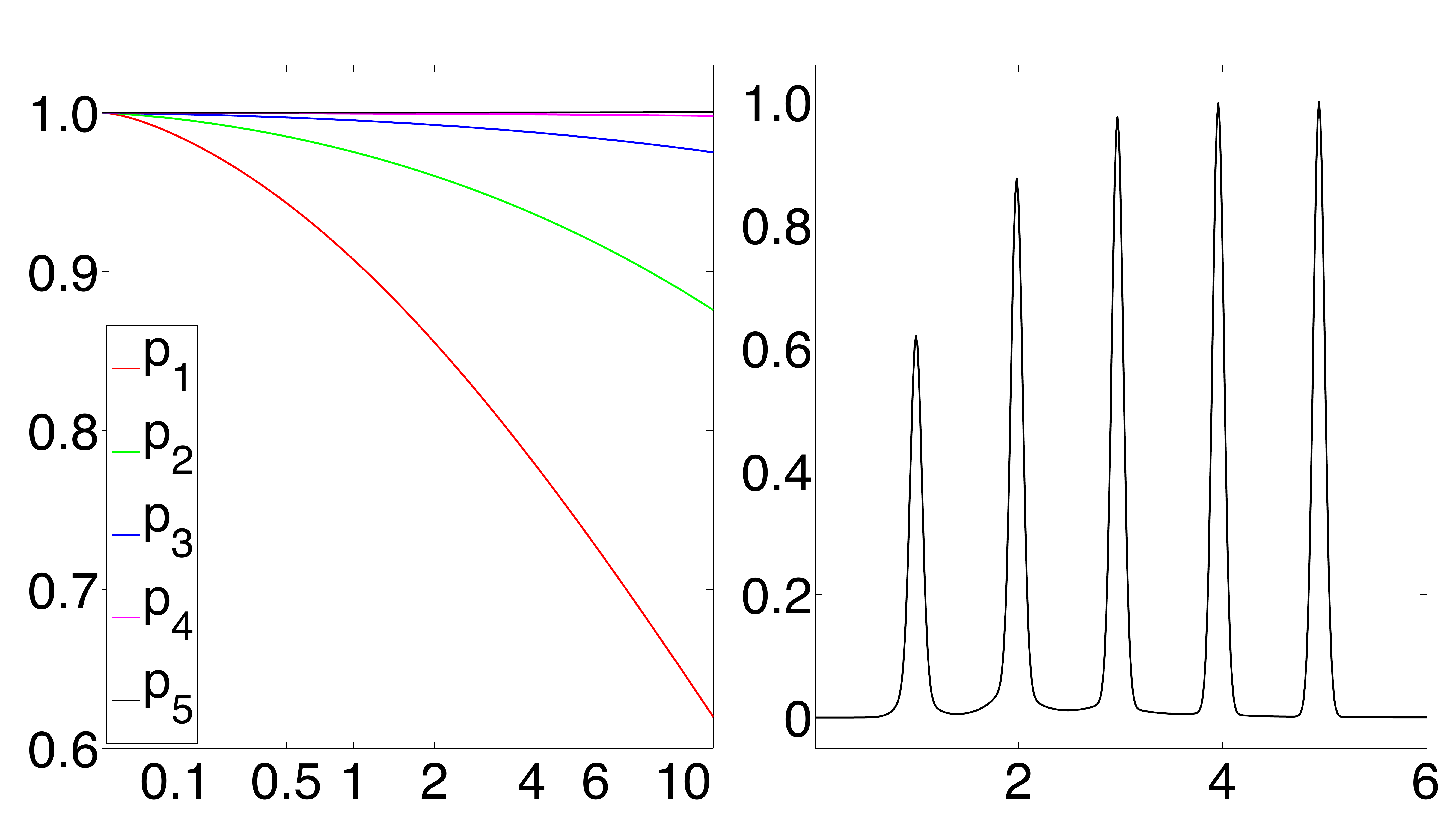}}  \end{tabular} \\ \Cline{0.3pt}{1-3}
 \multicolumn{1}{|c|}{$\!\!\varsigma\!\!$}  &  \hspace{-0.4cm}\begin{tabular}{c} \rule{0pt}{3.8cm}{\includegraphics[width=0.52\textwidth]{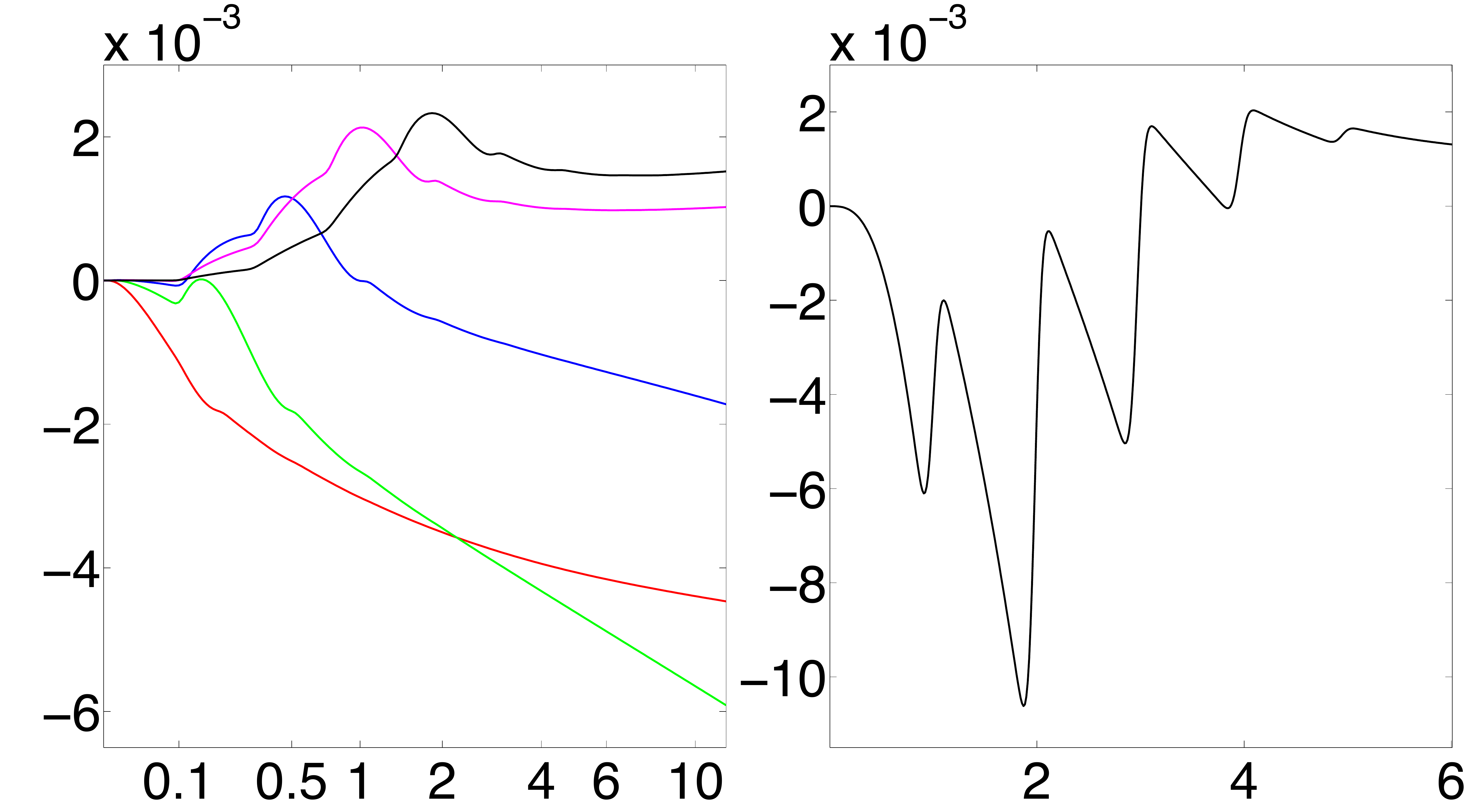}}  \end{tabular} & \hspace{-0.5cm} \begin{tabular}{c} \rule{0pt}{3.8cm}{\includegraphics[width=0.52\textwidth]{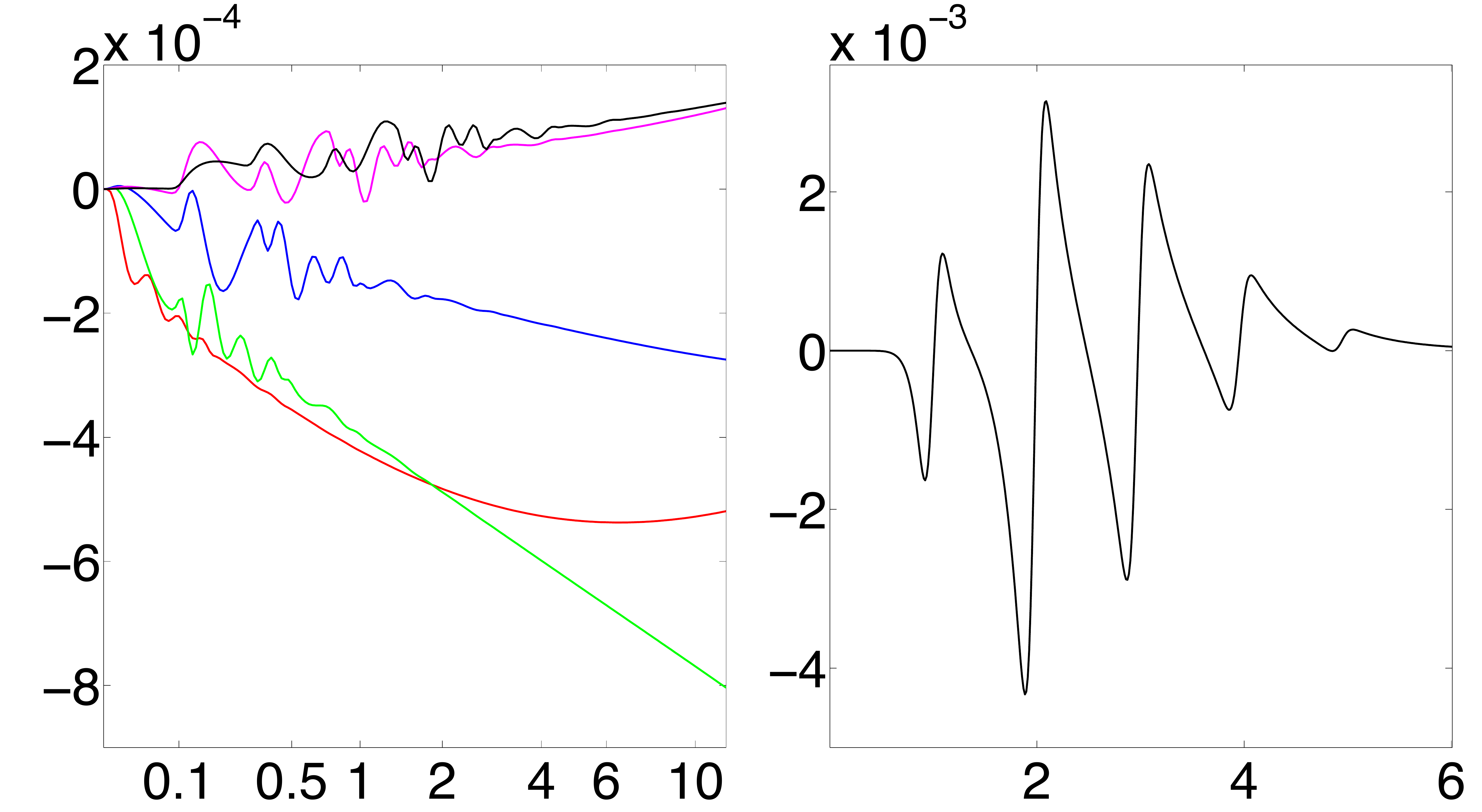}}  \end{tabular} \\ \Cline{0.3pt}{1-3}
  \multicolumn{1}{|c|}{$\!\!\chi\!\!$}  &   \hspace{-0.4cm}\begin{tabular}{c} \rule{0pt}{3.8cm}{\includegraphics[width=0.52\textwidth]{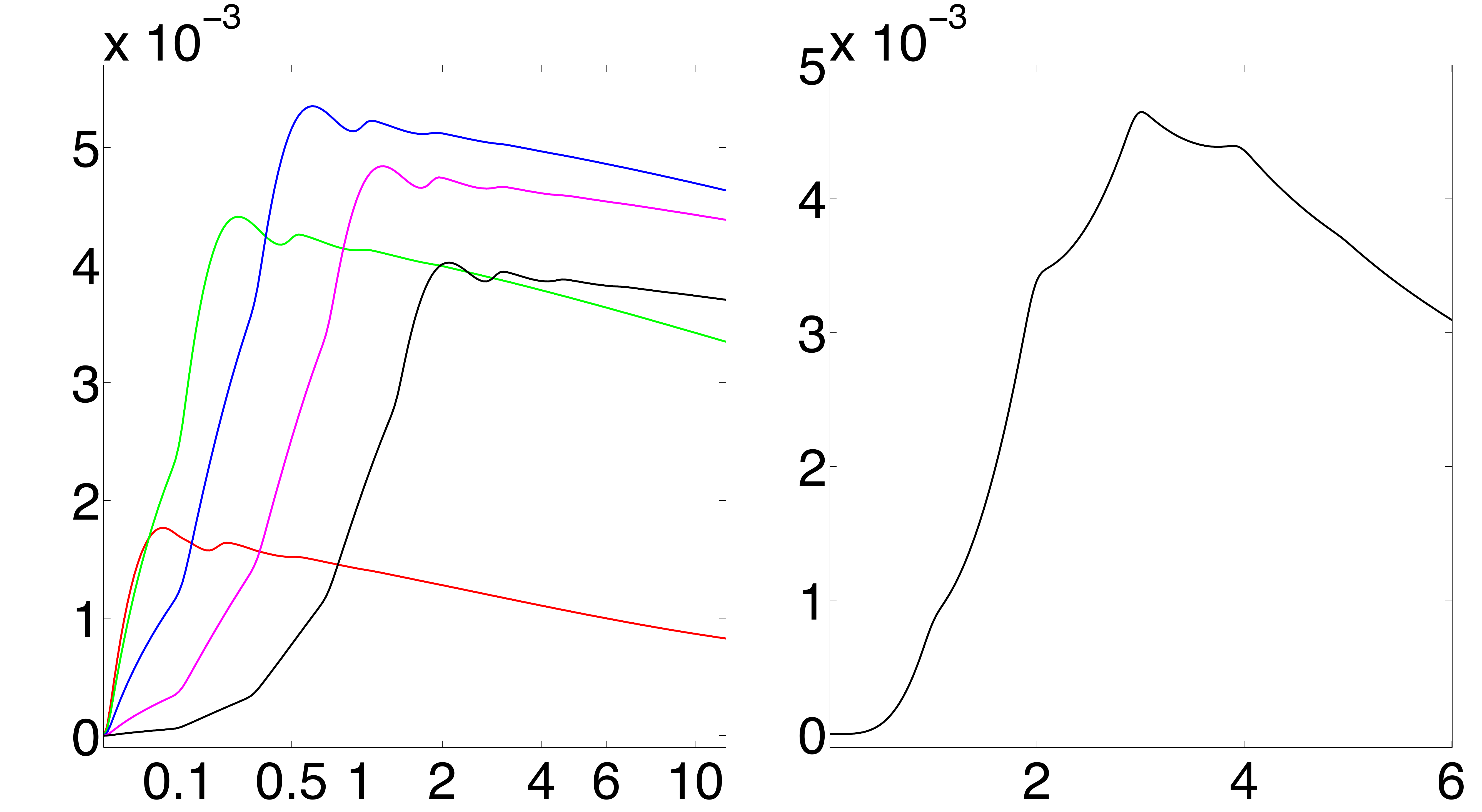}}  \end{tabular} & \hspace{-0.5cm} \begin{tabular}{c} \rule{0pt}{3.8cm}{\includegraphics[width=0.52\textwidth]{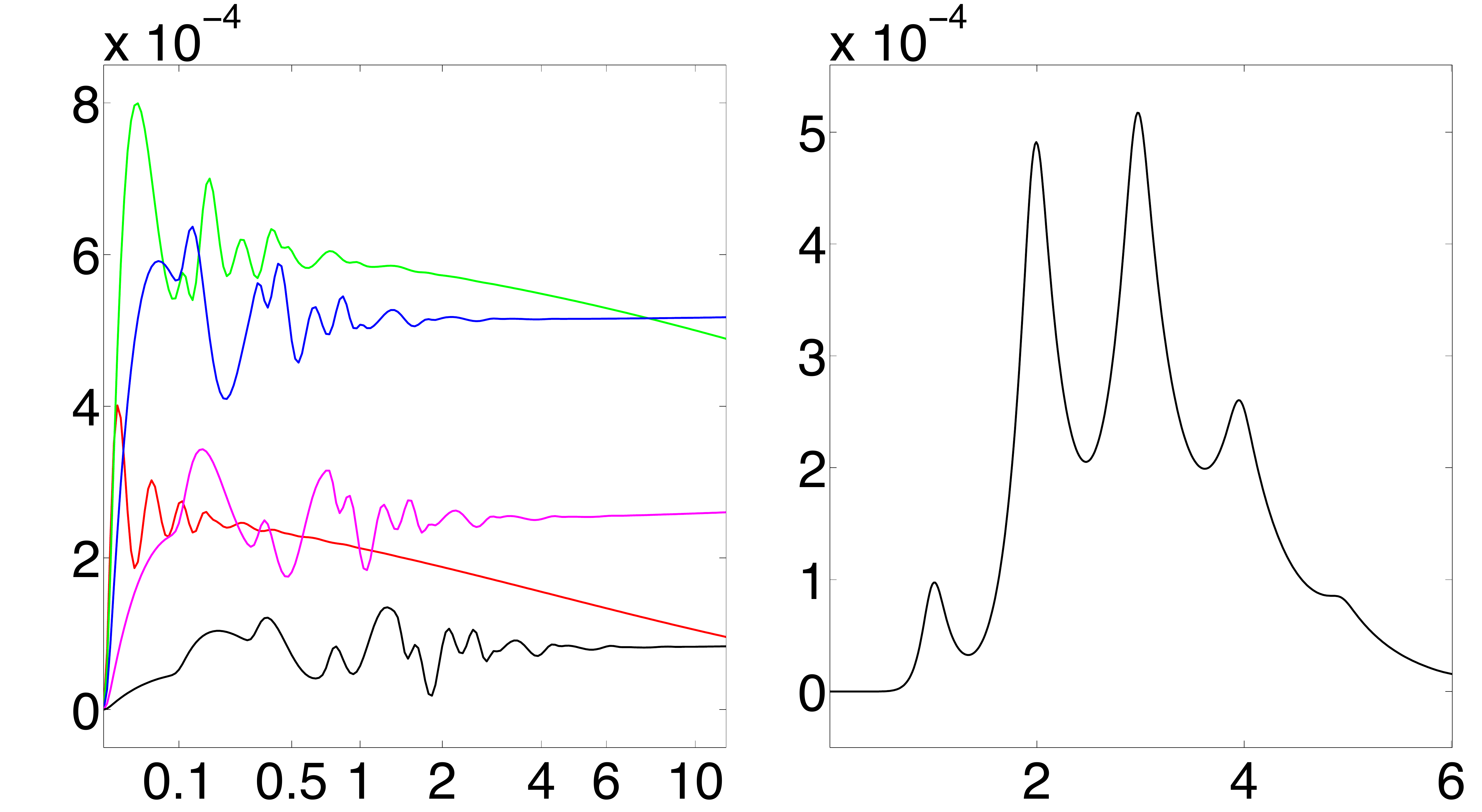}}  \end{tabular} \\ \Cline{0.3pt}{1-3}
 \multicolumn{1}{c|}{}  &  \multicolumn{1}{c|}{\hspace{0.1cm}\small{}Cosmic time (Gyr)\hspace{0.9cm}Distance (Gpc)} & \multicolumn{1}{c|}{\hspace{0.1cm}\small{}Cosmic time (Gyr)\hspace{0.9cm}Distance (Gpc)} \\ \Cline{0.3pt}{2-3}
 \multicolumn{1}{c}{} & \multicolumn{2}{c}{}\\[-3.0ex]
\end{tabular}
\caption{Temporal and spatial slices through the spacetime evolution observed in the top panel of Fig.~\ref{Tab:Case1and2Tab}. The variable $\varphi$ is clearly unaffected on the outskirts of the void in which the spacetime is almost Einstein-de Sitter. Since $\varsigma$ and $\chi$ are relatively sub-percent in amplitude, on each radial shell $\varphi$ more or less behaves as expected in an open, dust-dominated FLRW model, i.e. decays with time.}
\label{Tab:Case1and2Tab}
\end{figure*}
On the 2D plot, Fig.~\ref{Tab:MasterTab}, $\chi$ behaves like a propagating degree of freedom, evolving along the characteristics of the spacetime, and radiating energy away from each pulse. The behaviour of $\varsigma$ is more difficult to qualitatively describe because it is a mixture of frame dragging and gravitational wave degrees of freedom~-- the combination of non-propagating decay with some radiation can be seen in the figures. It is proportional to the trace-free part of the magnetic Weyl curvature \Eref{deltaHTF}, and thus represents a true relativistic degree of freedom.  Then, $\varphi$  follows a standard evolution throughout the spacetime: staying constant around the EdS region, while decaying faster deep inside the void.

The top panel of Fig.~\ref{Tab:Case1and2Tab} presents the profile of $\varphi$ today for these cases, as well as its time evolution along selected radii. As expected, $\varphi$ remains constant in the outer, quasi-FLRW regions of the void, given that it essentially satisfies the Bardeen equation there. Deep inside the void, $\varphi$ decreases for the most part as the usual Bardeen potential would in an open FLRW dust model, but there is evidence of influence from $\varsigma$ and $\chi$ at least at the sub-percent-level, as can be seen by the amplitudes of the latter in the middle and bottom panels of Fig.~\ref{Tab:Case1and2Tab}; see also section \ref{Subsection:CoupvsUnCoup} for a discussion on the importance of the couplings.

We show the spacetime configuration of $\Delta$ in Fig.~\ref{Tab:DerivedQuantities}: its growth appears to follow the peaks were $\varphi$ is concentrated, suggesting that the tiny $\chi$ and $\varsigma$ generated by the dynamics have little impact on the profile of density perturbations. Also included in Fig.~\ref{Tab:DerivedQuantities} is the resulting radial peculiar velocity $w$ and the trace-free part of the electric Weyl curvature.

\item[Cases 3 and 4:] Here, we initialise $\varsigma$, and set $\varphi=\chi=0$ initially. 

From the middle panels of Fig.~\ref{Tab:MasterTab} we see that $\varsigma$ decays very quickly $-$ in fact, roughly proportional to $\apar^{-2}$ $-$ along the peaks from where it is initially located, while sourcing $\varphi$ and $\chi$. As expected, $\chi$ is very well described as a propagating degree of freedom, but one also sees that the sourced $\varphi$ has a propagating component that follows the characteristics of the background spacetime and escapes the void.

\begin{figure*}[!t]
\centering
\begin{tabular}{c|m{6.35cm}|m{6.35cm}|}
 \multicolumn{1}{c}{} & \multicolumn{1}{c}{\small{}\textbf{CASE 3}} &  \multicolumn{1}{c}{\small{}\textbf{CASE 4}}\\
\Cline{0.3pt}{2-3}
   &  \multicolumn{1}{c|}{\small{}\hspace{0.05cm}Time evolution\hspace{1.35cm}Profile today}  & \multicolumn{1}{c|}{\small{}\hspace{0.05cm}Time evolution\hspace{1.35cm}Profile today} \\ \Cline{0.3pt}{1-3} 
 \multicolumn{1}{|c|}{$\!\!\varphi\!\!$}  &  \hspace{-0.5cm} \begin{tabular}{c} \rule{0pt}{3.8cm}{\includegraphics[width=0.52\textwidth]{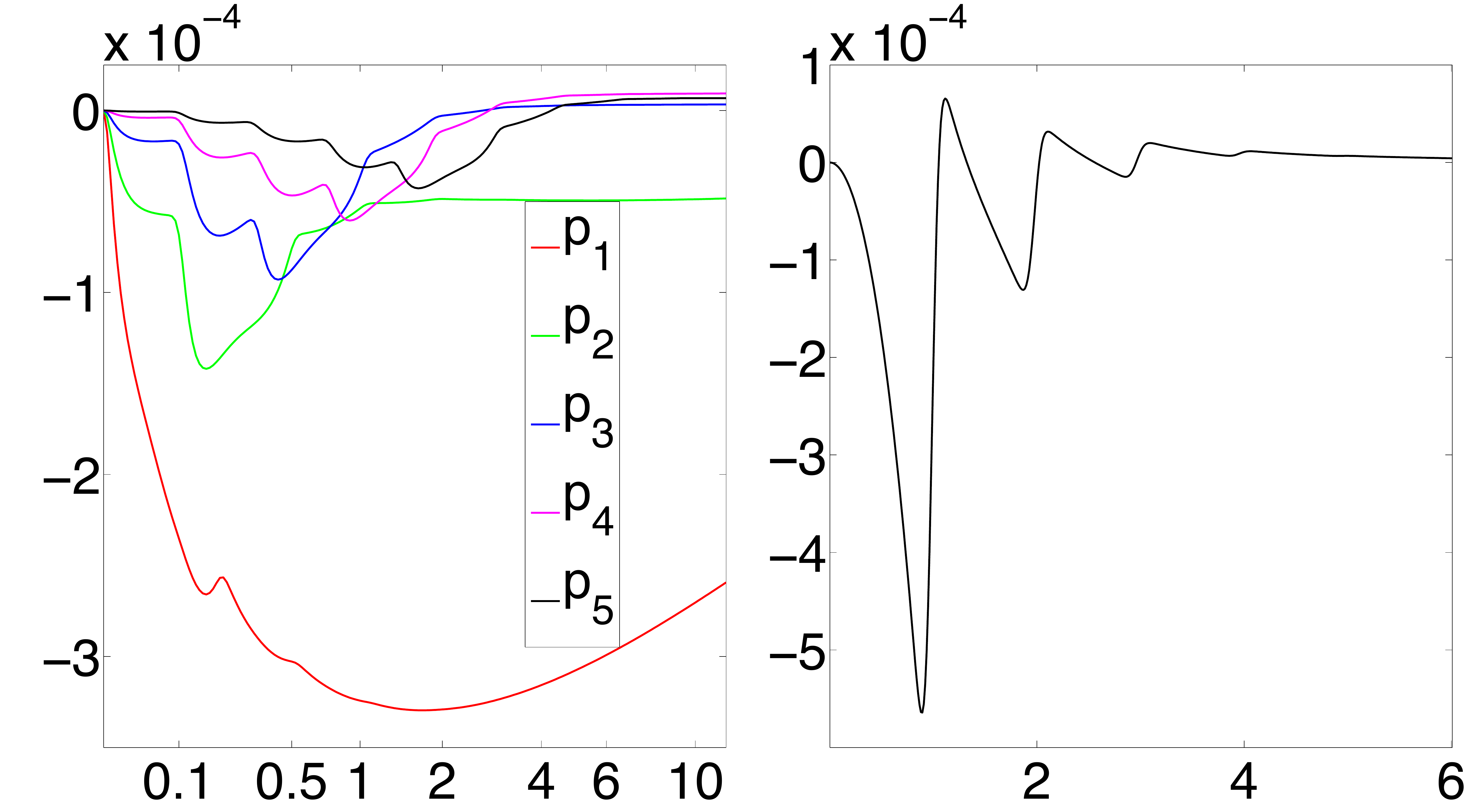}}  \end{tabular} &   \hspace{-0.5cm} \begin{tabular}{c} \rule{0pt}{3.8cm}{\includegraphics[width=0.52\textwidth]{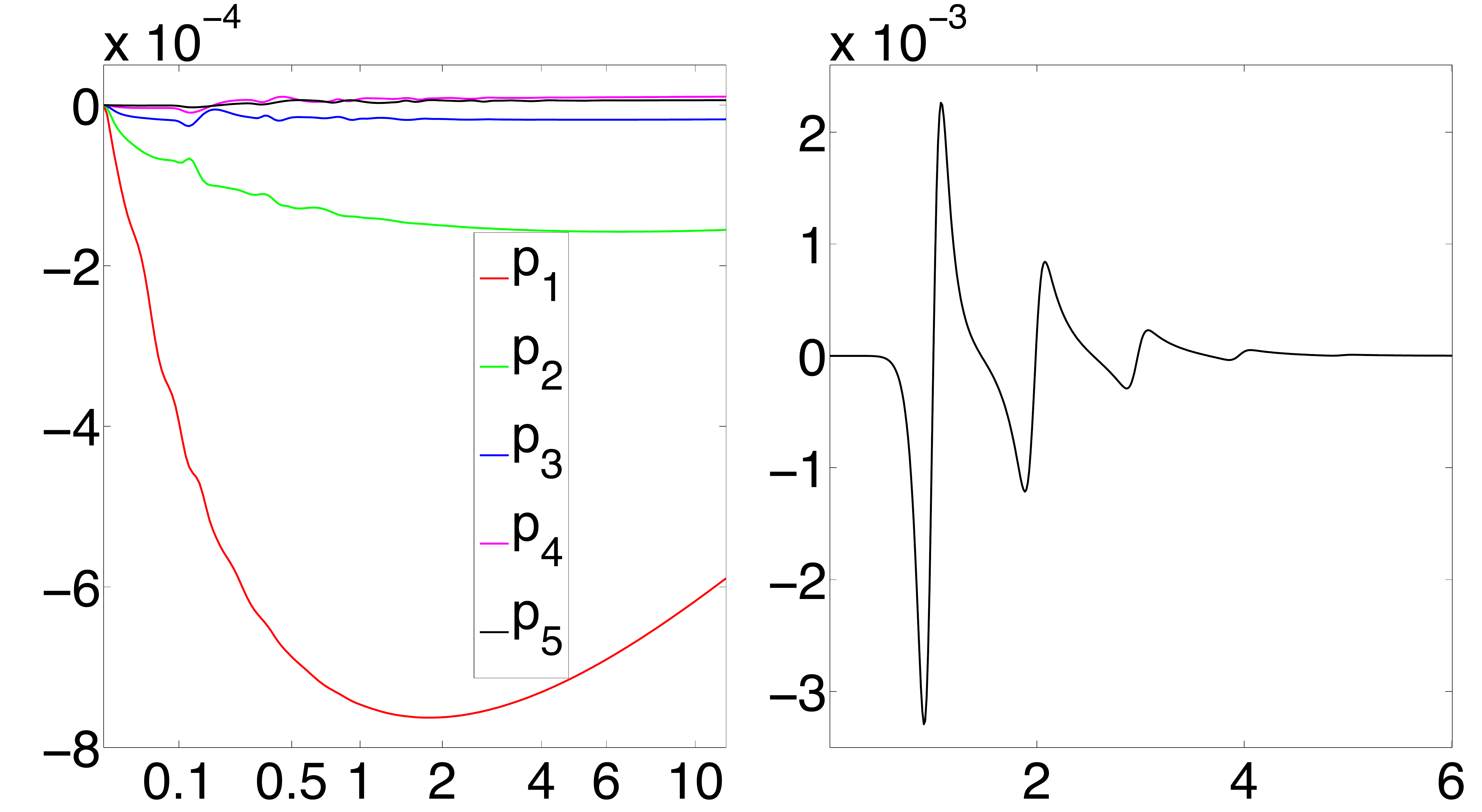}}  \end{tabular} \\ \Cline{0.3pt}{1-3}
 \multicolumn{1}{|c|}{$\!\!\varsigma\!\!$}  &  \hspace{-0.4cm}\begin{tabular}{c} \rule{0pt}{3.9cm}{\includegraphics[width=0.52\textwidth]{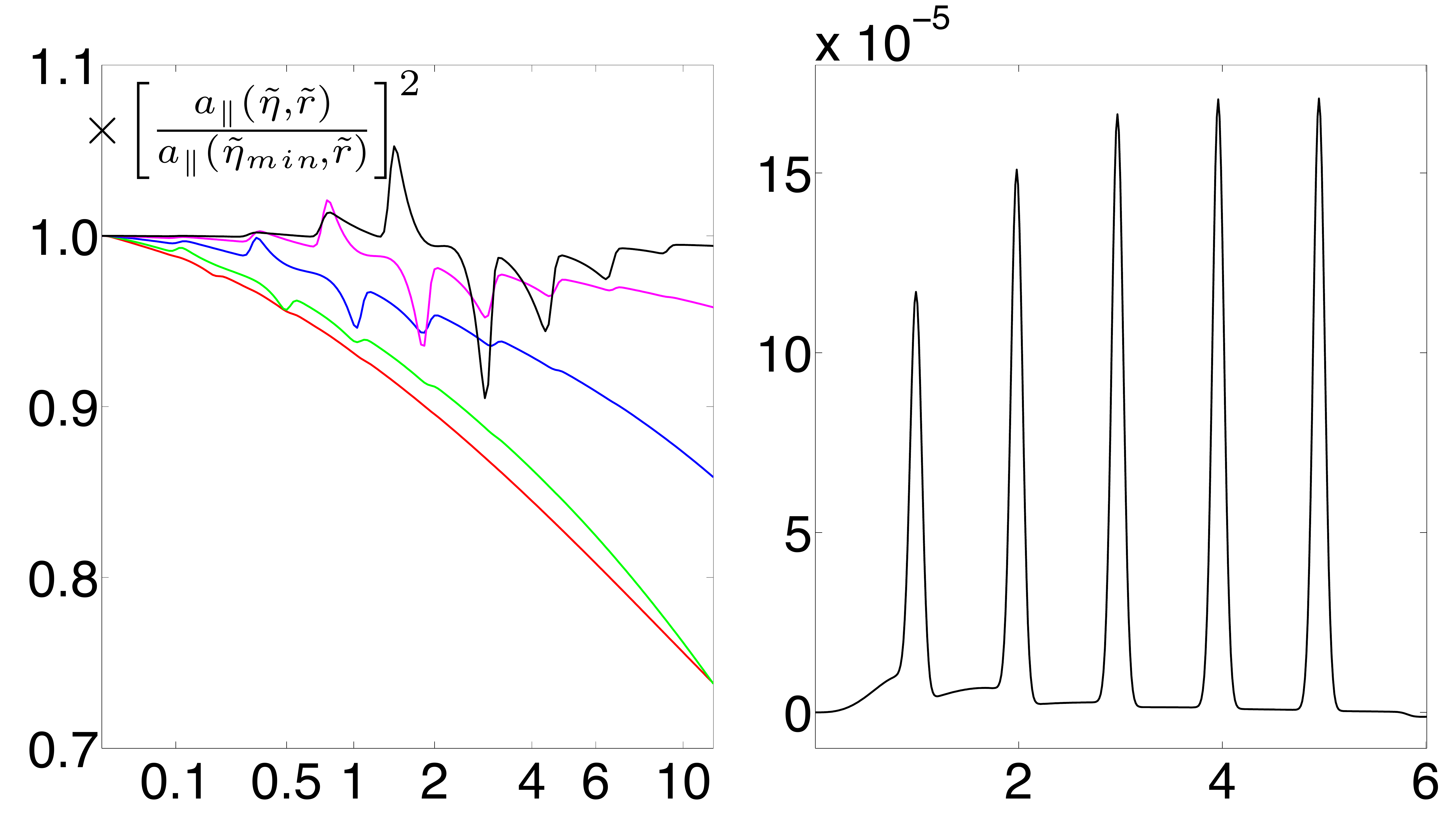}}  \end{tabular} & \hspace{-0.5cm} \begin{tabular}{c} \rule{0pt}{3.9cm}{\includegraphics[width=0.52\textwidth]{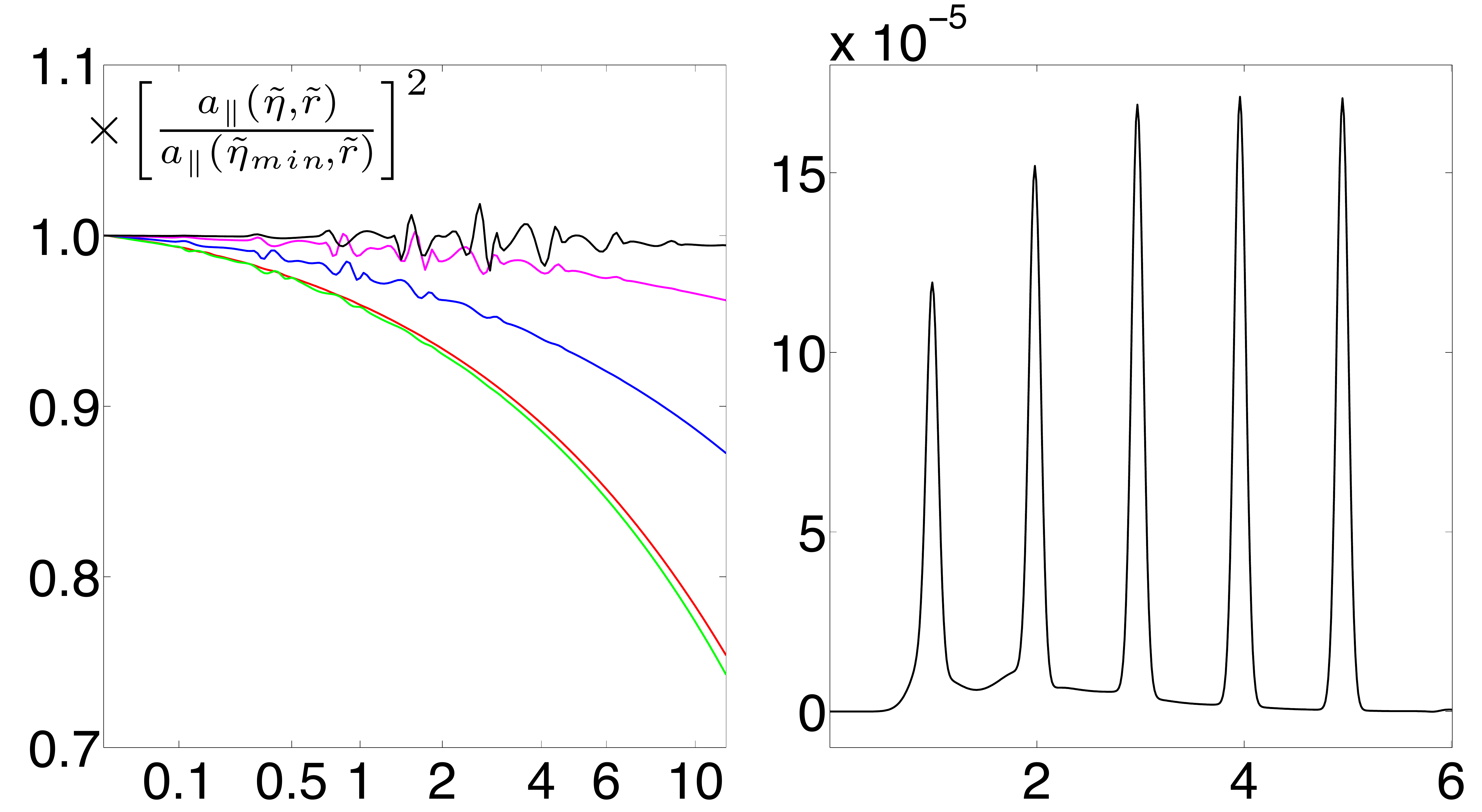}}  \end{tabular} \\ \Cline{0.3pt}{1-3}
  \multicolumn{1}{|c|}{$\!\!\chi\!\!$}  &   \hspace{-0.4cm}\begin{tabular}{c} \rule{0pt}{3.8cm}{\includegraphics[width=0.52\textwidth]{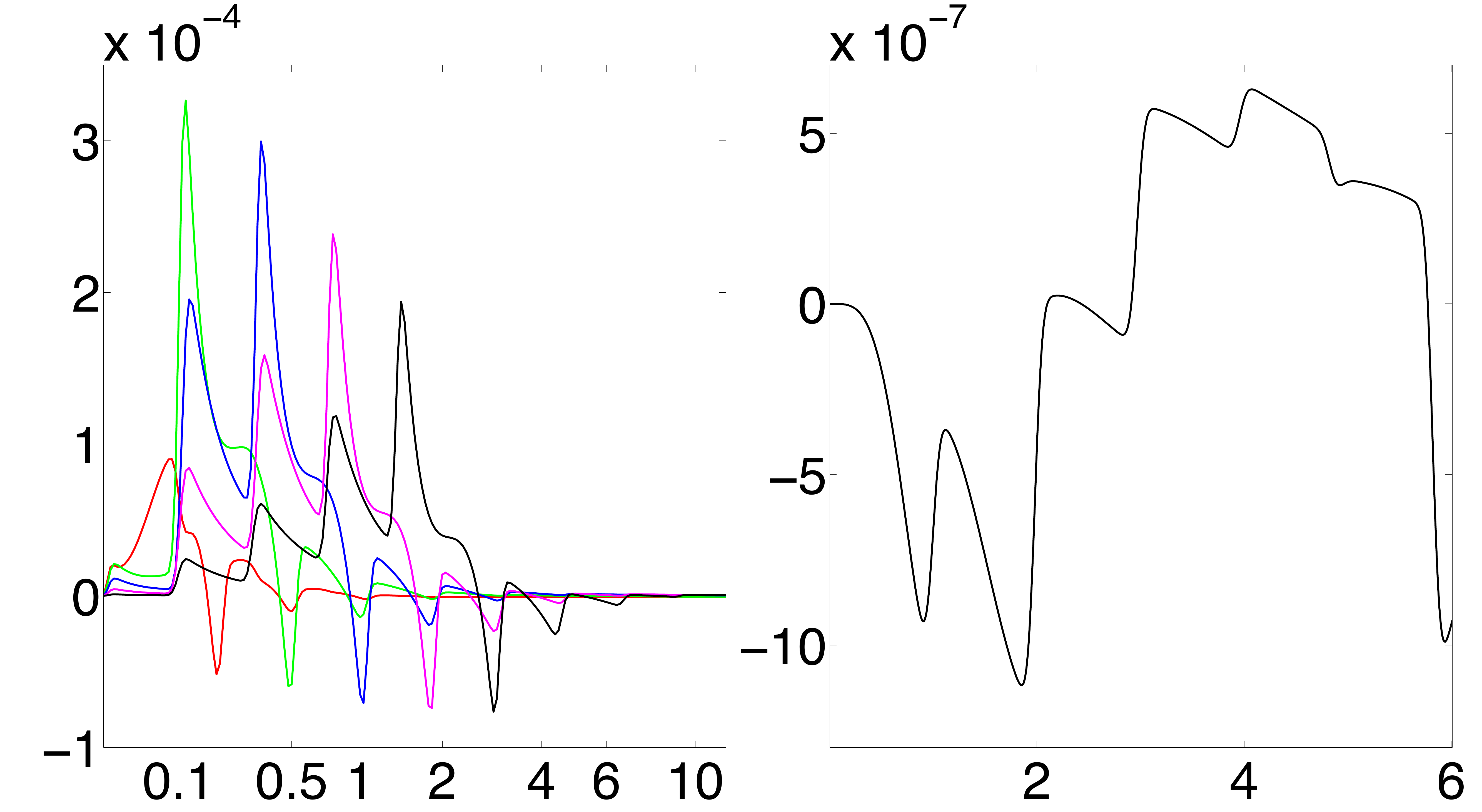}}  \end{tabular} & \hspace{-0.5cm} \begin{tabular}{c} \rule{0pt}{3.8cm}{\includegraphics[width=0.52\textwidth]{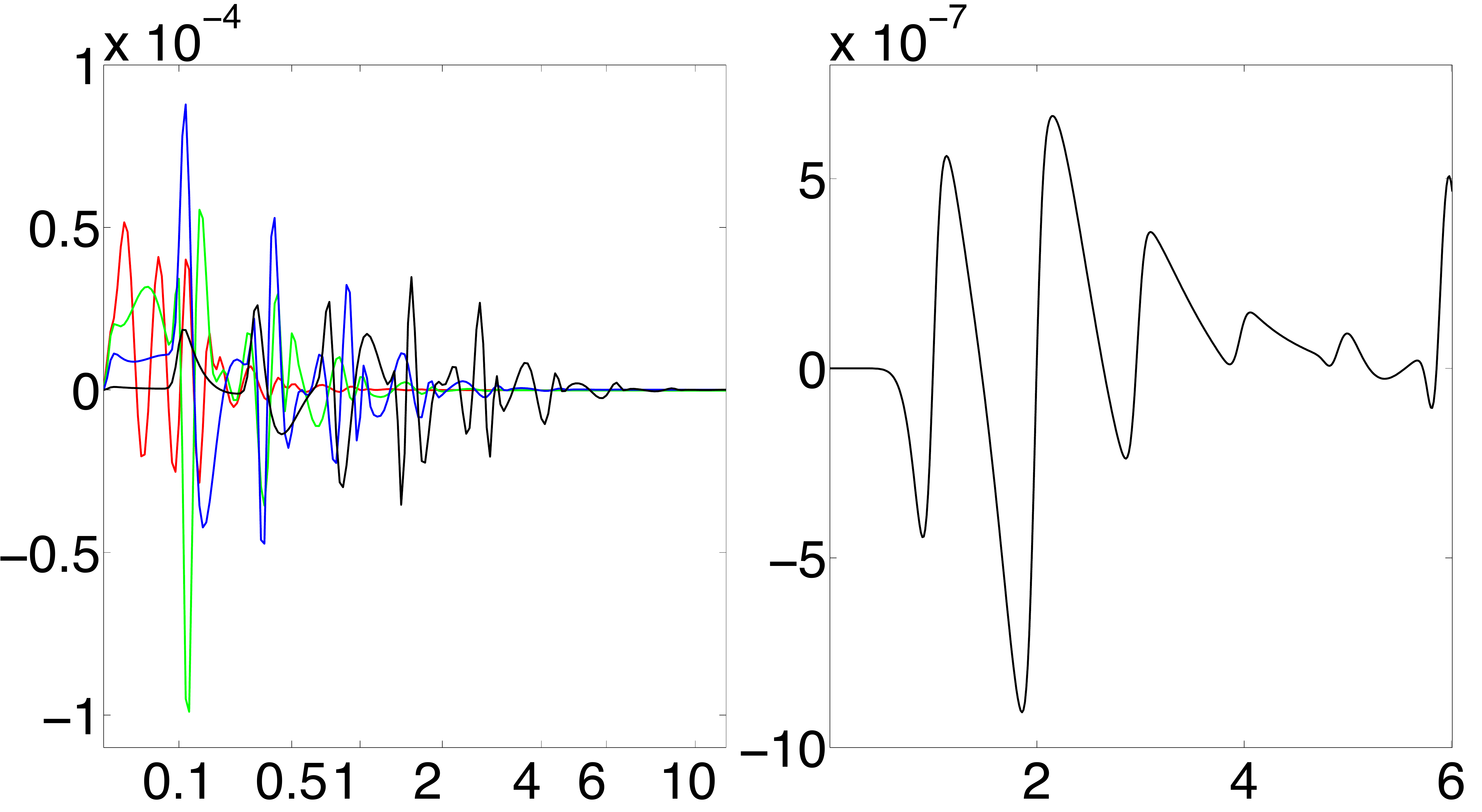}}  \end{tabular} \\ \Cline{0.3pt}{1-3}
 \multicolumn{1}{c|}{}  &  \multicolumn{1}{c|}{\hspace{0.1cm}\small{}Cosmic time (Gyr)\hspace{0.9cm}Distance (Gpc)} & \multicolumn{1}{c|}{\hspace{0.1cm}\small{}Cosmic time (Gyr)\hspace{0.9cm}Distance (Gpc)} \\ \Cline{0.3pt}{2-3}
 \multicolumn{1}{c}{} & \multicolumn{2}{c}{}\\[-3.0ex]
\end{tabular}
\caption{Temporal and spatial slices through the spacetime evolution observed in the central panel of Fig.~\ref{Tab:Case1and2Tab}. The variable $\varsigma$ decays roughly $\propto \apar^{-2}$ in the quasi-FLRW regions, but decreases more quickly deep inside the void due to the faster expansion rate there. Compared to the initial amplitude of $\varsigma$, $\varphi$ and $\chi$ remains sub-percent in magnitude.}
\label{Tab:Case3and4Tab}
\end{figure*}

\begin{figure*}[!b]
\centering
\begin{tabular}{c|m{6.35cm}|m{6.35cm}|}
 \multicolumn{1}{c}{} & \multicolumn{1}{c}{\small{}\textbf{CASE 5}} &  \multicolumn{1}{c}{\small{}\textbf{CASE 6}}\\
\Cline{0.3pt}{2-3}
   &  \multicolumn{1}{c|}{\small{}\hspace{0.05cm}Time evolution\hspace{1.35cm}Profile today}  & \multicolumn{1}{c|}{\small{}\hspace{0.05cm}Time evolution\hspace{1.35cm}Profile today} \\ \Cline{0.3pt}{1-3} 
 \multicolumn{1}{|c|}{$\!\!\varphi\!\!$}  &  \hspace{-0.5cm} \begin{tabular}{c} \rule{0pt}{3.8cm}{\includegraphics[width=0.52\textwidth]{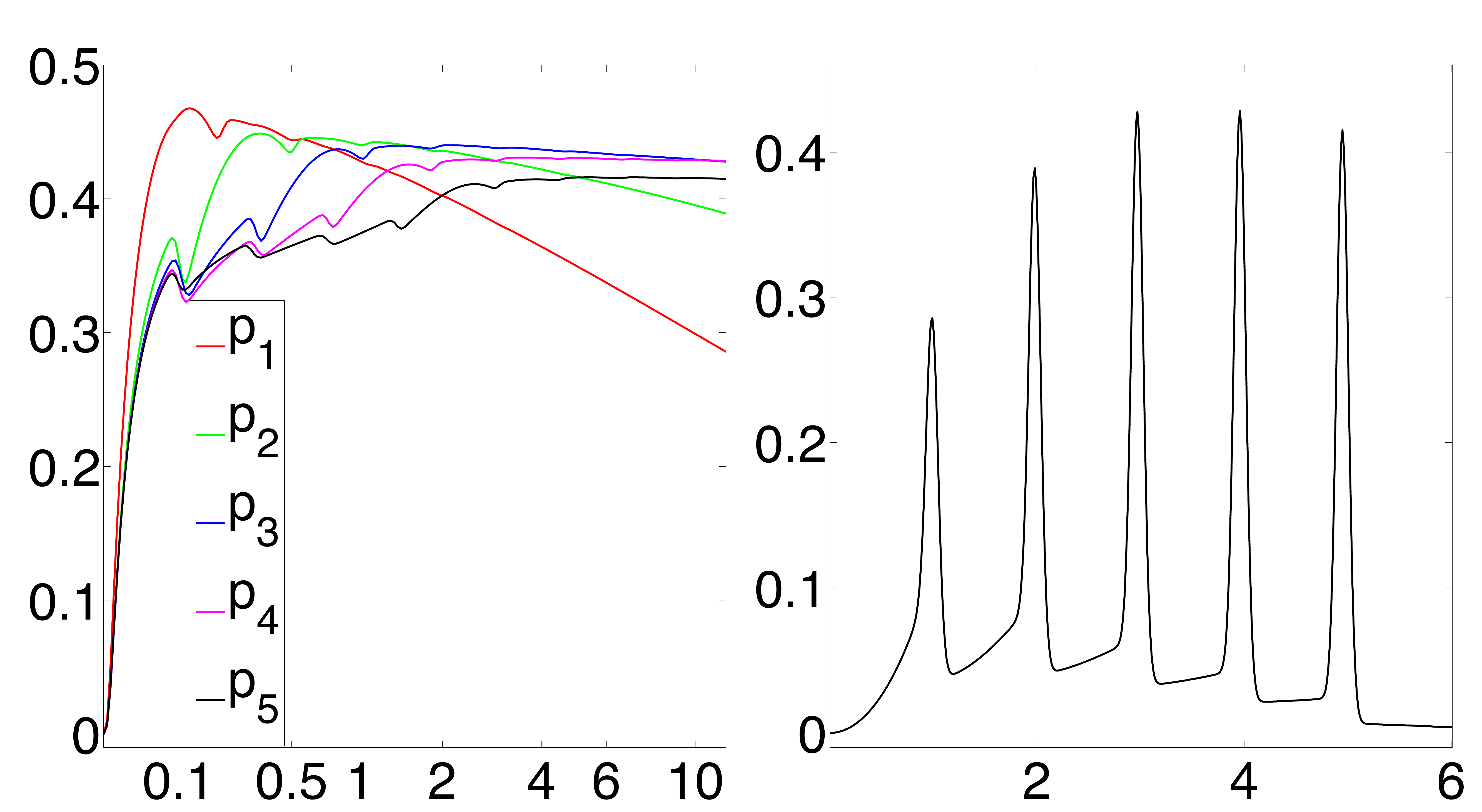}}  \end{tabular} &   \hspace{-0.5cm} \begin{tabular}{c} \rule{0pt}{3.8cm}{\includegraphics[width=0.52\textwidth]{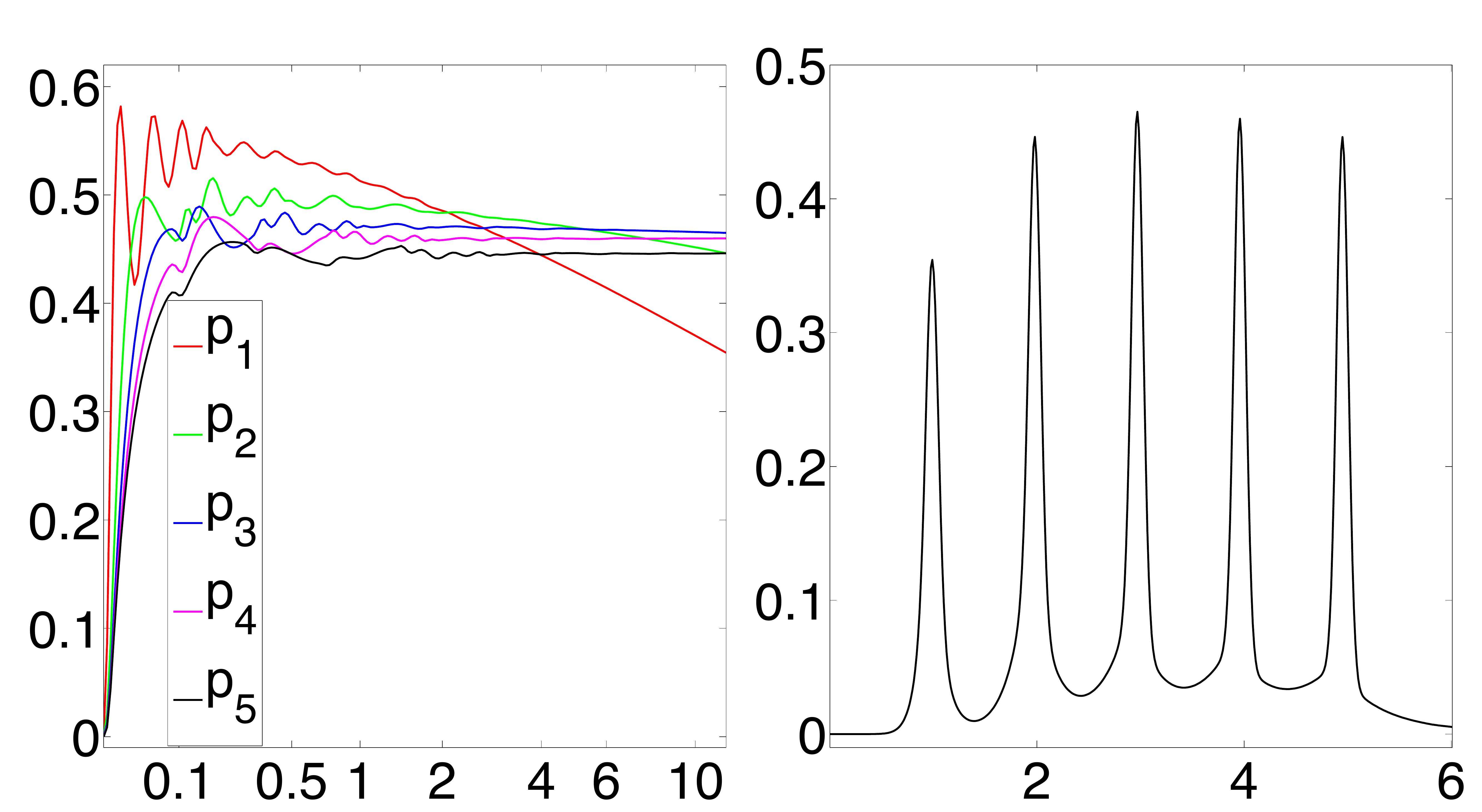}}  \end{tabular} \\  \Cline{0.3pt}{1-3}
 \multicolumn{1}{|c|}{$\!\!\varsigma\!\!$}  &  \hspace{-0.4cm}\begin{tabular}{c} \rule{0pt}{3.9cm}{\includegraphics[width=0.52\textwidth]{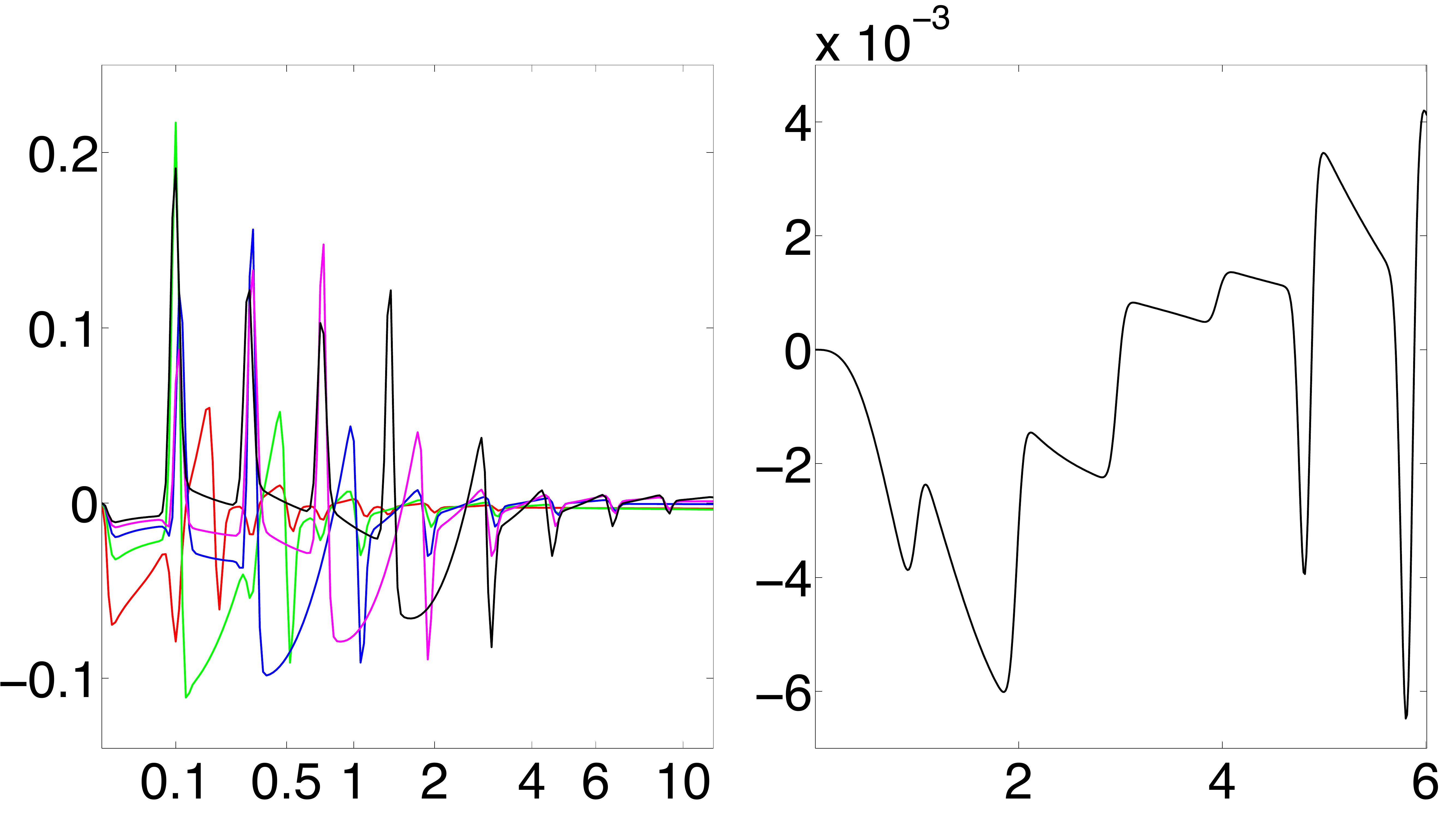}}  \end{tabular} & \hspace{-0.5cm} \begin{tabular}{c} \rule{0pt}{3.9cm}{\includegraphics[width=0.52\textwidth]{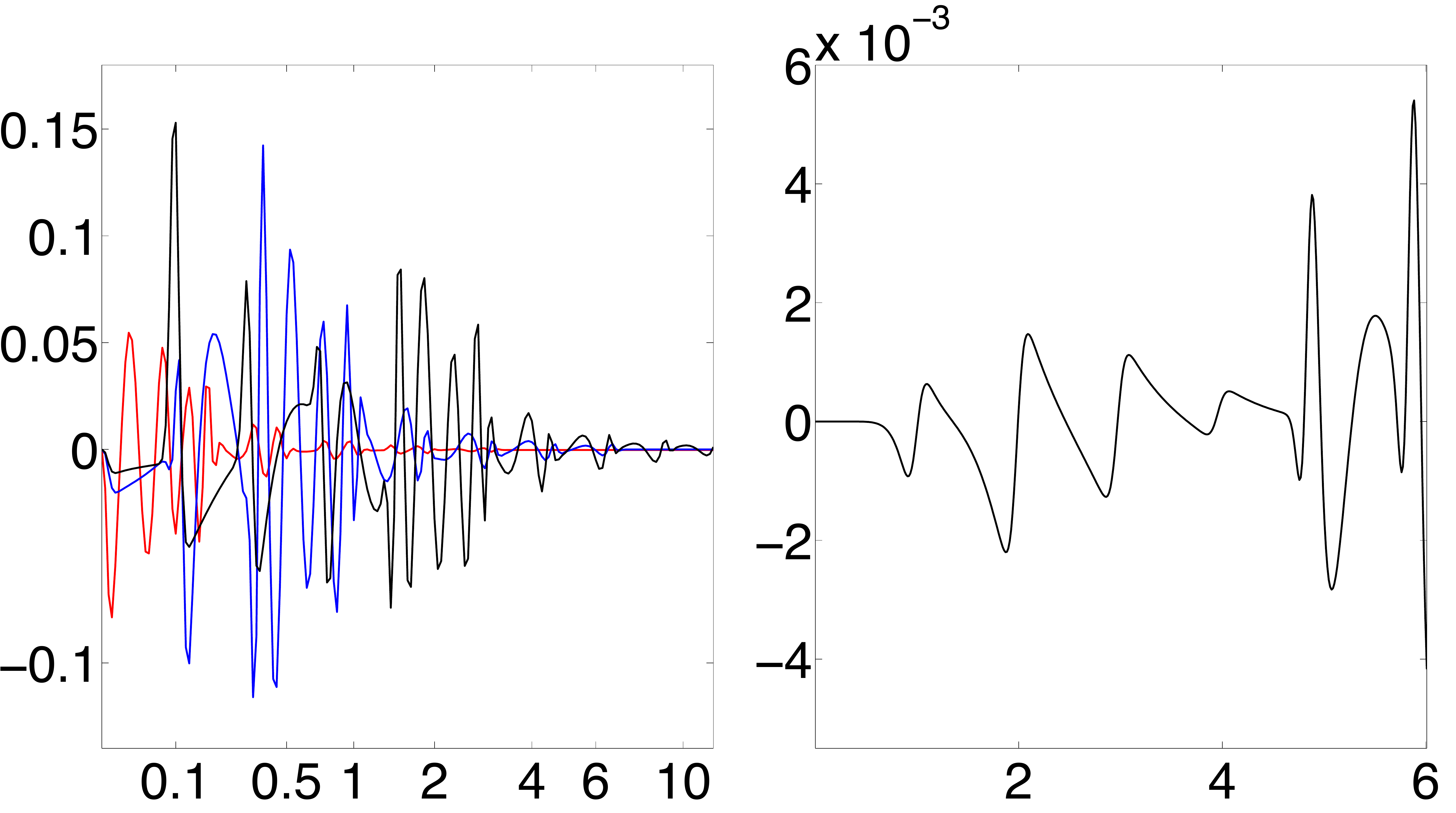}}  \end{tabular} \\ \Cline{0.3pt}{1-3}
  \multicolumn{1}{|c|}{$\!\!\chi\!\!$}  &   \hspace{-0.4cm}\begin{tabular}{c} \rule{0pt}{3.9cm}{\includegraphics[width=0.52\textwidth]{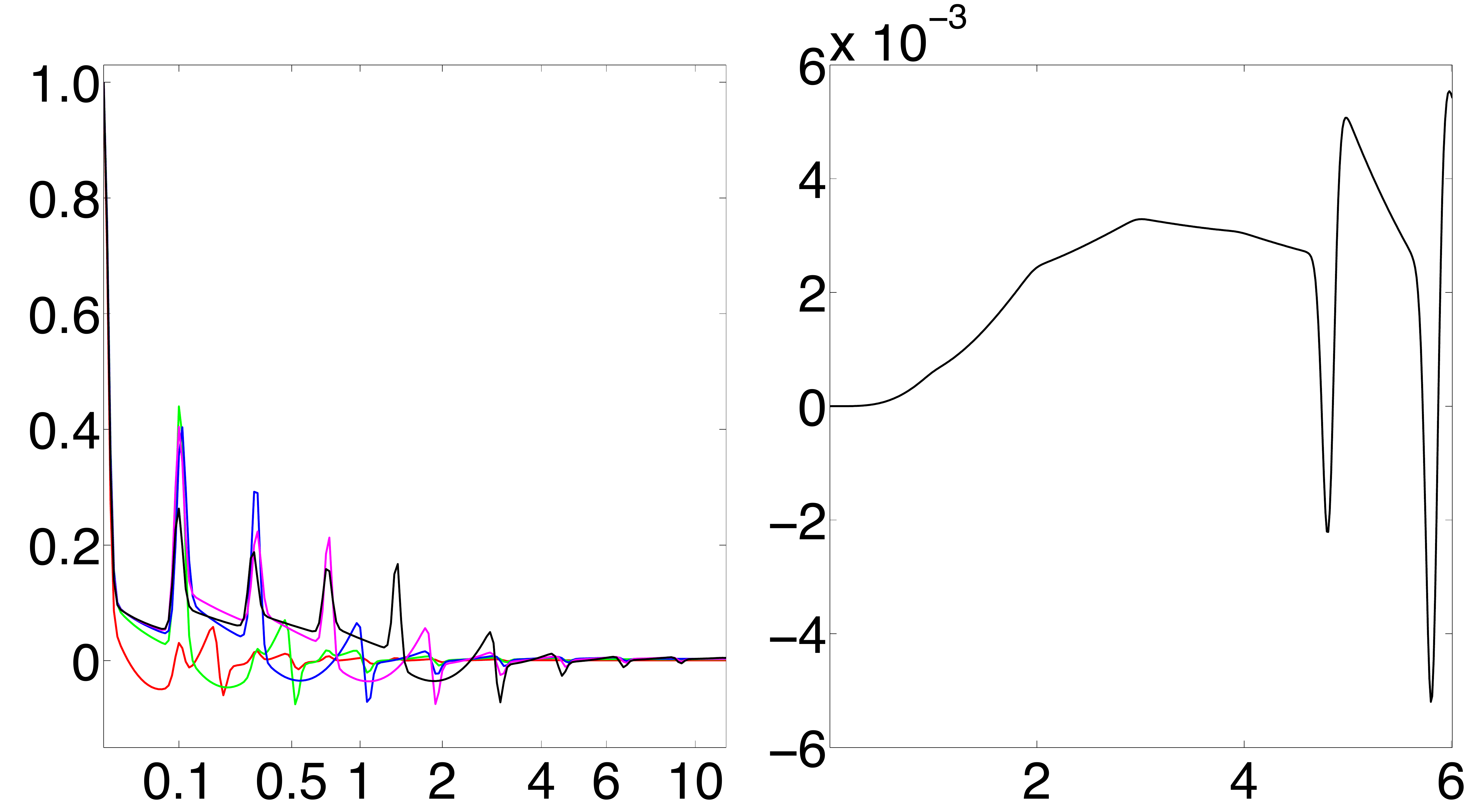}}  \end{tabular} & \hspace{-0.5cm} \begin{tabular}{c} \rule{0pt}{3.9cm}{\includegraphics[width=0.52\textwidth]{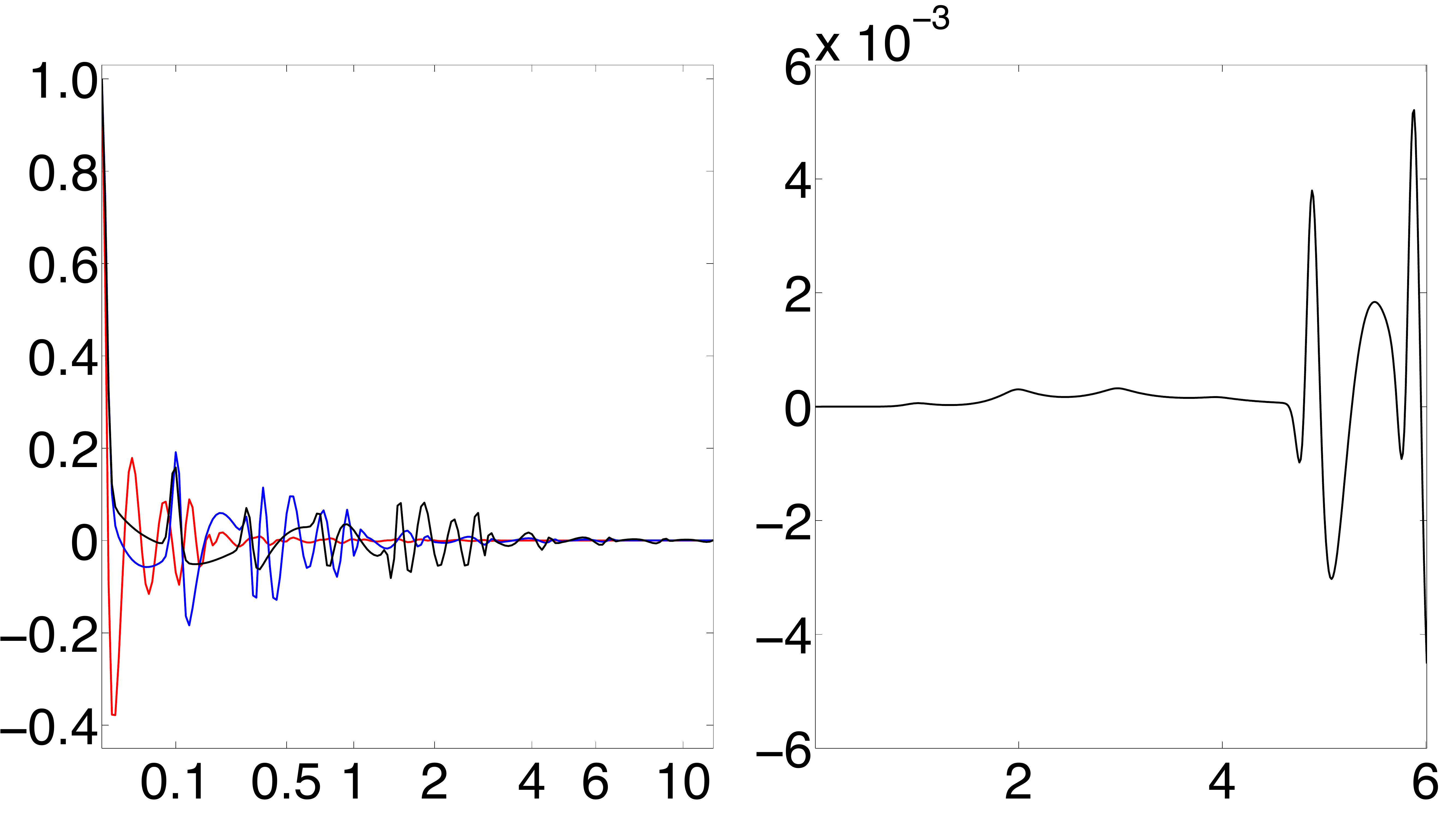}}  \end{tabular} \\ \Cline{0.3pt}{1-3}
 \multicolumn{1}{c|}{}  &  \multicolumn{1}{c|}{\hspace{0.1cm}\small{}Cosmic time (Gyr)\hspace{0.9cm}Distance (Gpc)} & \multicolumn{1}{c|}{\hspace{0.1cm}\small{}Cosmic time (Gyr)\hspace{0.9cm}Distance (Gpc)} \\ \Cline{0.3pt}{2-3}
 \multicolumn{1}{c}{} & \multicolumn{2}{c}{}\\[-3.0ex]
\end{tabular}
\caption{Temporal and spatial slices through the spacetime evolution observed in the bottom panel of Fig.~\ref{Tab:Case1and2Tab}. Here it is clear that the evolution of $\chi$ is dominated by a propagating mode. It is interesting to note that the level of $\varphi$ generated is of a similar order of magnitude than the initial $\chi$, and again showing a slower ($\sim$ 25\%) growth rate inside the void compared to the outskirts. The variable $\varsigma$ is also produced to a significant proportion early on, but nevertheless decays quickly with time.}
\label{Tab:Case5and6Tab}
\end{figure*}

\begin{figure*}[!b]
\flushleft
\begin{tabular}{cc|m{5.9cm}|m{5.9cm}|}
\Cline{0.3pt}{3-4}
  & &  \multicolumn{1}{c|}{\rule{0pt}{0.3cm}\hspace{-0.3cm}$\Delta/\normmax|\Delta\tinysubmin|$\hspace{0.35cm}$w/\normmax|w\tinysubmin|$\hspace{0.6cm}$\delta E_{(TF)}$}  & \multicolumn{1}{c|}{\rule{0pt}{0.3cm}\hspace{-0.3cm}$\Delta/\normmax|\Delta\tinysubmin|$\hspace{0.35cm}$w/\normmax|w\tinysubmin|$\hspace{0.6cm}$\delta E_{(TF)}$} \\
\Cline{0.3pt}{2-4} 
  \multicolumn{1}{c}{}  & \multicolumn{1}{|c|}{} &  \multicolumn{1}{c|}{\rule{0pt}{0.3cm}\hspace{-1.5cm}\tiny{}$[65]$\hspace{1.8cm}$[8.1]$\hspace{1.7cm}$[0]$}  & \multicolumn{1}{c|}{\rule{0pt}{0.3cm}\hspace{-0.5cm}\tiny{}$[59]$\hspace{1.8cm}$[8.1]$\hspace{1.7cm}$[1.4\!\times\!10^{-7}]$} \\ [-0.2ex]
& \multicolumn{1}{|c|}{\parbox[t]{1.5mm}{{\rotatebox[origin=c]{90}{\hspace{0.46cm}Initialised $\varphi$}}}} & \hspace{-0.4cm} \begin{tabular}{c} \rule{0pt}{3.7cm}{\includegraphics[width=0.47\textwidth]{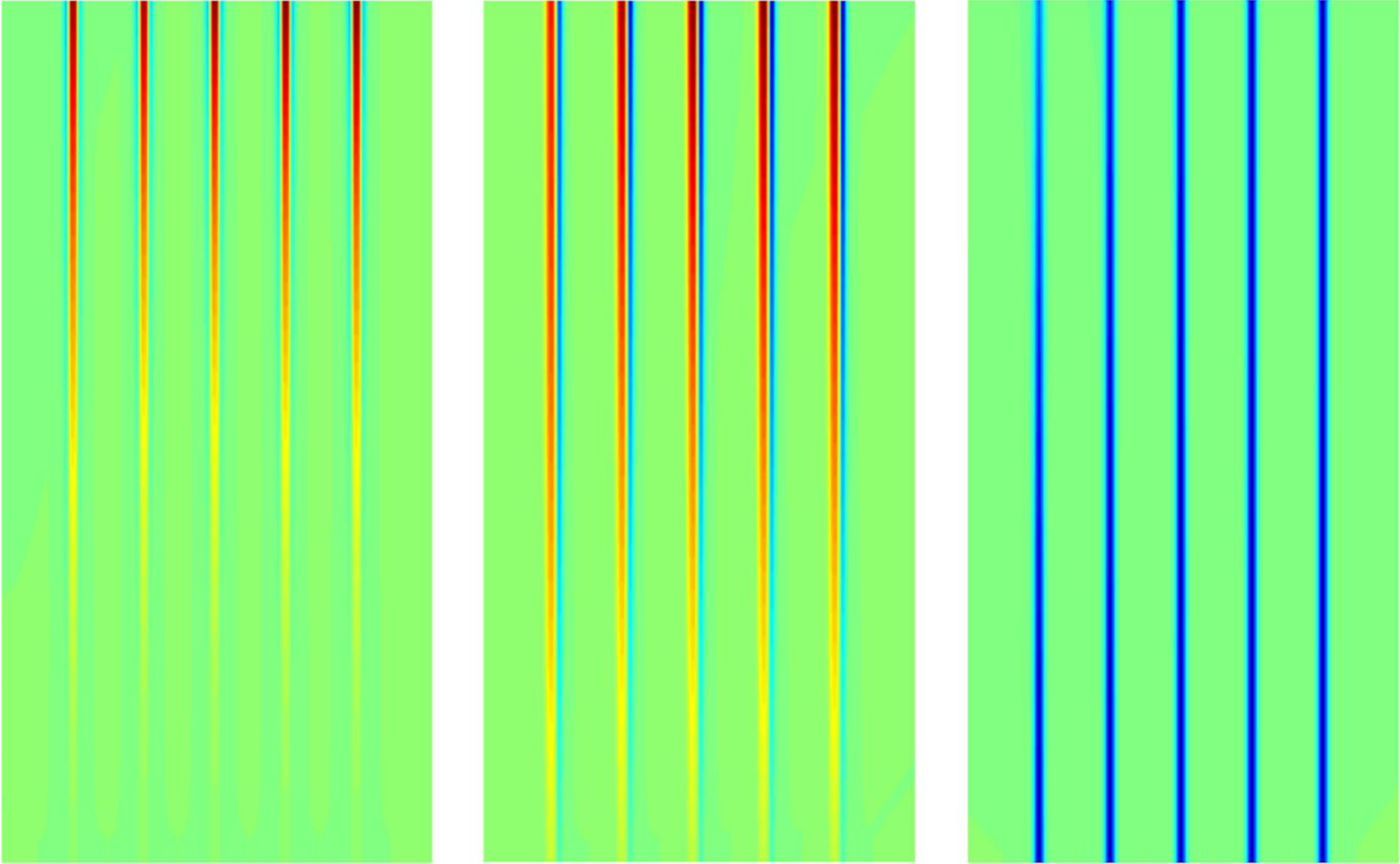}}  \end{tabular} &     \hspace{-0.3cm}\begin{tabular}{c} \rule{0pt}{3.7cm}{\includegraphics[width=0.47\textwidth]{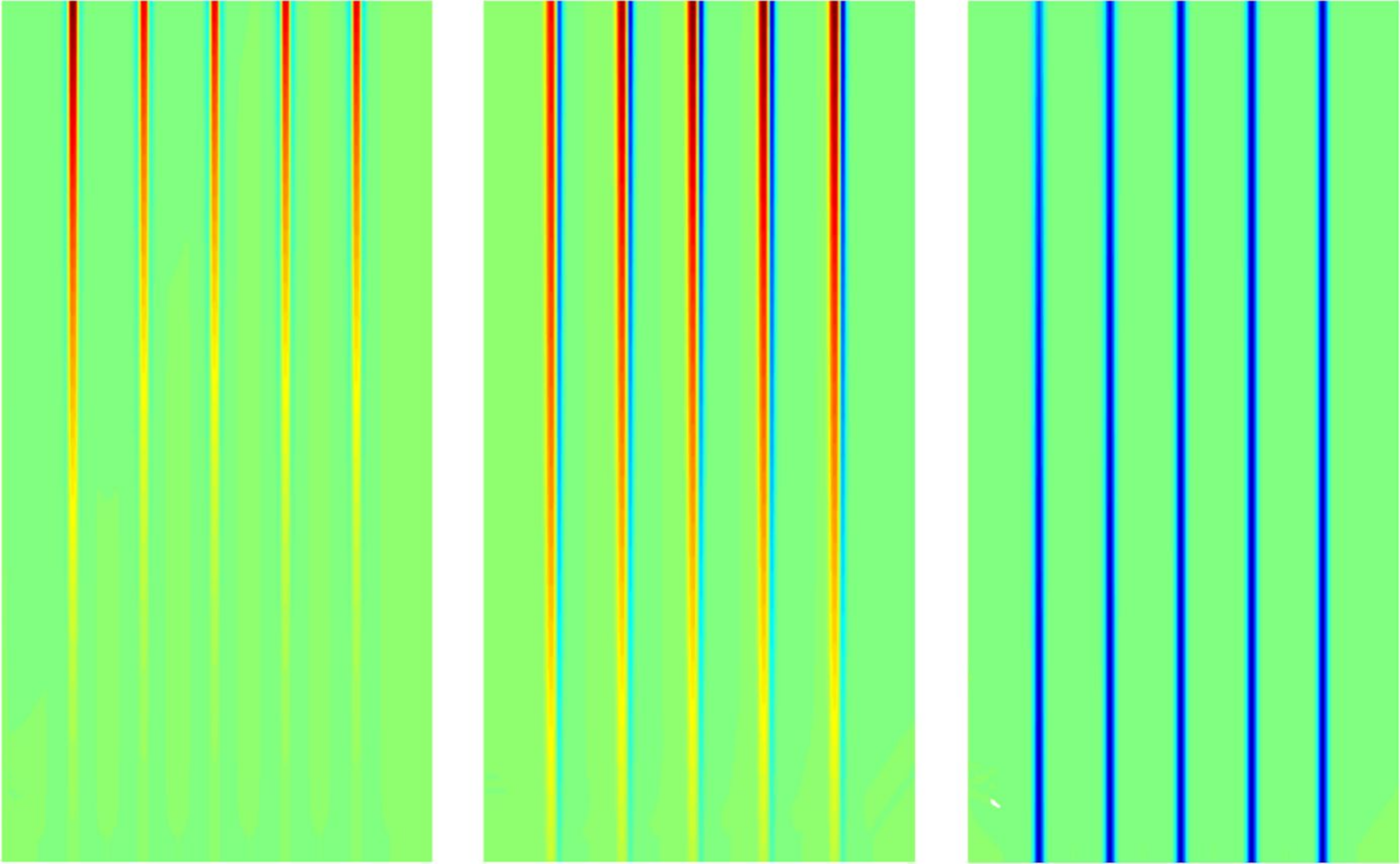}}  \end{tabular} \\ [-1.5ex]
  \multicolumn{1}{c}{}  & \multicolumn{1}{|c|}{} &  \multicolumn{1}{c|}{\rule{0pt}{0.3cm}\hspace{-1.0cm}\tiny{}$[-30]$\hspace{1.6cm}$[-8.1]$\hspace{1.5cm}$[-0.5]$}  & \multicolumn{1}{c|}{\rule{0pt}{0.3cm}\hspace{-0.9cm}\tiny{}$[-20]$\hspace{1.6cm}$[-8.1]$\hspace{1.5cm}$[-0.5]$} \\
& \multicolumn{1}{|c|}{}  &   \multicolumn{1}{c|}{\small{}\textbf{CASE 1}} & \multicolumn{1}{c|}{\small{}\textbf{CASE 2}} \\ \Cline{0.3pt}{2-4}
\multicolumn{1}{c}{}  & \multicolumn{1}{|c|}{} &  \multicolumn{1}{c|}{\rule{0pt}{0.3cm}\hspace{-0.4cm}\tiny{}$[1.0]$\hspace{1.7cm}$[1.0]$\hspace{1.7cm}$[3.7\!\times\!10^{-4}]$}  & \multicolumn{1}{c|}{\rule{0pt}{0.3cm}\hspace{-0.3cm}\tiny{}$[3.0]$\hspace{1.7cm}$[1.0]$\hspace{1.7cm}$[2.1\!\times\!10^{-3}]$} \\[-0.2ex]
\begin{tabular}{c} \parbox[t]{0.0mm}{{\rotatebox[origin=c]{90}{\hspace{0.44cm}\large{Cosmic time $\longrightarrow$ }}}} \end{tabular}  & \multicolumn{1}{|c|}{\parbox[t]{1.5mm}{{\rotatebox[origin=c]{90}{\hspace{0.46cm}Initialised $\varsigma$}}}}  &   \hspace{-0.35cm}\begin{tabular}{c} \rule{0pt}{3.7cm}{\includegraphics[width=0.47\textwidth]{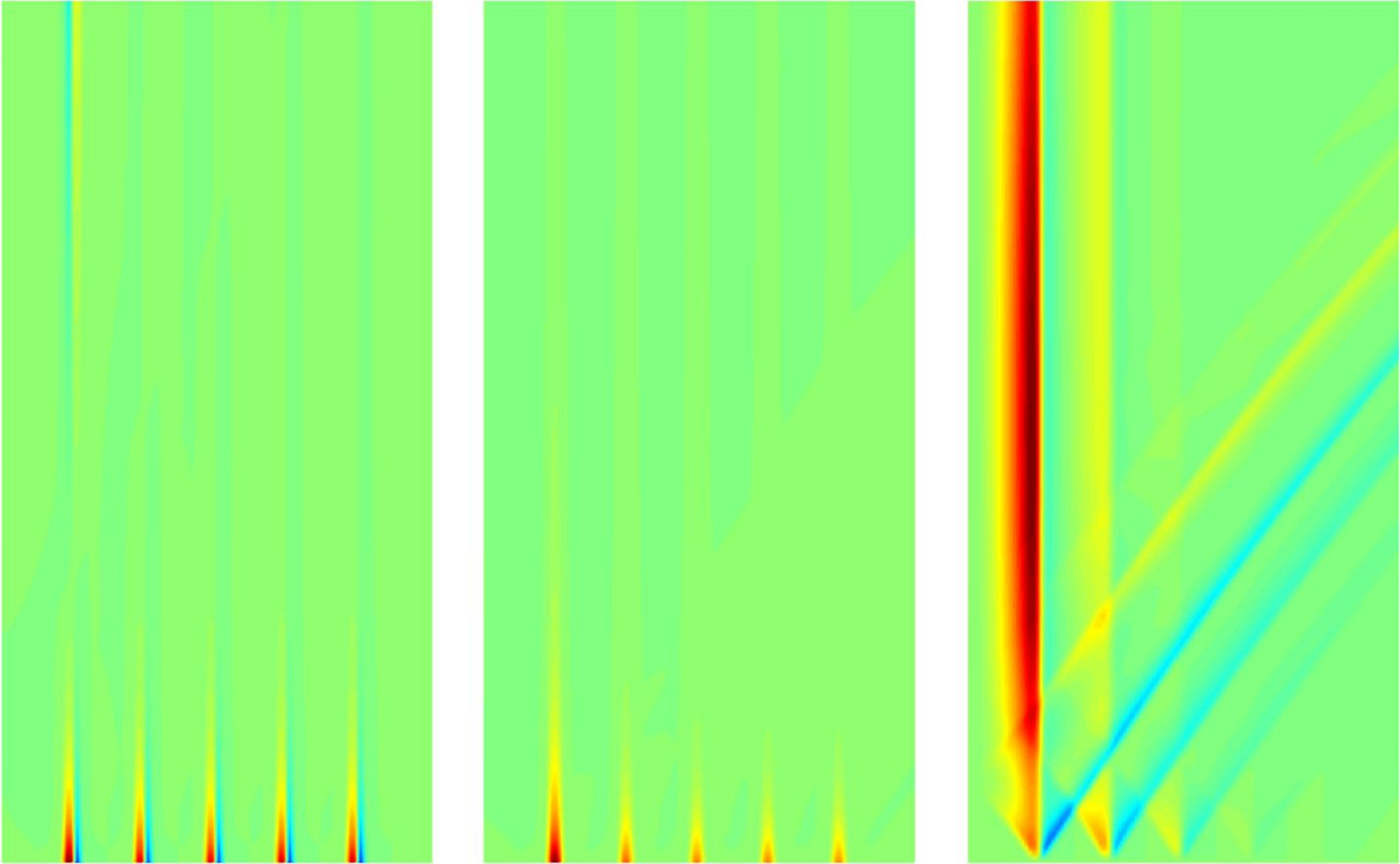}}  \end{tabular} &   \hspace{-0.3cm}\begin{tabular}{c} \rule{0pt}{3.7cm}{\includegraphics[width=0.47\textwidth]{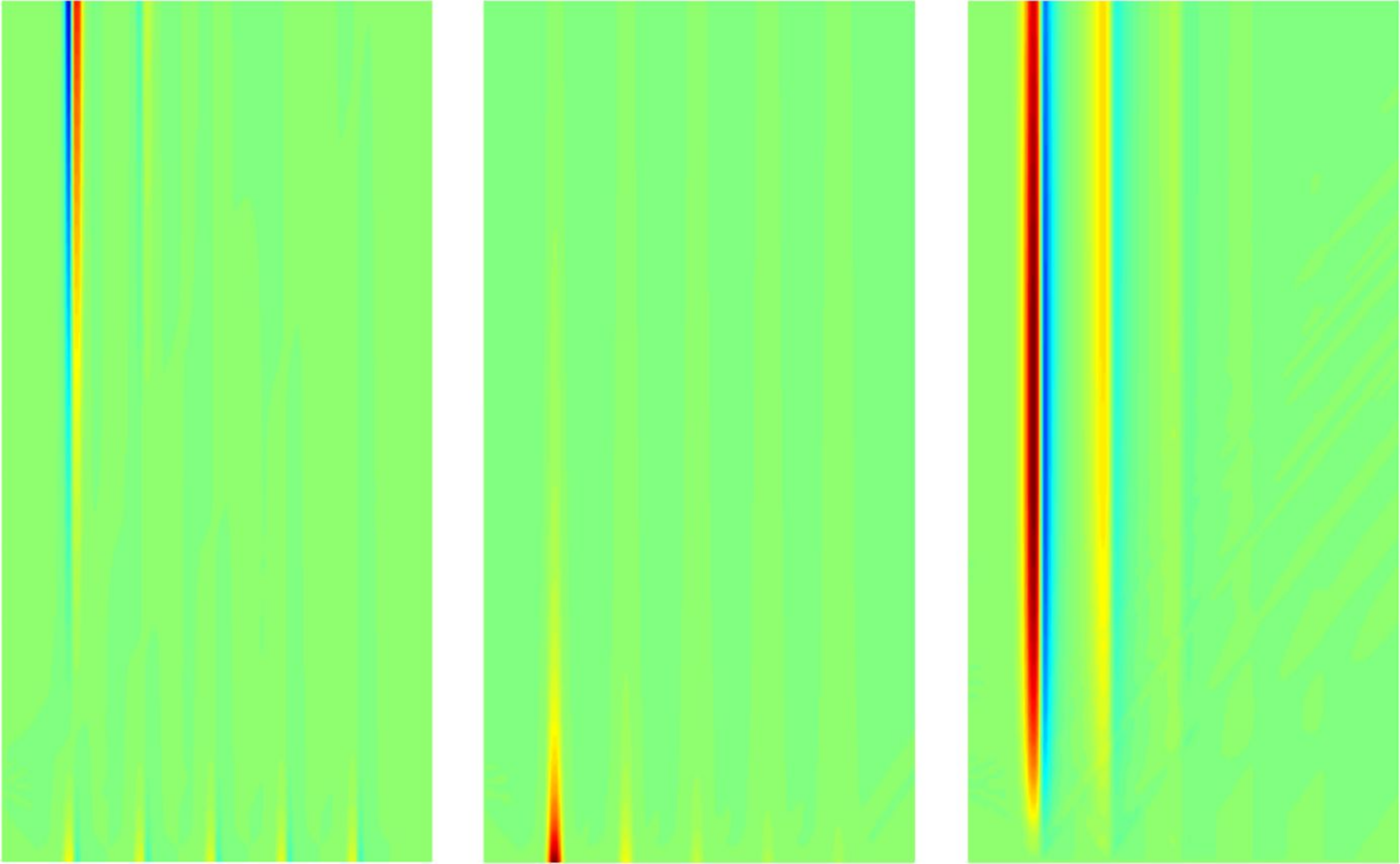}}  \end{tabular} \\ [-1.5ex]
\multicolumn{1}{c}{}  & \multicolumn{1}{|c|}{} &  \multicolumn{1}{c|}{\rule{0pt}{0.3cm}\hspace{-0.25cm}\tiny{}$[-0.8]$\hspace{1.5cm}$[-2.4\!\!\times\!\!10^{-3}]$\hspace{0.85cm}$[-2.0\!\times\!10^{-4}]$}  & \multicolumn{1}{c|}{\rule{0pt}{0.3cm}\hspace{-0.1cm}\tiny{}$[-4.4]$\hspace{1.5cm}$[-3.5\!\!\times\!\!10^{-3}]$\hspace{0.8cm}$[-1.4\!\times\!10^{-3}]$} \\
& \multicolumn{1}{|c|}{}  &   \multicolumn{1}{c|}{\small{}\textbf{CASE 3}} & \multicolumn{1}{c|}{\small{}\textbf{CASE 4}} \\ 
\Cline{0.3pt}{2-4}
\multicolumn{1}{c}{}  & \multicolumn{1}{|c|}{} &  \multicolumn{1}{c|}{\rule{0pt}{0.3cm}\hspace{-0.4cm}\tiny{}$[3.7\!\times\!10^{2}]$\hspace{1.1cm}$[3.2]$\hspace{1.7cm}$[3.5\!\times\!10^{-1}]$}  & \multicolumn{1}{c|}{\rule{0pt}{0.3cm}\hspace{-0.25cm}\tiny{}$[1.5\!\times\!10^{2}]$\hspace{1.1cm}$[3.2]$\hspace{1.65cm}$[1.7\!\times\!10^{-1}]$} \\[-0.2ex]
 & \multicolumn{1}{|c|}{\parbox[t]{1.5mm}{{\rotatebox[origin=c]{90}{\hspace{0.46cm}Initialised $\chi$}}}}  &    \hspace{-0.35cm}\begin{tabular}{c} \rule{0pt}{3.7cm}{\includegraphics[width=0.47\textwidth]{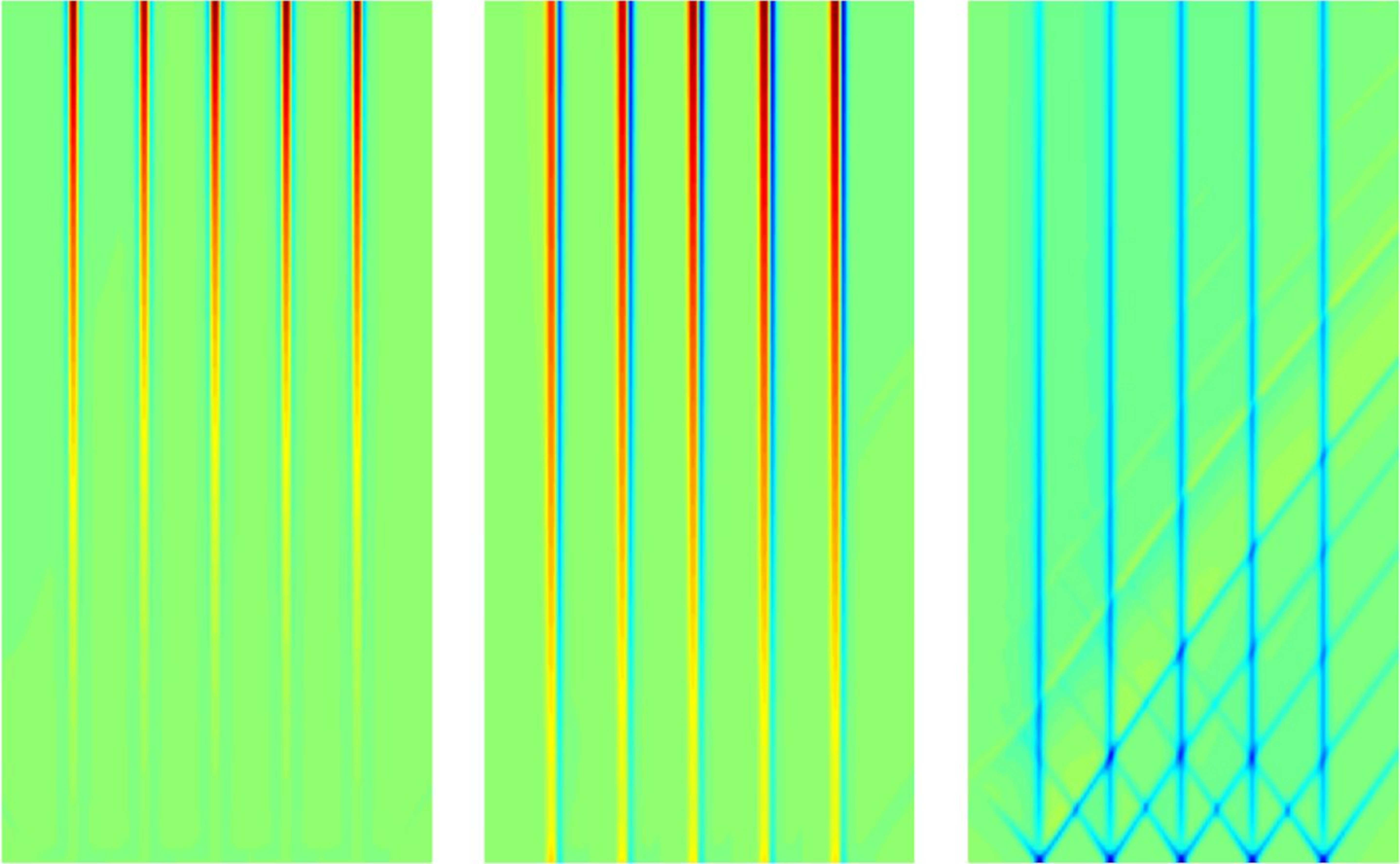}}  \end{tabular} &  \hspace{-0.38cm} \begin{tabular}{c} \rule{0pt}{3.7cm}{\includegraphics[width=0.47\textwidth]{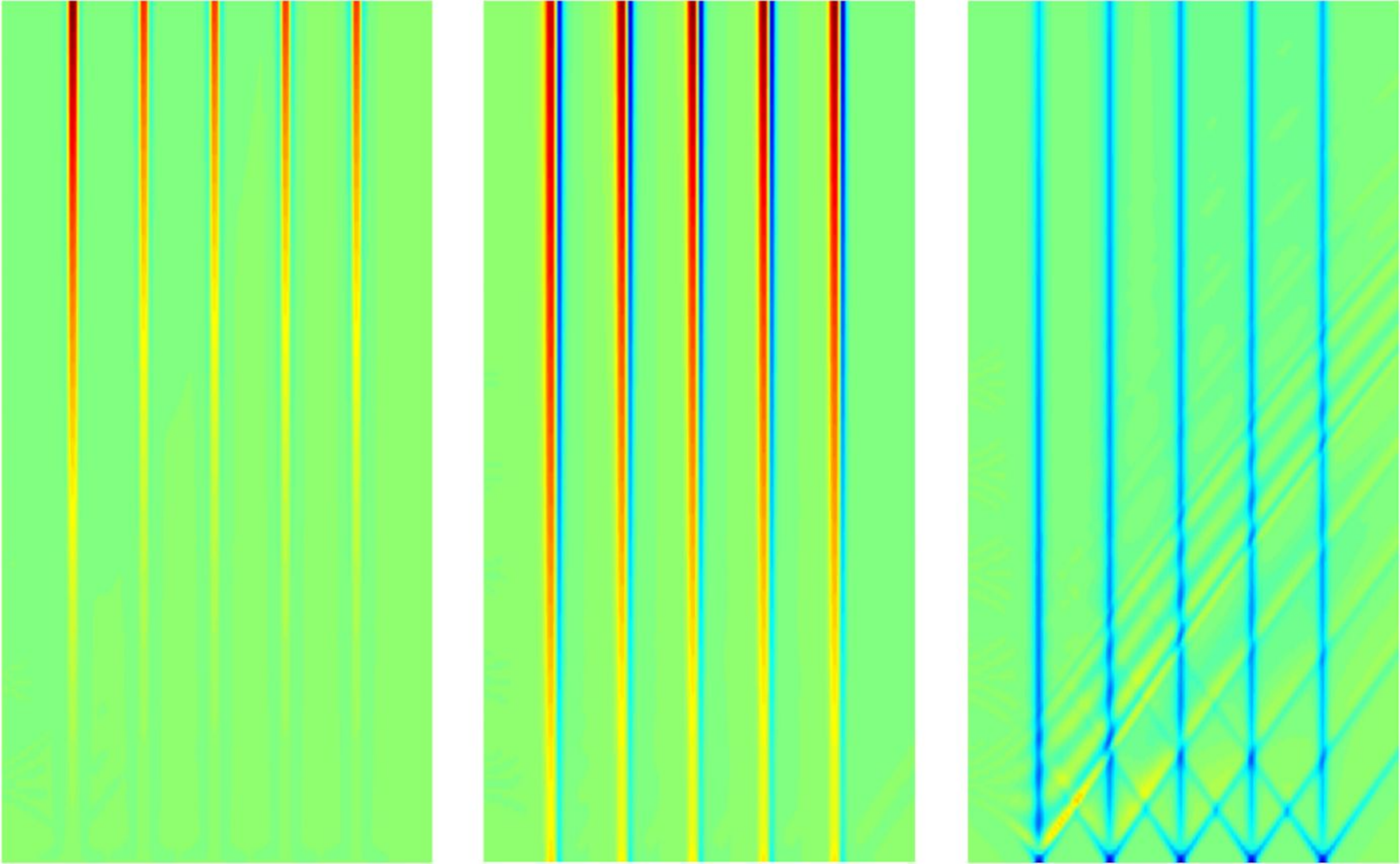}}  \end{tabular} \\[-1.5ex]
 \multicolumn{1}{c}{}  & \multicolumn{1}{|c|}{} &  \multicolumn{1}{c|}{\rule{0pt}{0.3cm}\hspace{-0.2cm}\tiny{}$[-1.7\!\times\!10^{2}]$\hspace{0.9cm}$[-3.3]$\hspace{1.5cm}$[-5.0\!\times\!10^{-1}]$}  & \multicolumn{1}{c|}{\rule{0pt}{0.3cm}\hspace{-0.1cm}\tiny{}$[-0.4\!\times\!10^{2}]$\hspace{0.9cm}$[-3.3]$\hspace{1.5cm}$[-5.0\!\times\!10^{-1}]$} \\
 & \multicolumn{1}{|c|}{}  &   \multicolumn{1}{c|}{\small{}\textbf{CASE 5}} & \multicolumn{1}{c|}{\small{}\textbf{CASE 6}} \\ 
 \Cline{0.3pt}{2-4}
  \multicolumn{1}{c}{}  & \multicolumn{1}{c}{} & \multicolumn{2}{c}{}\\[-2ex]
 \multicolumn{1}{c}{}  & \multicolumn{1}{c}{} & \multicolumn{2}{c}{\large{Radial distance $\longrightarrow$}} \\ [-3.0ex]
\end{tabular}
\caption{Spacetime evolution of selected quantities derived from those presented in Fig.~\ref{Tab:MasterTab}. We show: $\Delta$ and $w$ normalised to their maximum values (along the radial dimension) at the initial time ($t\submin$), as well as $\delta E_{(TF)}$, which describes the sum of $\varphi$ and $\chi$. Note that, due to the initial amplitude of unity chosen throughout, $\Delta$ here eventually becomes less than -1; while $\Delta$ doesn't exactly reduce to the usual density contrast in the FLRW limit $-$ it is sourced by propagating degrees of freedom $-$ an appropriate rescaling of the initial amplitudes (fluctuations in the standard Newtonian potential $\Phi$ are $\sim$ 10$^{-4}$ at $z=100$) in any case is sufficient to avoid any issues regarding the physical interpretation of $\Delta$ as a density contrast. The maximum and minimum values of the colour scale are indicated respectively in brackets above and below each 2D plot.}
\label{Tab:DerivedQuantities}
\end{figure*}

The middle panel of Fig.~\ref{Tab:Case3and4Tab} presents the profile of $\varsigma$ today for these cases, including the time evolution along selected radii. It's clear that $\varsigma$ decays, for the most part, approximately as $\apar^{-2}$. (In the FLRW limit this would be a pure vector mode with this exact decay rate.) The greater decay in $\varsigma$ seen in the central regions of the void can be attributed to the faster expansion rate there. The top and bottom panels of Fig.~\ref{Tab:Case3and4Tab} show the profiles for the other variables, $\varphi$ and $\chi$. 

We also show the spacetime configuration of $\Delta$ in Fig.~\ref{Tab:DerivedQuantities}. Remarkably, the density contrast generated initially by the presence of the perturbation $\varsigma$ also decays very rapidly at the peak locations, except deep inside the void (first few peaks) where it starts to grow at later times (much more at small angular scales than at large ones). This is associated with the potential $\varphi$ deepening in this region at the same time: the decay of $\varsigma$ into $\varphi$ is associated with a growth of structure deep within the void.
\begin{figure*}[!b]
\centering
\begin{tabular}{c|m{6.5cm}|m{6.5cm}|}
\Cline{0.3pt}{2-3}
   &  \multicolumn{1}{c|}{$l$=2}  &\multicolumn{1}{c|}{$l$=10} \\ \Cline{0.3pt}{1-3} 
  \multicolumn{1}{|c|}{}    &   \multicolumn{1}{c|}{}    &   \multicolumn{1}{c|}{}   \\[-2ex]
  \multicolumn{1}{|c|}{}    &  \multicolumn{1}{c|}{\footnotesize{}\hspace{1.1cm}\% diff.\hspace{1.9cm}Profile today}  & \multicolumn{1}{c|}{\footnotesize{}\hspace{1.0cm}\% diff.\hspace{1.8cm}Profile today} \\[-0.4ex]
\multicolumn{1}{|c|}{$\!\!\!$\parbox[t]{1.0mm}{{\rotatebox[origin=c]{90}{\hspace{0.15cm}$\varphi$}}}} & \hspace{-0.4cm} \begin{tabular}{c} \hspace{0.15cm} \rule{0pt}{3.5cm}{\includegraphics[width=0.53\textwidth]{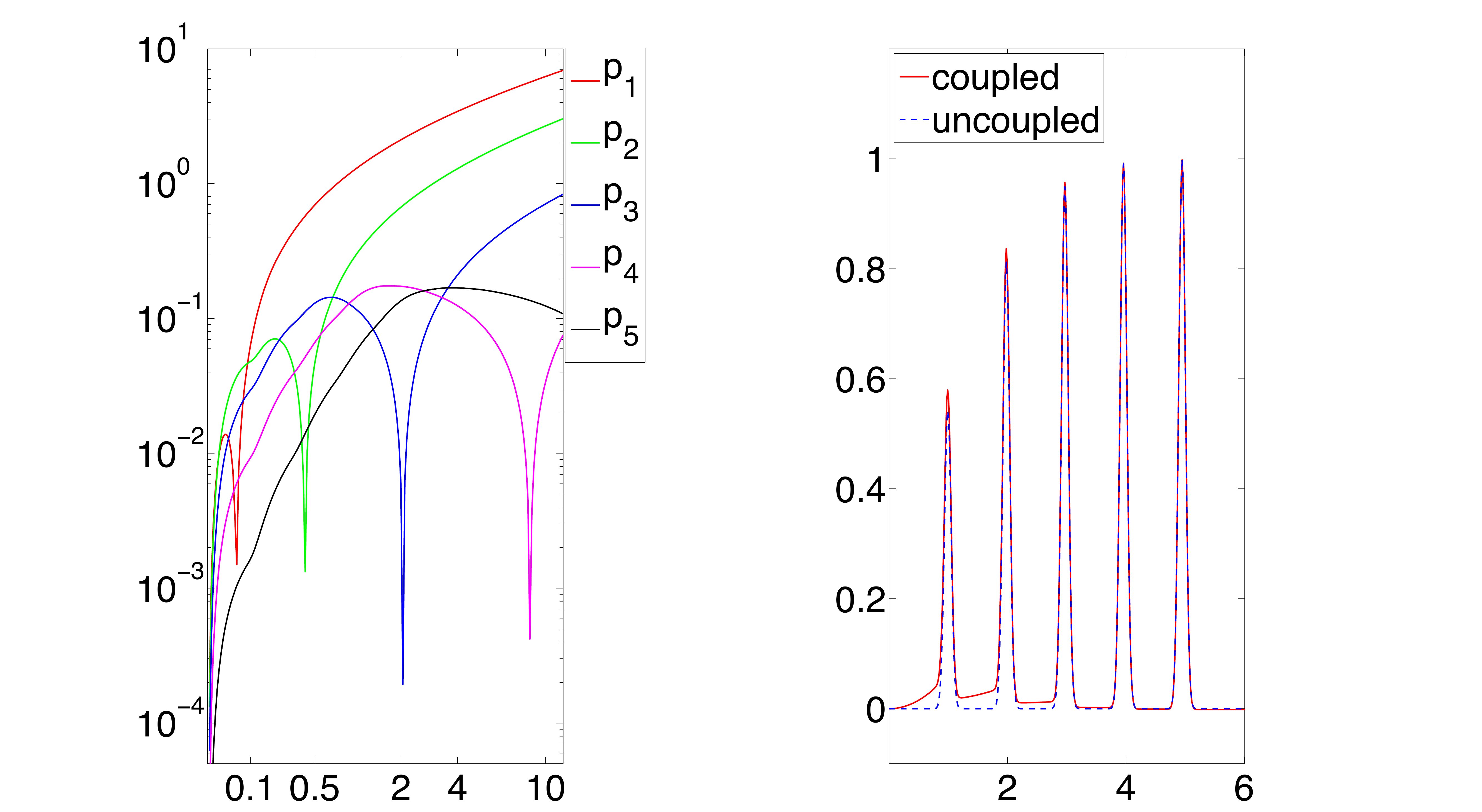}}  \end{tabular} &    \hspace{-0.4cm}\begin{tabular}{c} \hspace{0.15cm} \rule{0pt}{3.5cm}{\includegraphics[width=0.53\textwidth]{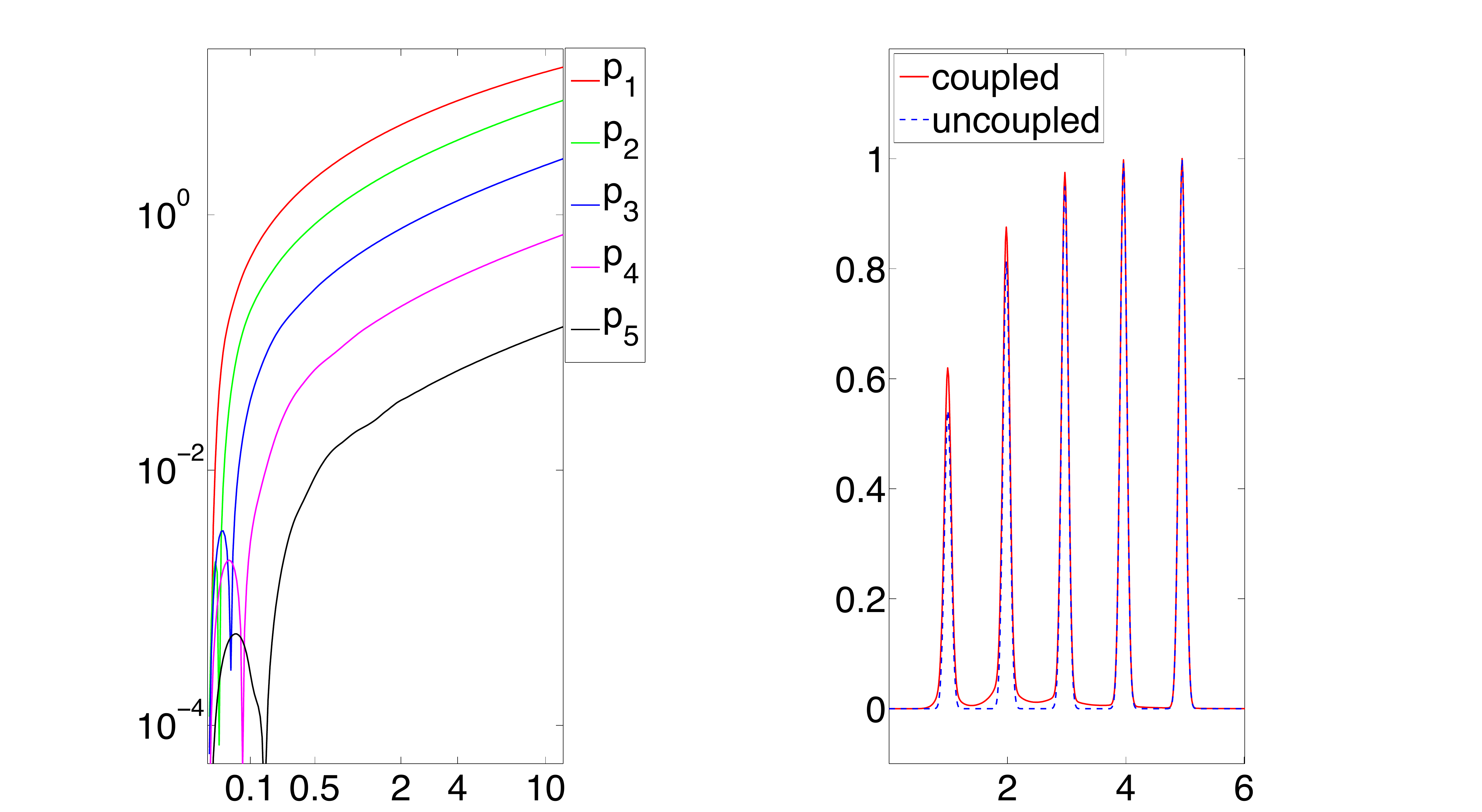}}  \end{tabular} \\ [-0.5ex]
 \multicolumn{1}{|c|}{}  &  \multicolumn{1}{c|}{\footnotesize{}\hspace{0.9cm}Time (Gyr)\hspace{1.5cm}Distance (Gpc)} & \multicolumn{1}{c|}{\footnotesize{}\hspace{0.8cm}Time (Gyr)\hspace{1.4cm}Distance (Gpc)} \\[0.5ex]\Cline{0.3pt}{1-3}
   \multicolumn{1}{|c|}{}    &   \multicolumn{1}{c|}{}    &   \multicolumn{1}{c|}{}   \\[-2ex]
   \multicolumn{1}{|c|}{}    &  \multicolumn{1}{c|}{\footnotesize{}\hspace{0.1cm}Initial profile\hspace{1.2cm}\% diff.\hspace{1.1cm}Profile today$\!\!$}  &  \multicolumn{1}{c|}{\footnotesize{}\hspace{0.1cm}Initial profile\hspace{1.2cm}\% diff.\hspace{1.1cm}Profile today$\!\!$}  \\[-0.4ex]
\multicolumn{1}{|c|}{$\!\!\!$\parbox[t]{2.0mm}{{\rotatebox[origin=c]{90}{\hspace{0.46cm}$\Delta/\textnormal{max}|\Delta_{\textnormal{min}}|$}}}}  &  \hspace{-0.4cm}\begin{tabular}{c} \rule{0pt}{3.2cm}{\includegraphics[width=0.53\textwidth]{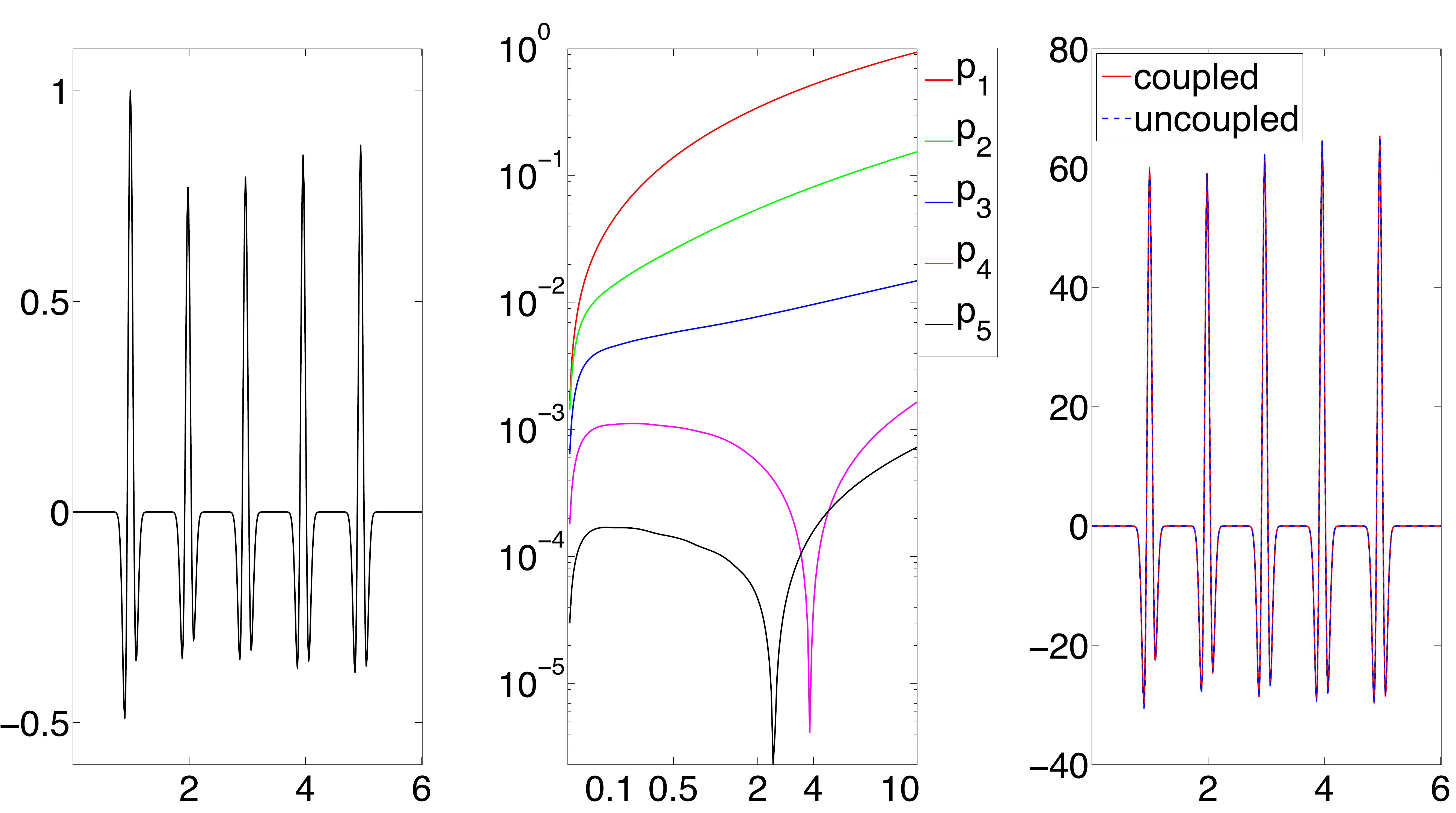}}  \end{tabular} &  \hspace{-0.4cm}\begin{tabular}{c} \rule{0pt}{3.2cm}{\includegraphics[width=0.53\textwidth]{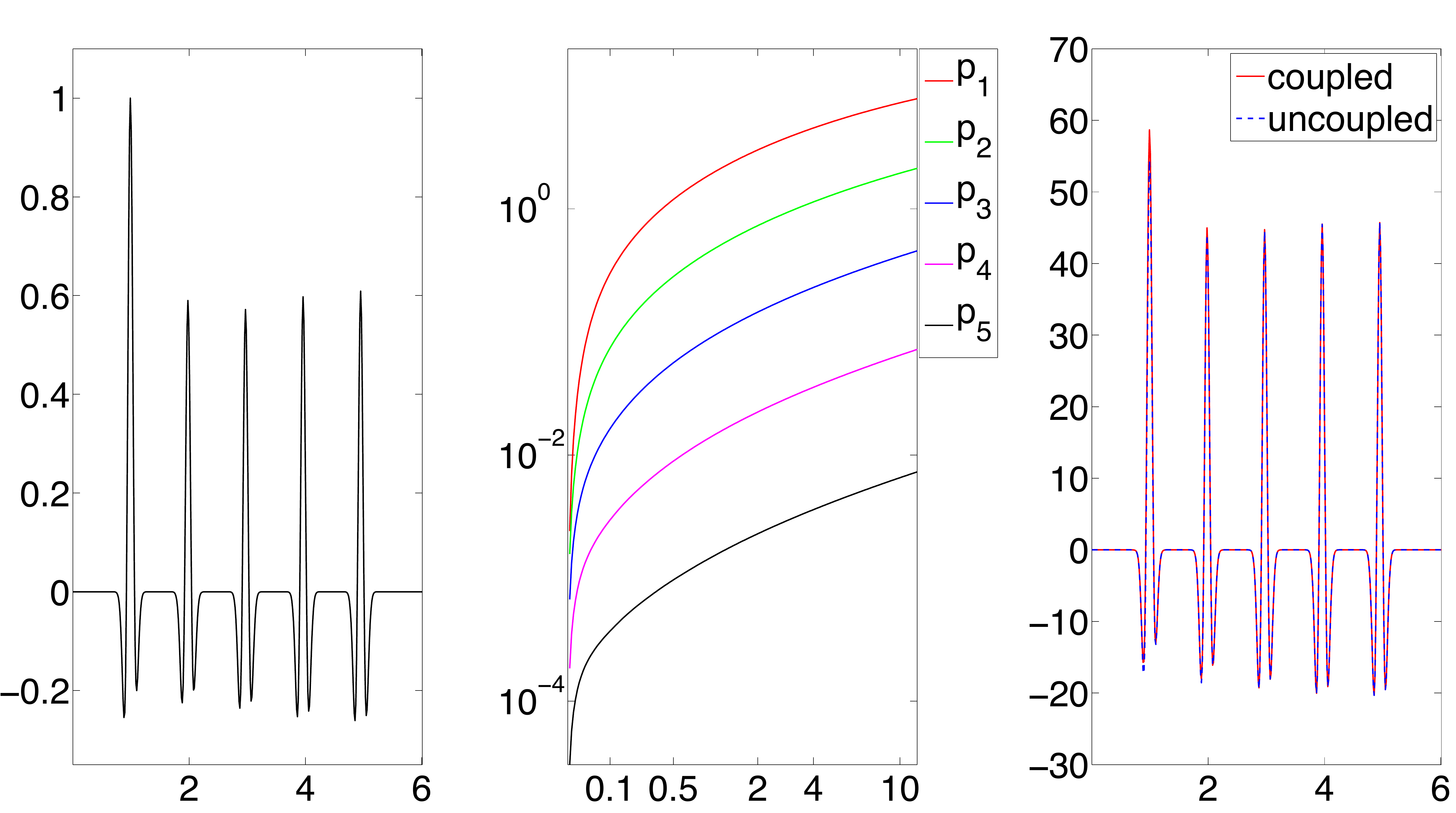}}  \end{tabular} \\ [-0.5ex]
 \multicolumn{1}{|c|}{}  &  \multicolumn{1}{c|}{\footnotesize{}\hspace{0.3cm}Distance\hspace{1.5cm}Time\hspace{1.5cm}Distance} & \multicolumn{1}{c|}{\footnotesize{}\hspace{0.3cm}Distance\hspace{1.5cm}Time\hspace{1.5cm}Distance} \\ \Cline{0.3pt}{1-3}
 \multicolumn{1}{c}{} & \multicolumn{2}{c}{}\\[-3.0ex]
\end{tabular}
\caption{Comparison of coupled to uncoupled runs, for cases 1 and 2. On the largest scales ($l=2$), and deep within the void ($p_1$), $\varphi$ (top panel) is enhanced by $\sim$ 10\% when the coupling is present, while $\Delta$ (bottom panel; normalised to its maximum value in space at the initial time, $t_{\textnormal{min}}$) is enhanced by $\sim$ 1\%. As we approach the outskirts of the void, the differences are sub-percent, as expected.}
\label{Tab:CoupledVsDecoupled}
\end{figure*}
\item[Cases 5 and 6:] Here, we initialise $\chi$, and set $\varphi=\varsigma=0$ initially.

According to the bottom panels of Fig.~\ref{Tab:MasterTab}, $\chi$ and the generated $\varsigma$ propagate to the outskirts of the void along the characteristics of the background, resulting in the localised generation of the potential $\varphi$, and the associated growth of density perturbations, as is shown in Fig.~\ref{Tab:DerivedQuantities}.

The bottom panel of Fig.~\ref{Tab:Case5and6Tab} presents the profile of $\chi$ today for these cases, as well as the time evolution along selected radii, while the profiles for the other variables, $\varphi$ and $\varsigma$, are shown in the top and middle panels.
\end{description}

All of these cases demonstrate that $\varphi$, $\chi$ and $\varsigma$ are much more difficult to interpret than on a FLRW background. As emphasised in \cite{Clarkson:2009sc}, they are mixtures of scalar, vector and tensor modes and therefore their coupling is an essential ingredient of first order perturbation theory around a LTB background: in principle, they cannot be treated as separate, independent modes that describe different physical aspects of perturbations. In the next subsection, we compare the behaviours of the fully coupled perturbation system to cases where they are decoupled ``by hand'' as has been done before in various ways to simplify the analysis~\cite{Zibin:2008vj,Dunsby:2010ts,February:2012fp}. In particular, we would like to analyse the errors in $\varphi$ and $\Delta$ when $\chi$ and $\varsigma$ are neglected, not only for initial data but also all during the evolution of the system.

\subsection{Coupled vs uncoupled dynamics}
\label{Subsection:CoupvsUnCoup}
In this section, we quantify the errors induced when assuming that the coupling of $\varphi$ to $\chi$ and $\varsigma$ is negligible  by considering models which initialise $\varphi$ only. We compare these to cases where Eqs.~\Eref{varphi_evol_eq}, \Eref{varsigma_evol_eq} and \Eref{chi_evol_eq} are solved retaining terms with no coupling between $\varphi$ and $\{\chi,\varsigma\}$, that is, by solving the reduced system:
\ba
\qquad \;\; \ddot{\varphi}  &=& - 4\Hperp\dot{\varphi}  + \bigg[\frac{2\kappa}{\aperp^2}  - \Lambda \bigg]\varphi \\
8\pi G \rhom \Delta &=&  - X^{-2}\varphi'' + X^{-2}\Bigg[\frac{{\apar}'}{\apar} + \frac{\kappa r + \frac{1}{2}r^2\kappa'}{1-\kappa r^2}- \; -2\frac{\apar}{\aperp r}\Bigg] \varphi'\nonumber \\
&&  + \Theta\dot{\varphi} +  \Bigg[3\Hperp\Big(\sigma^2 + \Hperp \Big)- \Big(1+2\frac{\aperp}{\apar}\Big)\frac{\kappa}{\aperp^2} - \frac{r\kappa'}{\aperp\apar} \nonumber \\
&&  + \frac{l(l+1)}{\aperp^2 r^2}\Bigg]\varphi. 
\ea
\begin{figure*}[!t]
\centering
\begin{tabular}{|m{6.7cm}|m{6.7cm}|}
\multicolumn{2}{c}{\large{Percentage errors on $\varphi$ and $\Delta$ when neglecting the full coupling}}\\
\Cline{0.3pt}{1-2}
\multicolumn{1}{|c|}{\hspace{0.3cm}\small{}1st peak\hspace{1.25cm}3rd peak \hspace{0.85cm} 5th peak} & \multicolumn{1}{|c|}{\hspace{0.3cm}\small{}1st peak\hspace{1.25cm}3rd peak \hspace{0.85cm} 5th peak} \\ \Cline{0.3pt}{1-2}    
\multicolumn{1}{|c|}{} & \multicolumn{1}{c|}{} \\[-2.2ex]
\multicolumn{1}{|c|}{\hspace{0.55cm}$\small{}\mathbf{t=\frac{1}{4}t_0}$} & \multicolumn{1}{c|}{\hspace{0.55cm}$\small{}\mathbf{t=\frac{1}{2}t_0}$} \\
   \hspace{-0.5cm}  \begin{tabular}{c} \rule{0pt}{3.0cm}{\includegraphics[width=0.55\textwidth]{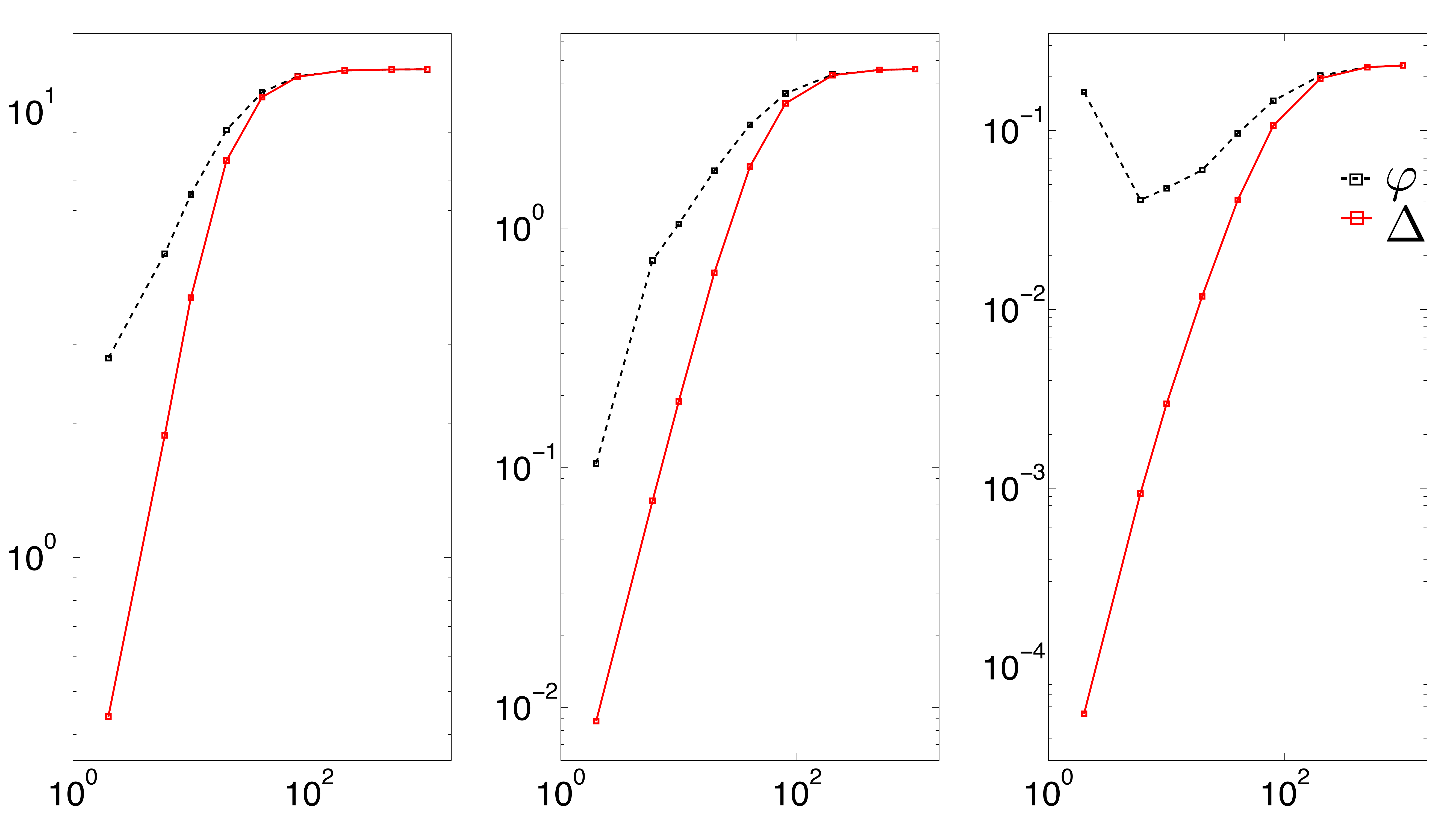}}  \end{tabular} &   \hspace{-0.5cm} \begin{tabular}{c} \rule{0pt}{3.0cm}{\includegraphics[width=0.55\textwidth]{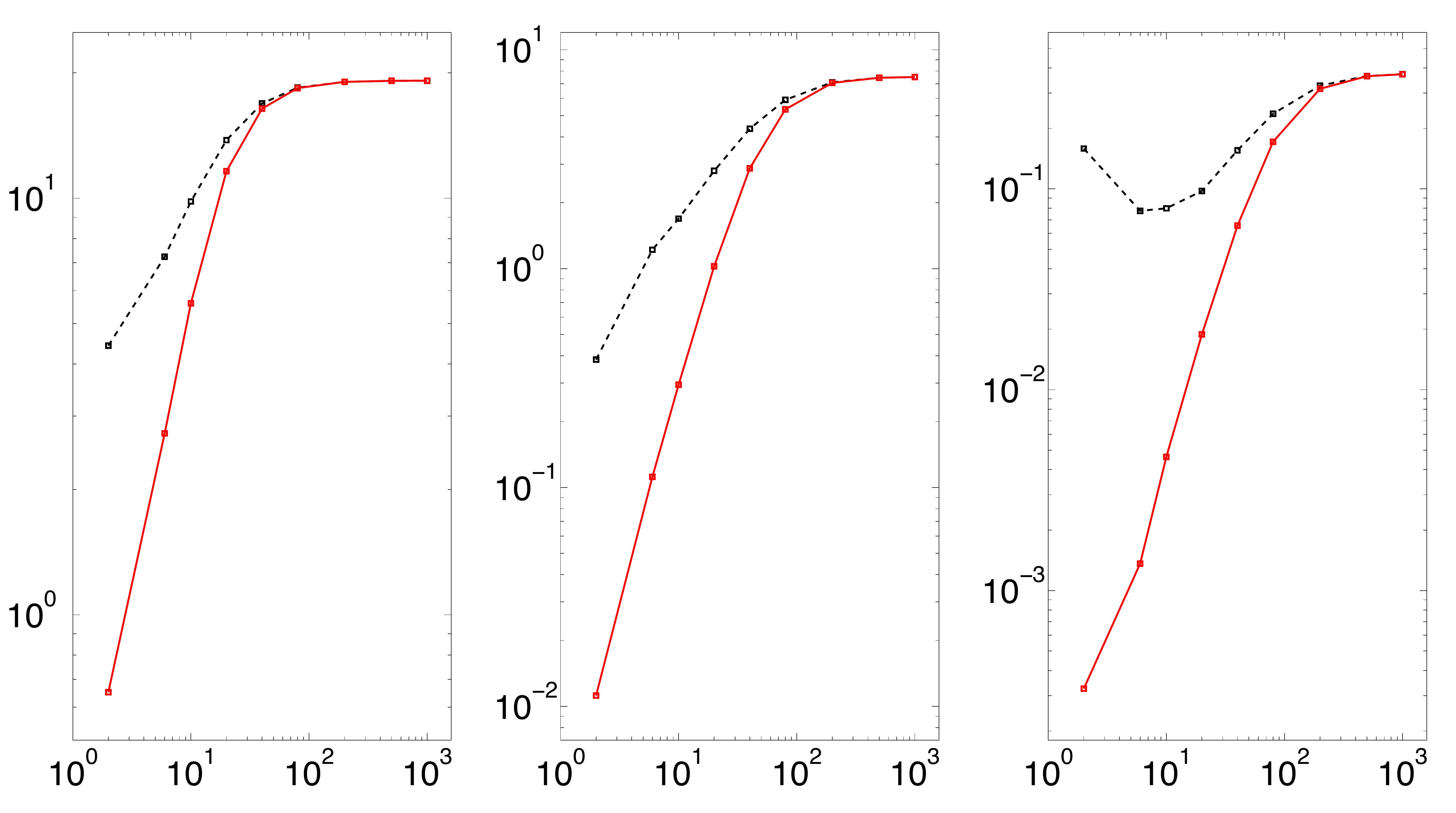}}  \end{tabular}  \\ \Cline{0.3pt}{1-2}
    \multicolumn{1}{|c|}{} & \multicolumn{1}{c|}{} \\[-2.2ex]
\multicolumn{1}{|c|}{\hspace{0.55cm}$\small{}\mathbf{t=\frac{3}{4}t_0}$} & \multicolumn{1}{c|}{\hspace{0.55cm}$\small{}\mathbf{t=t_0}$} \\
    \hspace{-0.5cm} \begin{tabular}{c} \rule{0pt}{3.5cm}{\includegraphics[width=0.55\textwidth]{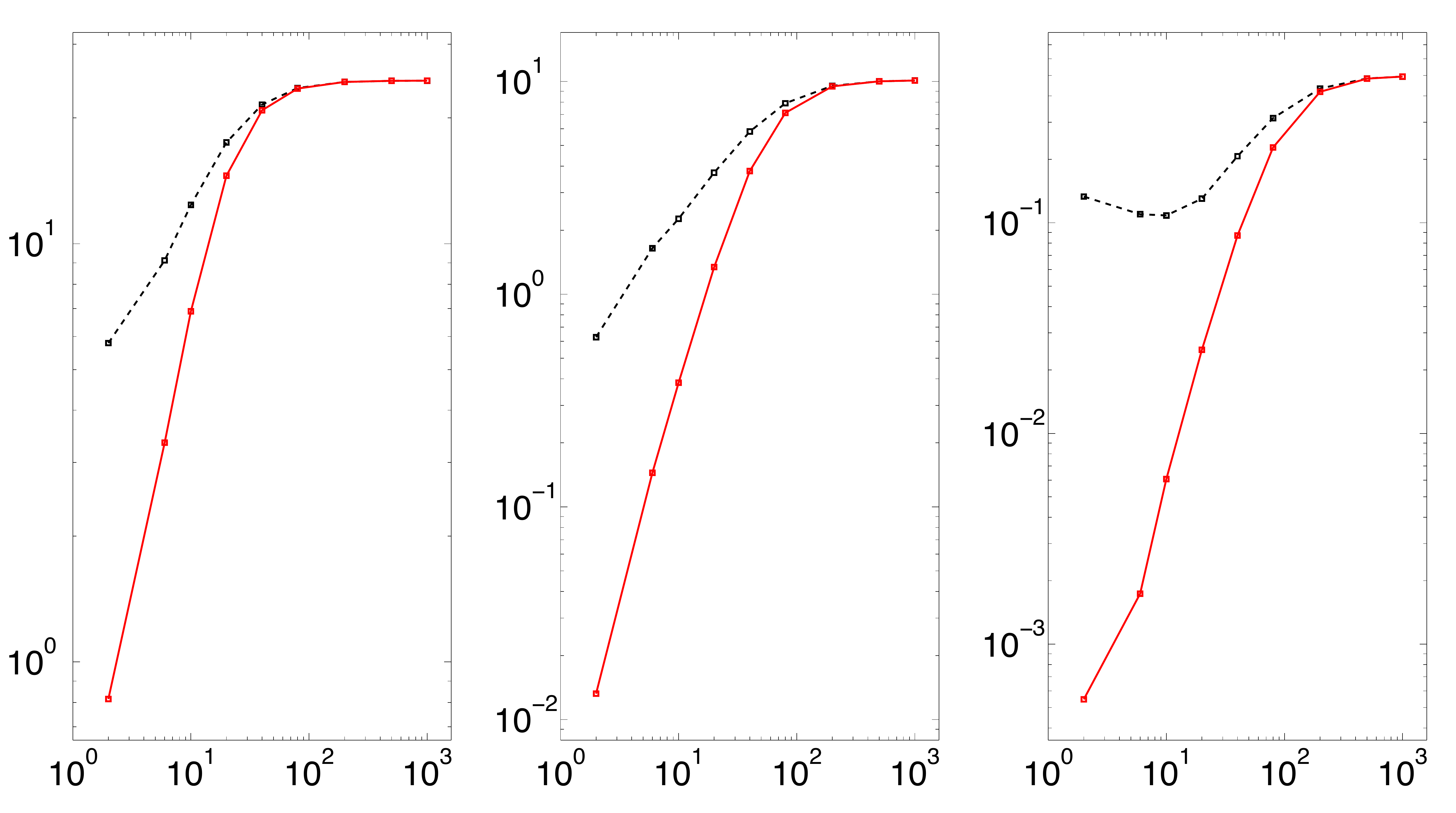}}  \end{tabular} &   \hspace{-0.5cm} \begin{tabular}{c} \rule{0pt}{3.5cm}{\includegraphics[width=0.55\textwidth]{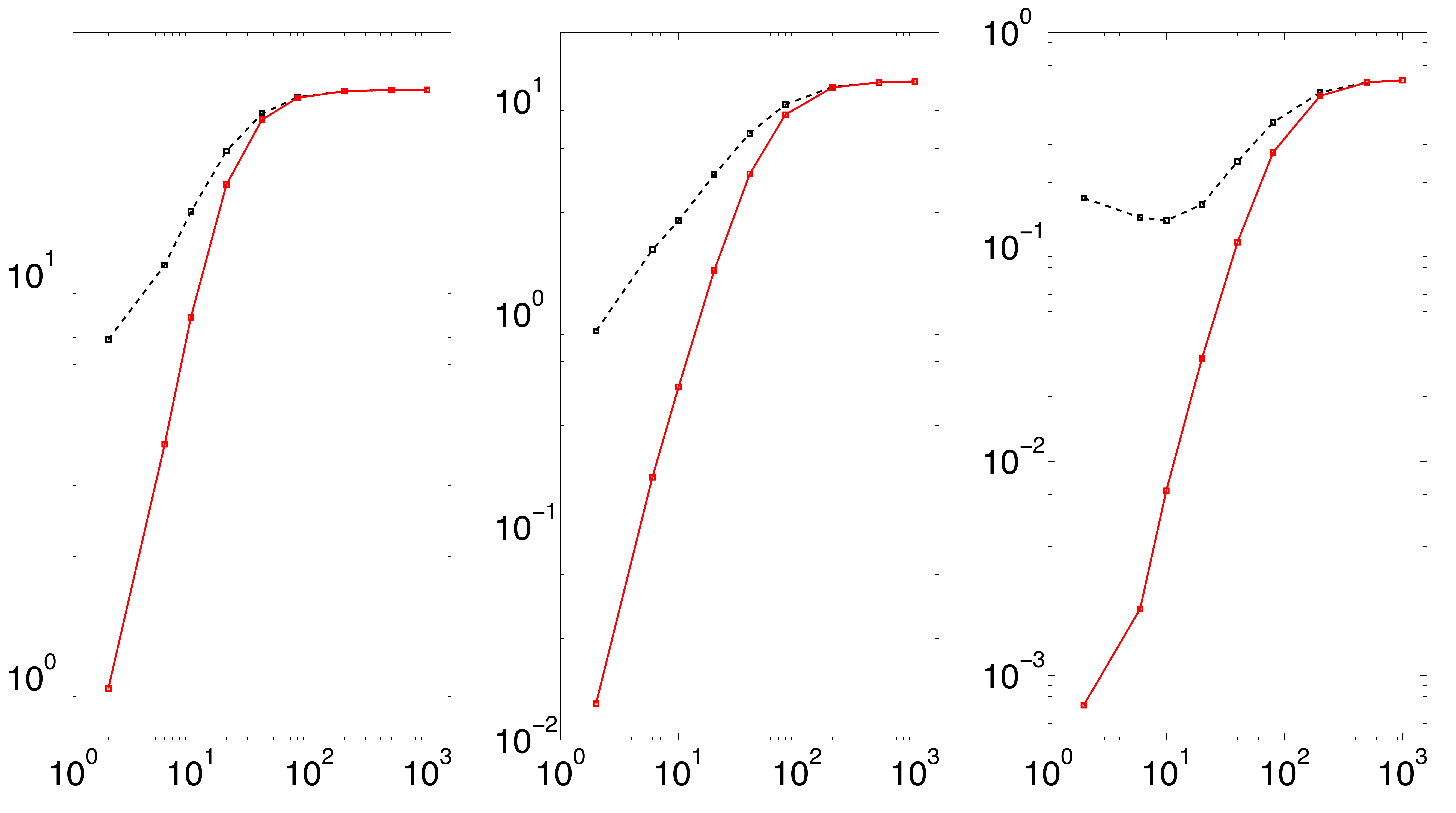}}  \end{tabular}  \\ \Cline{0.3pt}{1-2}
\multicolumn{1}{c}{} & \multicolumn{1}{c}{} \\[-1.5ex]
  \multicolumn{2}{c}{Multipole moment $l$ $\longrightarrow$}  \\ [-3.0ex]
\end{tabular}
\caption{Percentage errors acquired on $\varphi$ (black, dashed line and squares) and $\Delta$ (red, solid line and squares) from neglecting the coupling between the modes in the case of an initial $\varphi$, as a function of $l$ at selected times and radii. We see in general that the errors increase with time, as well as with increasing $l$, and are larger deep within the void than towards the outskirts. The errors on $\varphi$ and $\Delta$ converge on smaller scales since the term in Eq.~\Eref{poiss_LTB} proportional to $l(l+1) \varphi$ dominates. Note that by $t=t_0$ we reach errors of around 30\% well within the void, and thus in general can expect percent-level corrections to, say, the amplitude of the baryon acoustic oscillation (BAO) bump in the two-point correlation function of the galaxy distribution.}
\label{Tab:CouplingVsL}
\end{figure*}
As it turns out, the full coupling is seen to be important for the dynamics of $\varphi$, and also for the behaviour of $\Delta$ on small angular scales (large $l$) $-$ see Figs.~\ref{Tab:CoupledVsDecoupled} and~\ref{Tab:CouplingVsL}.

From Fig.~\ref{Tab:CoupledVsDecoupled} we see that deep inside the void (first peak) the differences in $\varphi$ are already of order $15\%$ for $l=10$ and $8\%$ for $l=2$. This could have a major impact down the central observer's past light-cone and therefore such couplings could be very important in determining observables accurately. On the other hand, $\Delta$ is well approximated by the uncoupled dynamics for large scales, with errors below 1\% for $l=2$; but, already for $l=10$, we see errors of order 7 to 8\%.

Including a few more angular scales, all the way up to $l=1000$, as well as intermediate snapshots in time, an overall picture of the error in neglecting the couplings is captured in Fig.~\ref{Tab:CouplingVsL}. Regardless of where (in radial distance) we choose to observe $\varphi$ and $\Delta$, as we go to smaller scales their expected errors approach some equivalent maximum value $-$ equivalent due to their relation via the analogue of the Poisson equation \Eref{poiss_LTB} which has $\Delta \propto l^2 \varphi$ on small scales (large $l$). 

As for observable quantities such as the two-point correlation function of the galaxy distribution, we should expect corrections of a few percent in the amplitude of the BAO bump when including the full coupling (because this quantity is of order the square of $\Delta$) $-$ see \cite{February:2012fp} for the particular case in which the coupling is neglected.

\section{Conclusions and perspectives}
We have developed a numerical scheme to solve the system of coupled, linear partial differential equations describing the evolution of (polar) perturbations on a background LTB spacetime. The implementation is numerically consistent, attaining the expected 2nd-order convergence with resolution over a wide range of scales.

To illustrate the nature of the coupling between the three master variables in the problem, in separate runs we initialised the
data by several Gaussian peaks in each variable, spanning regions both inside and outside the void while setting the remaining two variables to zero initial amplitude. Initial pulses in $\varphi$ result in growth of $\varsigma$ and $\chi$ at the sub-percent level, implying that the variable $\varphi$ $-$ commonly ascribed to the analogue of the Bardeen/Newtonian potential $-$ nevertheless contains relativistic degrees of freedom. Initialising non-zero $\varsigma$ induces a sub-percent signal in $\varphi$ and $\chi$, all while decaying roughly as $\apar^{-2}$ $-$ analogous, but not equivalent, to the vector mode in a FLRW spacetime. Finally, a non-zero $\chi$ induces a $\varphi$ to the level of nearly 50\% today, while inducing only a sub-percent level of $\varsigma$ (from a maximum level of $\sim$ 20\% at earlier times). The propagating nature of $\chi$ is clearly seen in this case.

We also investigated whether the coupling between the master variables may be safely ignored. In particular, we focused on the case of an initialised $\varphi$, and considered how much error we expect to obtain on $\Delta$ and $\varphi$ when neglecting the coupling of $\varphi$ to $\varsigma$ and $\chi$. Our results indicate that, well inside the void and on the largest scales, the errors picked up on $\Delta$ are at the sub-percent level, and so neglecting the coupling in that case is not an unreasonable assumption.  However, the corresponding corrections to $\varphi$ itself will be more important, and contributions from lensing and integrated Sachs-Wolfe effects may be enhanced at around the 10\% level when taking the coupling into account. On smaller scales though, corrections to the assumption of negligible coupling can grow to a few tens of percent for both $\varphi$ and $\Delta$ for regions well inside the void. For an observable such as the galaxy-galaxy correlation function, we estimate corrections to the amplitude of the BAO peak at the percent-level. Of course, since we have considered aspects of structure formation only valid in the linear regime, we expect that any non-linear effects --- the details of which is not clear at this point --- will modify small-scale corrections in some non-trivial way. In any case, as we approach the outskirts of the void corrections are well below the percent-level on all scales, as expected in regions of spacetime close to FLRW.

Having performed such a calculation for the case of a cosmological-sized void, our analysis can be easily adapted to smaller astrophysical-sized voids, and even halos. This will be left for future work.

\ack{We thank Bishop Mongwane and Roy Maartens for useful discussions. SF is supported by the South African Square Kilometre Array (SKA) Project. Computations were performed using facilities provided by the University of Cape Town ICTS High Performance Computing team.}

\appendix
\section{Runge-Kutta algorithm}
\label{sec:algorithm}
Let's write our system of coupled PDEs in the following compact way
\ba
\ddot{\varphi}  &=& \mathcal{F}^{\varphi}(t,\dot{\varphi},\varphi) + \mathcal{S}^{\varphi}(t,\chi',\dot{\chi},\chi,\varsigma) \,, \label{orig_pde_varphi} \\
\dot{\varsigma}  &=& \mathcal{F}^{\varsigma}(t,\varsigma) + \mathcal{S}^{\varsigma}(\chi') \,,\label{orig_pde_varsigma}\\
\ddot{\chi} &=& \mathcal{F}^{\chi}(t,\chi'',\chi',\dot{\chi},\chi) + \mathcal{S}^{\chi}(t,\dot{\varphi},\varphi,\varsigma',\varsigma) \label{orig_pde_chi}\,,
\ea where $\varphi$, $\varsigma$ and $\chi$ are functions of time ($t$) and radial coordinate ($r$), and a prime denotes partial differentiation with respect to $r$. Splitting \Eref{orig_pde_varphi} and \Eref{orig_pde_chi} into two pairs of first-order equations we get
\ba
\dot{\varphi} &=& \overline{\varphi} \label{varphi_split1} \,,\\
\dot{\overline{\varphi}}  &=&  \mathcal{F}^{\varphi}(t,\overline{\varphi},\varphi) + \mathcal{S}^{\varphi}(t,\chi',\overline{\chi},\chi,\varsigma) \,, \label{varphi_split2}\\ \nonumber \\
\dot{\chi} &=& \overline{\chi}  \label{chi_split1}\,,\\
\dot{\overline{\chi}}  &=&  \mathcal{F}^{\chi}(t,\chi'',\chi',\overline{\chi},\chi) + \mathcal{S}^{\chi}(t,\overline{\varphi},\varphi,\varsigma',\varsigma)  \label{chi_split2}\,.
\ea 

Discretising \Eref{orig_pde_varsigma}, \Eref{varphi_split1}, \Eref{varphi_split2}, \Eref{chi_split1} and \Eref{chi_split2} on a spacetime grid in a Runge-Kutta fashion for the time dependence we have
\ba
\varphi_{i+1,j} &=& \varphi_{i,j} + \frac{\Delta t}{6}\Big({a_1}_{i,j} + 2{a_2}_{i,j} + 2{a_3}_{i,j} + {a_4}_{i,j}\Big)\,, \label{varphi_sol}\\
\overline{\varphi}_{i+1,j} &=& \overline{\varphi}_{i,j} + \frac{\Delta t}{6}\Big({b_1}_{i,j} + 2{b_2}_{i,j} + 2{b_3}_{i,j} + {b_4}_{i,j}\Big)\,,\\\nonumber\\
\varsigma_{i+1,j} &=& \varsigma_{i,j} + \frac{\Delta t}{6}\Big({c_1}_{i,j} + 2{c_2}_{i,j} + 2{c_3}_{i,j} + {c_4}_{i,j}\Big)\,, \label{varsigma_sol}
\ea
\ba
\chi_{i+1,j} &=& \chi_{i,j} + \frac{\Delta t}{6}\Big({d_1}_{i,j} + 2{d_2}_{i,j} + 2{d_3}_{i,j} + {d_4}_{i,j}\Big)\,, \label{chi_sol}\\
\overline{\chi}_{i+1,j} &=& \overline{\chi}_{i,j} + \frac{\Delta t}{6}\Big({e_1}_{i,j} + 2{e_2}_{i,j} + 2{e_3}_{i,j} + {e_4}_{i,j}\Big)\,, 
\ea where
\ba
{a_1}_{i,j} &=& \overline{\varphi}_{i,j} \,,\\
{b_1}_{i,j} &=& \mathcal{F}^{\varphi}\Big(t_i,\overline{\varphi}_{i,j},\varphi_{i,j}\Big) + \mathcal{S}^{\varphi}\Big(t,\chi'_{i,j},\overline{\chi}_{i,j},\chi_{i,j},\varsigma_{i,j}\Big) \,,\\
{c_1}_{i,j} &=& \mathcal{F}^{\varsigma}\Big(t_i,\varsigma_{i,j}\Big) + \mathcal{S}^{\varsigma}\Big(t,\chi'_{i,j} \Big) \,,\\
{d_1}_{i,j} &=& \overline{\chi}_{i,j} \,,\\
{e_1}_{i,j} &=& \mathcal{F}^{\chi}\Big(t_i,\chi''_{i,j},\chi'_{i,j},\overline{\chi}_{i,j},\chi_{i,j}\Big) + \mathcal{S}^{\chi}\Big(t_i,\overline{\varphi}_{i,j},\varphi_{i,j},\varsigma'_{i,j},\varsigma_{i,j}\Big)  \,,\\ \nonumber\\
{a_2}_{i,j} &=& {a_1}_{i,j} + {b_1}_{i,j}\Delta t/2\,,\\
{d_2}_{i,j} &=& {d_1}_{i,j} + {e_1}_{i,j}\Delta t/2 \,,\\
{b_2}_{i,j} &=& \mathcal{F}^{\varphi}\Big(t_i + \Delta t/2,{a_2}_{i,j},\varphi_{i,j}+{a_1}_{i,j}\Delta t/2\Big)+ \mathcal{S}^{\varphi}\Big(t_i+\Delta t/2, \nonumber \\
&& \chi'_{i,j}+{d_1}'_{i,j}\Delta t/2,{d_2}_{i,j},\chi_{i,j} +{d_1}_{i,j}\Delta t/2 ,\varsigma_{i,j} + {c_1}_{i,j}\Delta t/2\Big) \,,\\
{c_2}_{i,j} &=& \mathcal{F}^{\varsigma}\Big(t_i + \Delta t/2,\varsigma_{i,j}+{c_1}_{i,j}\Delta t/2\Big) + \mathcal{S}^{\varsigma}\Big(t_i + \Delta t/2,\nonumber\\
&& \chi'_{i,j} + {d_1}'_{i,j}\Delta t/2 \Big) \,,
\ea
\ba
{e_2}_{i,j} &=& \mathcal{F}^{\chi}\Big(t_i + \Delta t/2,\chi''_{i,j}+{d_1}''_{i,j}\Delta t/2,\chi'_{i,j}+{d_1}'_{i,j}\Delta t/2,{d_2}_{i,j}, \nonumber \\
&&\chi_{i,j}+{d_1}_{i,j}\Delta t/2\Big) +  \mathcal{S}^{\chi}\Big(t_i + \Delta t/2,{a_2}_{i,j},\varphi_{i,j}+{a_1}_{i,j}\Delta t/2, \nonumber \\
&&\varsigma'_{i,j}+{c_1}'_{i,j}\Delta t/2,\varsigma_{i,j}+{c_1}_{i,j}\Delta t/2\Big) \,,
\ea
\ba
{a_3}_{i,j} &=& {a_1}_{i,j} + {b_2}_{i,j}\Delta t/2\,,\\
{d_3}_{i,j} &=& {d_1}_{i,j} + {e_2}_{i,j}\Delta t/2 \,,\\
{b_3}_{i,j} &=& \mathcal{F}^{\varphi}\Big(t_i + \Delta t/2,{a_3}_{i,j},\varphi_{i,j}+{a_2}_{i,j}\Delta t/2\Big) + \mathcal{S}^{\varphi}\Big(t_i + \Delta t/2,\nonumber\\
&&\chi'_{i,j}+{d_2}'_{i,j}\Delta t/2,{d_3}_{i,j},\chi_{i,j} +{d_2}_{i,j}\Delta t/2 ,\varsigma_{i,j} + {c_2}_{i,j}\Delta t/2\Big) \,,\\
{c_3}_{i,j} &=& \mathcal{F}^{\varsigma}\Big(t_i + \Delta t/2,\varsigma_{i,j}+{c_2}_{i,j}\Delta t/2\Big) + \mathcal{S}^{\varsigma}\Big(t_i + \Delta t/2,\nonumber\\
&&\chi'_{i,j} + {d_2}'_{i,j}\Delta t/2 \Big) \,,\\
{e_3}_{i,j} &=& \mathcal{F}^{\chi}\Big(t_i + \Delta t/2,\chi''_{i,j}+{d_2}''_{i,j}\Delta t/2,\chi'_{i,j}+{d_2}'_{i,j}\Delta t/2,{d_3}_{i,j},\nonumber\\
&&\chi_{i,j}+{d_2}_{i,j}\Delta t/2\Big) + \mathcal{S}^{\chi}\Big(t_i + \Delta t/2,{a_3}_{i,j},\varphi_{i,j}+{a_2}_{i,j}\Delta t/2,\nonumber\\
&&\varsigma'_{i,j}+{c_2}'_{i,j}\Delta t/2,\varsigma_{i,j}+{c_2}_{i,j}\Delta t/2\Big) \,,\\ \nonumber\\
{a_4}_{i,j} &=& {a_1}_{i,j} + {b_3}_{i,j}\Delta t\,,\\
{d_4}_{i,j} &=& {d_1}_{i,j} + {e_3}_{i,j}\Delta t \,,\\
{b_4}_{i,j} &=& \mathcal{F}^{\varphi}\Big(t_i + \Delta t,{a_4}_{i,j},\varphi_{i,j}+{a_3}_{i,j}\Delta t\Big) + \mathcal{S}^{\varphi}\Big(t_{i}+\Delta t,\nonumber\\
&&\chi'_{i,j}+{d_3}'_{i,j}\Delta t/2,{d_4}_{i,j},\chi_{i,j} +{d_3}_{i,j}\Delta t ,\varsigma_{i,j} + {c_3}_{i,j}\Delta t\Big) \,,\\
{c_4}_{i,j} &=& \mathcal{F}^{\varsigma}\Big(t_i + \Delta t,\varsigma_{i,j}+{c_3}_{i,j}\Delta t\Big) + \mathcal{S}^{\varsigma}\Big(t_i + \Delta t,\chi'_{i,j} + {d_3}'_{i,j}\Delta t \Big) \,,\\
{e_4}_{i,j} &=& \mathcal{F}^{\chi}\Big(t_i + \Delta t,\chi''_{i,j}+{d_3}''_{i,j}\Delta t,\chi'_{i,j}+{d_3}'_{i,j}\Delta t,{d_4}_{i,j},\nonumber\\
&&\chi_{i,j}+{d_3}_{i,j}\Delta t\Big) + \mathcal{S}^{\chi}\Big(t_i + \Delta t,{a_4}_{i,j},\varphi_{i,j}+{a_3}_{i,j}\Delta t,\nonumber\\
&&\varsigma'_{i,j}+{c_3}'_{i,j}\Delta t,\varsigma_{i,j}+{c_3}_{i,j}\Delta t\Big) \,.
\ea

\end{document}